\documentclass[a4paper,11pt]{article}
\pdfoutput=1 % if your are submitting a pdflatex (\emph{i.e.} if you have
\usepackage{jheppub} % for details on the use of the package, please
\usepackage{caption} 
\usepackage[]{color}
\usepackage{commath}

\usepackage[bottom]{footmisc}

\renewcommand{\thesection}{\arabic{section}}
\title{\boldmath Stirring a black hole}
 \newcommand\smallO{
  \mathchoice
    {{\scriptstyle\mathcal{O}}}% \displaystyle
    {{\scriptstyle\mathcal{O}}}% \textstyle
    {{\scriptscriptstyle\mathcal{O}}}% \scriptstyle
    {\scalebox{.7}{$\scriptscriptstyle\mathcal{O}$}}%\scriptscriptstyle
  }

 \author{Julija Markevi\v{c}i\={u}t\.e, }
 \author{Jorge E. Santos}
 \affiliation{Department of Applied Mathematics and Theoretical Physics, \\
 University of Cambridge, \\ Cambridge, CB3 0WA, UK}
 \emailAdd{j.markeviciute@damtp.cam.ac.uk}
 \emailAdd{j.e.santos@damtp.cam.ac.uk}
\abstract{We present novel asymptotically global AdS$_4$ solutions, constructed by turning on a dipolar differential rotation at the conformal boundary. At fixed energy and boundary profile, we find two different geometries: a horizonless spacetime, and a deformed, hourglass shaped black hole with zero net angular momentum. Both solutions exist up to some maximum amplitudes of the boundary profile, and develop an ergoregion attached to the boundary before the maximum amplitude is reached. We show that both spacetimes develop hair as soon as the ergoregion develops. Furthermore, we discuss the full phase diagram, including the possibility of phases with disconnected horizons, by  considering the  Mathisson-Papapetrou equations for a spinning test particle. Finally, we provide a first principle derivation of the superradiant bound purely from CFT data, and outline possible scenarios for the late time evolution of the system.}
\begin{document} 
\maketitle
%%%%%%%%%%%%%%%%%%%%%%%%%
%%%%%%%%%%%%%%%%%%%%%%%%%  
\section{\label{sec:intro}Introduction}
%%%%%%%%%%%%%%%%%%%%%%%%%
%%%%%%%%%%%%%%%%%%%%%%%%%  
\noindent\indent The behaviour of quantum field theories (QFTs) on a fixed curved spacetime is largely unknown territory, posing both conceptual and technical challenges. Perhaps the only known universal phenomenon is the emission of Hawking radiation around asymptotically flat black holes, and its concomitant information paradox \cite{Hawking:1974rv}. However, even in the absence of horizons, interesting novel phenomenology can arise when considering quantum field theory on curved spacetimes, such as vacuum polarisation or particle production.

Much of our current understanding of QFTs on fixed nontrivial backgrounds comes from perturbation theory. The reasons for this are twofold: we do not know how to analytically study generic strongly coupled theories and we do not yet have a lattice formulation for a QFT on a generic curved spacetime. This in turn implies that questions that involve strongly coupled phenomenology, such as confinement, are out of reach and might bring novel effects. The gauge/gravity duality \cite{Maldacena:1997re,Gubser:1998bc,Witten:1998qj,Aharony:1999ti} offers a unique opportunity to study such scenarios.

The gauge/gravity duality maps a class of $d-$dimensional string theories with anti de-Sitter boundary conditions (AdS), to certain non-gravitational conformal quantum field theories (CFTs) formulated in $\tilde{d}$ dimensions, with $\tilde{d}<d$. CFTs typically come with two parameters: a dimensionless coupling $\lambda$ which measures the interaction of the microscopic constituents of the theory, and a measure of its degrees of freedom, which we will denote by $N$. One of the realisations in \cite{Maldacena:1997re} was the fact that in the limit where $\lambda\to+\infty$ and $N\to+\infty$, the corresponding string theory duals become classical supergravity theories. This is precisely the regime where the dual CFT becomes strongly coupled, and where we expect interesting new phenomena to arise.

A seemingly unrelated topic is that of \emph{superradiance}, the study of which was pioneered by \cite{Zeldovich:1971,Zeldovich:1972,Press:1972zz,Starobinsky:1973scalar,Starobinsky:1973,Detweiler:1973zz,Zouros:1979iw,Detweiler:1980uk} in the asymptotically flat context, and the first quantitative study involving rotating black holes in AdS featured in \cite{Kunduri:2006qa,Dias:2011at,Dias:2013sdc,Cardoso:2013pza}. The idea behind the so called superradiant instability is simple: rotating black holes in AdS with sufficient angular velocity $\Omega_H$ can have ergoregions, which can act as reservoirs for negative energy if waves with azimuthal quantum number $m$ and frequency $0<\omega<m\,\Omega_H$ are scattered. If a rotating black hole with an ergoregion is surrounded by a confining potential (in our case this is provided by the AdS potential, but it could be equally provided by a mass term \cite{Detweiler:1973zz,Herdeiro:2014goa} for asymptotically flat spacetimes), then this process will be recurrent and will eventually lead to an instability. The endpoint of this instability seems to be outside the scope of current numerical methods in AdS, however, in \cite{Dias:2011at,Dias:2015rxy,Niehoff:2015oga} it has been argued that the instability is likely to lead to a pointwise violation of the weak cosmic censorship conjecture in AdS.

One might have expected \emph{superradiance} to produce an interesting phase diagram for CFTs in compact spacetimes, via the gauge/gravity duality. While this is certainly true in the context of the microcanonical ensemble, the grand-canonical ensemble seems rather innocuous. The reason for this is that the region of moduli space unstable to superradiance is cloaked by a first order Hawking-Page phase transition \cite{Hawking:1982dh}. This implies that, for a fixed temperature and angular velocity that render the Kerr-AdS black hole unstable, the preferred phase is that of thermal AdS!

\begin{figure}
\centering
  \begin{minipage}[t]{0.27\textwidth}
    \includegraphics[width=\textwidth]{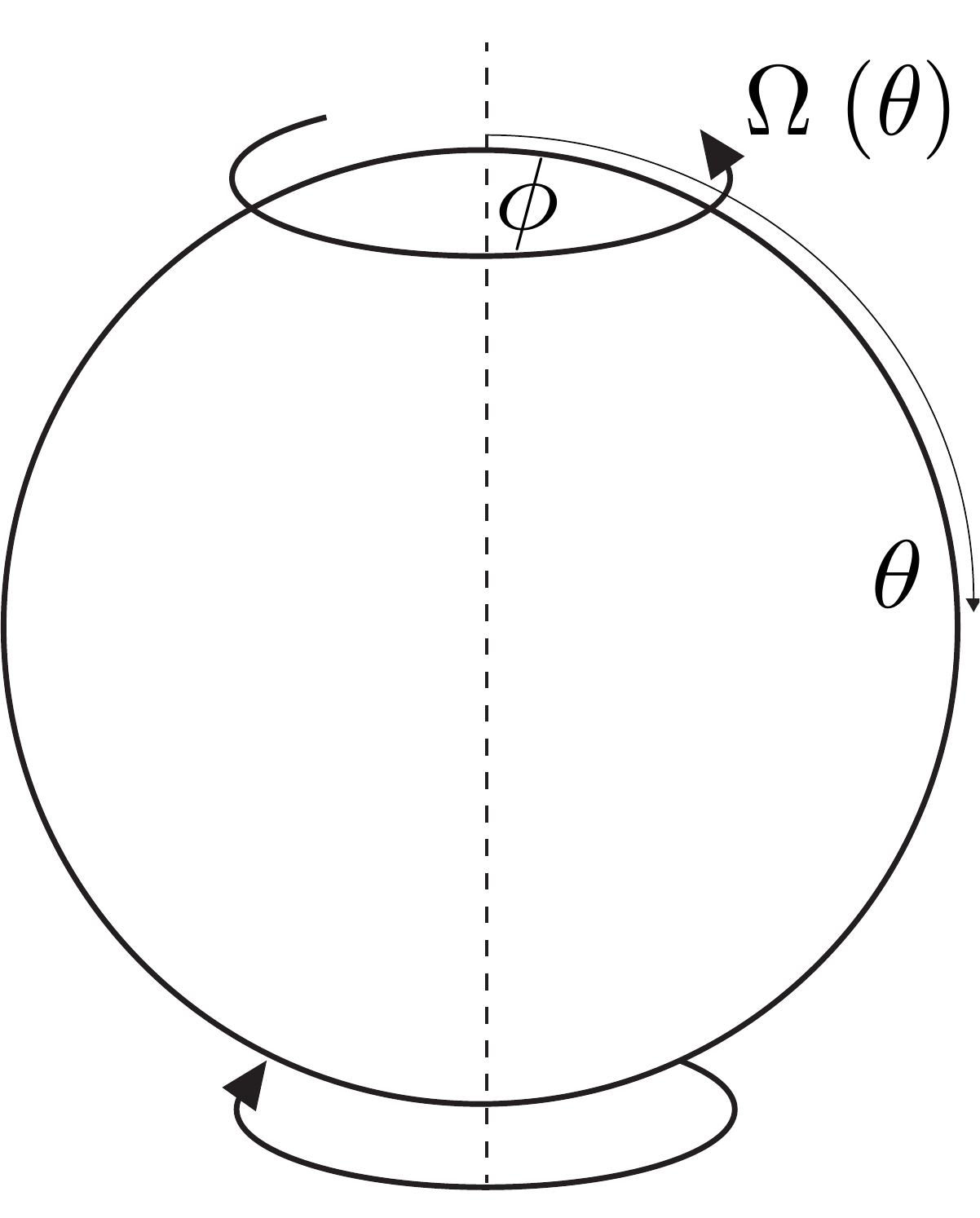}
  \end{minipage}
 \hspace{+10em}
    \begin{minipage}[t]{0.27\textwidth}
    \includegraphics[width=\textwidth]{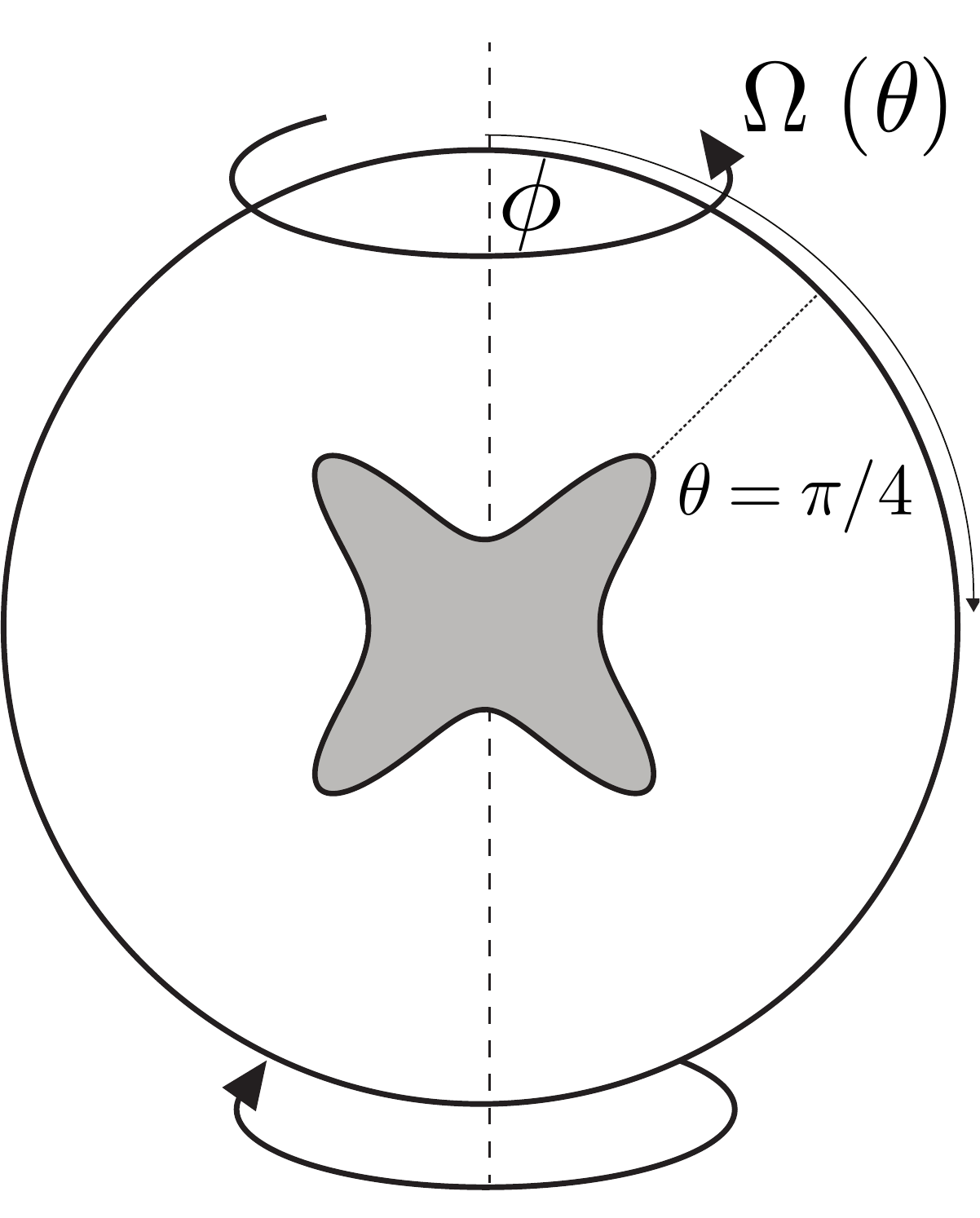}
  \end{minipage}
  \caption{Illustration of the novel AdS$_4$ geometries. \textit{Left}: Solitonic solutions with a differentially rotating boundary. \textit{Right}: Black hole solutions. The rotational pull force is maximal at $\theta=\pi/4$, deforming the black hole horizon into an hourglass shape.}
  \label{fig:pic}
\end{figure}

We shall see that this is no longer the case if we consider nontrivial boundary metrics that have differential rotation. For each boundary profile metric, we find two classes of solutions: a solitonic horizonless spacetime and black hole solutions, which are schematically pictured in Fig.~\ref{fig:pic}.
The solitonic solutions have no horizon, and correspond to deformations of global AdS$_4$, while black hole solutions can be seen as continuous deformations of Schwarzschild-AdS$_4$ black holes. We will find that both phases develop hair due to superradiance
if the boundary deformation is sufficiently large. In addition, we shall see that the phase diagram of solutions is rather intricate, including certain regions of moduli space with up to six-fold degeneracy.

One might wonder about the interpretation of our results in terms of CFT data. We shall see that we will be able to reconstruct the critical value of the amplitude for which the system develops hair from pure CFT data. This will allow us to identify which CFT observables might be sensitive to this phenomenon and to the hypothetical violation of weak cosmic censorship in the bulk.

For simplicity, our work will focus on four-dimensional bulk physics and can easily be embedded into a full string theory embedding, as for instance in ABJM \cite{Aharony:2008ug}. While some of what we describe might follow through to the higher-dimensional case, there are some important differences. For instance, for $d>4$, the Gregory-Laflamme instability might also play a role, and render some of our solutions unstable before they have a chance of becoming superradiant. We note however, that such an instability is also likely to lead to a violation of the weak cosmic censorship conjecture~\cite{Lehner:2010pn,Figueras:2015hkb, Figueras:2017zwa}. In this sense, we expect the relevant physics to be easier to dissect in $d=4$.

This paper is organised as follows. Section~\ref{sec:num} details the construction of soliton and black hole solutions. In  section~\ref{sec:res} we present results, and carry out thermodynamical analysis. Section~\ref{sec:sta} presents stability analysis in which we consider a massless scalar field. Section~\ref{sec:fie} discusses the CFT interpretation of our results and attempts to identify the field theory variables that best describe it. In section~\ref{sec:eqv}, we study possible multi horizon equilibrium configurations for spinning test particles. Finally, in section~\ref{sec:noergo}, we briefly discuss the holographic dual of theories living on a fixed rotating background geometry containing no ergoregions. We conclude in section~\ref{sec:con} with discussion and an outline for further work.

%%%%%%%%%%%%%%%%%%%%%%%%%
%%%%%%%%%%%%%%%%%%%%%%%%%
\section{\label{sec:setup}Setup}
%%%%%%%%%%%%%%%%%%%%%%%%%
%%%%%%%%%%%%%%%%%%%%%%%%%
\noindent\indent We start with the four-dimensional Einstein-Hilbert action supplemented by the Gibbons-Hawking-York boundary term \cite{York:1972sj,Gibbons:1976ue}:
\begin{align}
 S=\frac{1}{16\pi G_N}\int_{\small{\mathcal{M}}} \mathrm{d}^4x&\sqrt{-g}\left(R[g]+\frac{6}{L^2}\right)+
 \frac{1}{8\pi G_N}\int_{\small{\partial\mathcal{M}}} \mathrm{d}^3x \sqrt{-\gamma}\,K,
 \label{eq:action}
\end{align}
where $\partial \mathcal{M}$ is the boundary of AdS$_4$, $\gamma_{\mu\nu}$ is the induced metric on the boundary, $K$ is the trace of the extrinsic curvature of the boundary, $L$ is the length scale of AdS$_4$ and $R[g]$ the Ricci scalar. The equations of motion derived from (\ref{eq:action}) read
\begin{align*}
R_{\mu\nu}+\dfrac{3}{L^2}g_{\mu\nu}=0,
\end{align*} 
and we are looking for asymptotically AdS$_4$ solutions, with a rotating conformal boundary given by
\begin{align}
\mathrm{d}s^2_\partial=-\mathrm{d}t^2+\mathrm{d}\theta^2+\sin^2{\theta}\left[\mathrm{d}\phi+\Omega(\theta)\mathrm{d}t\right]^2,
\label{eq:boundary}
\end{align} 
\noindent with a dipolar differential rotation profile $\Omega(\theta)=\varepsilon\cos{\theta}$. We will be interested in finding solutions where $\partial_t$ and $\partial_\phi$ extend as Killing fields into the bulk. Close to the poles and on the equator, $\partial_t$ is always timelike for any value of $\varepsilon$. However, its norm
\begin{equation}
\norm{\partial_t}^2=-1+\frac{\varepsilon^2}{4}\sin^2{2\theta} 
\end{equation}
is maximal at $\theta=\pi/4$, and $\partial_t$ becomes spacelike for certain regions of $\theta$ if ${\varepsilon>2}$. We shall see that both black holes and solitons develop hair due to superradiance if $\varepsilon>2$. One might wonder whether this could have been anticipate using the results of Green, Hollands, Ishibashi and Wald~\cite{Green:2015kur}. However, the theorems presented in \cite{Green:2015kur} were deduced under the assumption that the boundary metric was metrically that of the Einstein static universe, which is not the case in our setup. Our results suggest that a generalisation of the results in \cite{Green:2015kur} should exist even for nontrivial boundary metrics such as ours.
%%%%%%%%%%%%%%%%%%%%%%%%%
%%%%%%%%%%%%%%%%%%%%%%%%%
\section{\label{sec:num}Numerical Construction}
%%%%%%%%%%%%%%%%%%%%%%%%%
\subsection{\label{subsec:nummet}Numerical Method}
%%%%%%%%%%%%%%%%%%%%%%%%%
To solve the Einstein equation we employ the DeTurck method~\cite{Headrick:2009pv,Wiseman:2011by} (for a review see~\cite{2016CQGra..33m3001D}) where we instead solve the equivalent Einstein-DeTurck equation
\begin{equation}
\label{eq:deturck} 
R_{\mu\nu}+\dfrac{3}{L^2}g_{\mu\nu}-\nabla_{(\mu}\xi_{\nu)}=0, 
\end{equation}
where $\xi^{\mu}=g^{\nu\rho}\left(\Gamma^\mu_{\nu\rho}[g]-\Gamma^\mu_{\nu\rho}[\tilde{g}]\right)$ is the DeTurck vector and $\tilde{g}$ is an appropriate reference metric of our choice (as detailed in the following subsection). Provided that $\xi^\mu=0$, solutions to~(\ref{eq:deturck}) will also be solutions of the Einstein equation. Fortunately, it has recently been shown that, for spacetimes possessing a $(t,\phi)\to-(t,\phi)$ reflection symmetry, solutions with $\xi\neq0$ cannot exist \cite{Figueras:2016nmo} - our spacetimes enjoy such a symmetry. We can thus monitor the norm of $\xi$ to provide a global measure of our numerical integration scheme (see Appendix~\ref{sec:numval}). 

The DeTuck method has the advantage of rendering the equations elliptic, and conveniently fixing the gauge. We solve~\eqref{eq:deturck} using the Newton-Raphson method with pseudospectral collocation on a square half--half Chebyshev grid. Because our ansatz is written in such a way that all unknown functions are even about two of the grid edges (see the next subsection), we can halve the Chebyshev grid while effectively interpolating over the full grid. This in turn allows us to use a smaller grid size to achieve a desired numerical accuracy. Convergence of our solutions is presented in Appendix~\ref{sec:numval} (Fig.~\ref{fig:solc}).
%%%%%%%%%%%%%%%%
%%%%%%%%%%%%%%%%
\subsection{\label{subsec:ans}Ans\"atze}
%%%%%%%%%%%%%%%%
\subsubsection{\label{subsubsec:sol}Soliton}
%%%%%%%%%%%%%%%%
In order to find solitonic solutions we consider the following numerical ansatz
\begin{multline}
\mathrm{d}s^2=\frac{L^2}{(1-y^2)^2}\Bigg\{-Q_1\,\mathrm{d}t^2+\frac{4\, Q_2\,\mathrm{d}y^2}{2-y^2}+y^2 (2 - y^2) \Bigg[\frac{4\,Q_3}{2-x^2}\left(\mathrm{d}x+\frac{x}{y} \sqrt{2-x^2}\,\,Q_4\, \mathrm{d}y\right)^2 \\+(1-x^2)^2 Q_5\,\left(\mathrm{d}\phi+y x\sqrt{2-x^2}\,Q_6\,\mathrm{d}t\right)^2 \Bigg]\Bigg\}\,,
\label{eq:ansatzsol}	
\end{multline}
\noindent with $Q_i$, $i\in\{1,\ldots,6\}$, being functions of $(x,y)$. If $Q_1=Q_2=Q_3=Q_5=1$ and $Q_4=Q_6=0$, the line element (\ref{eq:ansatzsol}) reduces to that of pure AdS$_4$ in global coordinates, with $y$ being related to the usual radial coordinate in AdS$_4$ via $r = L\,y \sqrt{2 - y^2}/(1 - y^2)$, and $x$ parametrising the standard polar angle on $S^2$ with the identification $\sin\theta = 1-x^2$. The powers of $y$ and $1-x^2$ appearing in the ansatz ensure that the metric is regular at the origin $y=0$ and axis of rotation $x=\pm1$ for smooth functions $Q_i$.

The reference metric $\tilde{g}$ used in the DeTurck method is chosen by setting $Q_6=\varepsilon$, and $Q_1=Q_2=Q_3=Q_5=Q_4+1=1$, everywhere in the bulk. The parameter $\varepsilon$ controls the rotation on the boundary and in turn the stirring of the bulk. 

As we are considering a dipolar problem we have reflection symmetry $x\to -x$ and thus both coordinates take values in the square domain $[0,1]\times[0,1]$. Because the functions~$Q_i$ are constructed to be even around $x=0$ (the equator) and $y=0$ (the origin), we can use a square half--half Chebyshev grid clustering around the $x=1$ and $y=1$ boundaries and thus effectively interpolate the functions with half the number of points. This is especially advantageous as spectral convergence is exponential with grid size. In addition, it fixes smoothness of the geometry at the reflection boundary.

This leaves us with two boundaries, one at the conformal boundary located at $y=1$, and one at $x=1$. The latter is a physical boundary, whereas the former is a fictitious boundary where regularity must be imposed. Expanding the equations of motion near $x=1$ as a power series in $(1-x)$ gives $Q_3(1,y)=Q_5(1,y)$ and $Q_4(1,y)=0$, as well as $\partial_x Q_1=\partial_x Q_2=\partial_x Q_3=\partial_x Q_4=\partial_x Q_5=0$.

At the conformal boundary, we demand the metric \eqref{eq:ansatzsol} approaches global AdS$_4$ with a rotating conformal boundary metric, that is to say, we require
\begin{equation}
Q_1(x,1)=Q_2(x,1)=Q_3(x,1)=Q_5(x,1)=Q_4(x,1)+1=1\,, \quad\text{and}\quad Q_6(x,1)=\varepsilon\,.
\label{eq:BCsyp}
\end{equation}
Comparing the conformal boundary metric obtained from our boundary conditions (\ref{eq:BCsyp}) and (\ref{eq:boundary}), gives $\Omega(\theta)=\varepsilon\cos\theta$, in standard polar coordinates on $S^2$.
%%%%%%%%%%%%%%%%
\subsubsection{\label{subsubsec:bla}Black Holes}
%%%%%%%%%%%%%%%%
For the black hole solutions we use the ansatz 
\begin{subequations}
\begin{multline} 
\mathrm{d}s^2=\frac{L^2}{(1-y^2)^2}\Bigg\{-y^2\tilde{\Delta}(y)Q_1\mathrm{d}t^2+\frac{4\,y_+^2 Q_2\,\mathrm{d}y^2}{\tilde{\Delta}(y)}+y_+^2 \Bigg[\frac{4\,Q_3}{2-x^2}\left(\mathrm{d}x+x \sqrt{2-x^2}\,y\,Q_4\, \mathrm{d}y\right)^2\\+(1-x^2)^2 Q_5\,\left(\mathrm{d}\phi+y^2x\sqrt{2-x^2}\,Q_6\,\mathrm{d}t\right)^2 \Bigg]\Bigg\},
\label{eq:ansatzbh}
\end{multline}
\noindent where
\begin{equation}
\Delta(y)=(1- y^2)^2 + y_+^2 (3 - 3 y^2 + y^4)\,,\quad\text{and}\quad \tilde{\Delta}(y)= \Delta(y) \delta + y_+^2 (1 - \delta)\,,
\end{equation}
\end{subequations}
\noindent with $Q_i$, $i\in\{1,\ldots,6\}$, being functions of $(x,y)$. If $Q_1=Q_2=Q_3=Q_5=\delta=1$ and $Q_4=Q_6=0$, the line element (\ref{eq:ansatzsol}) reduces to that of Schwarzschild-AdS$_4$ black hole in global coordinates, with $y$ being related to the usual radial coordinate via  $r=L\,y_+/(1-y^2)$, and $x$ parametrising the standard polar angle on $S^2$ with the identification $\sin\theta = 1-x^2$.

The reference metric is chosen to have $Q_6 = \varepsilon$ and, similarly to the soliton case, we work on a unit square grid. Regularity at $x=1$ demands $\partial_x Q_i(x,y)=0$ ($i\ne 4$), $Q_4(x,y)=0$, and $Q_3(x,y)=Q_5(x,y)$. At the conformal boundary, we set $Q_i(x,1)=1$ for $i=1,2,3,5$, $Q_4(x,1)=0$ and $Q_6(x,1)=\varepsilon$.

In the ansatz~(\ref{eq:ansatzbh}) we have three parameters: $\varepsilon$, which sets the amplitude of the boundary rotation, and $(y_+,\delta)$, which set the black hole temperature. The Hawking temperature is computed in the usual way by requiring smoothness of the Euclidean spacetime and is given by 
\begin{equation}
T=\frac{(1-\delta )y_+^2+\delta  \left(3y_+^2+1\right)}{4 \pi  y_+}.
\label{eq:tem}
\end{equation}
For $\delta =1$, the temperature has a minimum at $y_+=1/\sqrt{3}$, coinciding with the minimal temperature of a Schwarzschild-AdS$_4$, occurring at $T_{\min}^{\mathrm{Schw}}\equiv\sqrt{3}/(2\pi)$. In general we will have two branches of solutions with the same temperature which we will refer to as \textit{large} and \textit{small} black holes. However, when $\varepsilon\neq0$, our new deformed black holes can have a minimum temperature below that of the Schwarzschild-AdS$_4$ black hole. In order to bypass this, we introduced a parameter $\delta$ which allows us to get arbitrarily close to zero temperature\footnote{This ansatz allows us to reach zero temperature, for instance, see~\cite{Markeviciute:2016ivy}, where it was used to construct hairy AdS$_5$ black holes, which in the $T\rightarrow 0$ limit approach a smooth BPS soliton.}. Computations with $T>T_{\min}^{\mathrm{Schw}}$ were performed using the ansatz where $\delta=1$, and the results with $\delta\neq 1$ enjoy similarly sufficient convergence properties as those with $\delta=1$ (see Appendix~\ref{sec:numval}).
 
%%%%%%%%%%%%%%%%%%%%%%%%%
%%%%%%%%%%%%%%%%%%%%%%%%%
\section{\label{sec:res}Numerical Results}
%%%%%%%%%%%%%%%%%%%%%%%%%
For simplicity of presentation, in all of our plots we will set $L=1$.
\subsection{\label{subsec:sol}Soliton}
%%%%%%%%%%%%%%%%%%%%%%%%%
 \begin{figure}[b]
 \centering
 \includegraphics[width=0.45\textwidth]{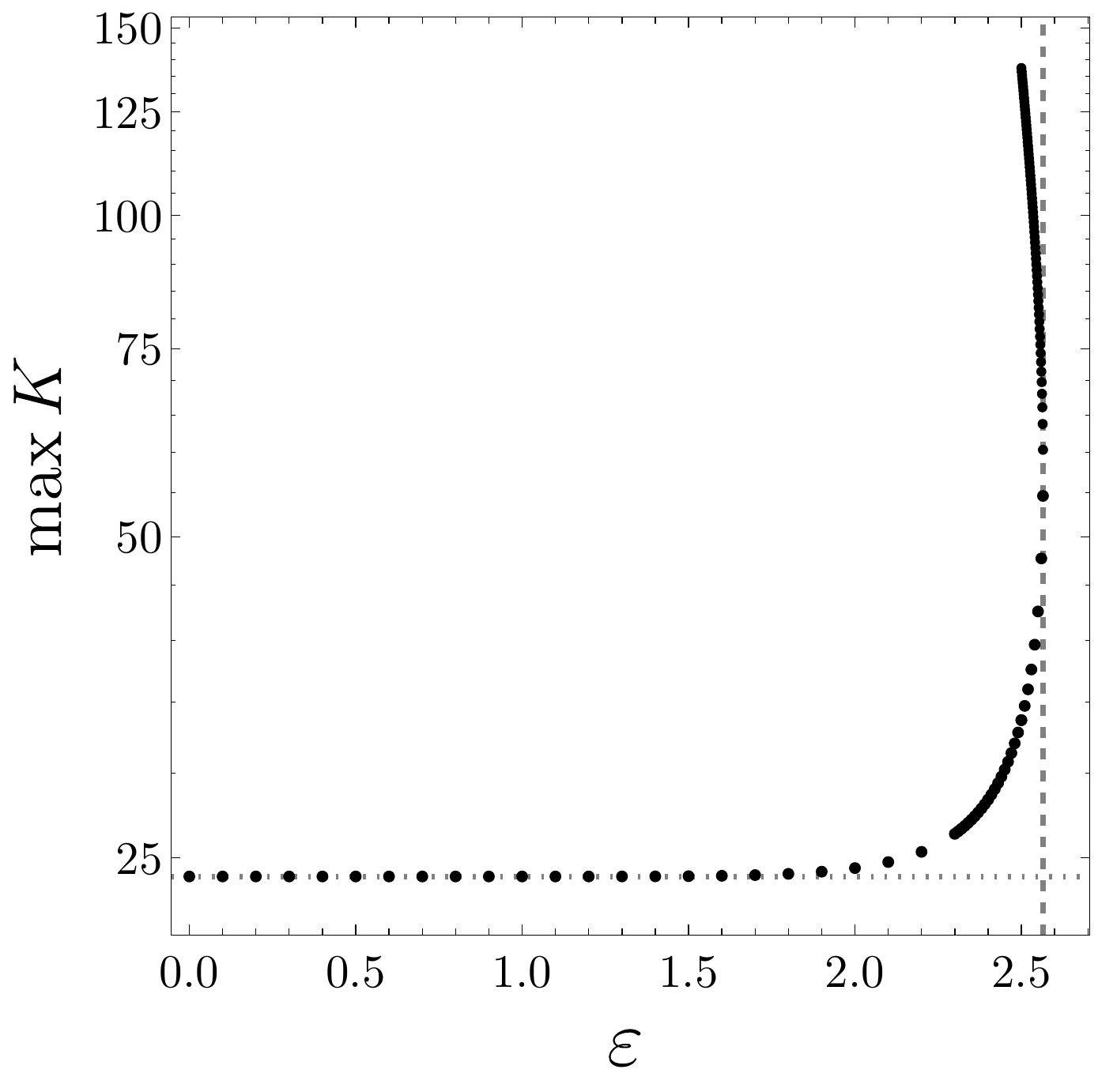}
 \caption{The log-linear plot of the maximum of the Kretschmann scalar against the rotation parameter for the soliton solutions. The dashed gridline marks $\varepsilon_c$, and the dotted line shows the $\max K=24/L^4$ for pure AdS$_4$.}
 \label{fig:kresol}
 \end{figure} 
 
One of the most interesting results of our manuscript is that we find no soliton solutions for $\varepsilon>\varepsilon_c \simeq 2.565$. In fact, we find no stationary axisymmetric solutions for $\varepsilon>\varepsilon_c$. Furthermore, there is a range $\varepsilon\in(2,\varepsilon_c)$ in which two solitons\footnote{We cannot exclude the possibility that more than two solutions might exist at some values of $\varepsilon$ as we approach $\varepsilon=2$.} exist for each value of $\varepsilon$, demonstrating non-uniqueness within the soliton family.

We have monitored the behaviour of the maximum of the Kretschmann scalar as a function of $\varepsilon$. Since we have no matter fields, besides a negative cosmological constant, the Kretschmann curvature invariant $K=R_{abcd}R^{abcd}$ is related to the norm of the Weyl tensor in a simple manner
$$
L^4\,W_{abcd}W^{abcd}=L^4\,K-24\,,
$$
that is to say, the local tidal forces scale like $K^{1/2}$.

The Kretschmann scalar $K$ is maximal at the equator $x=0$, and slowly deforms from the pure AdS$_4$ value of $24/L^4$ as we increase $\varepsilon\rightarrow\varepsilon_{c}$, forming two large extrema. The minima grows only slightly slower than the maxima, and the latter is plotted against $\varepsilon$ in Fig~\ref{fig:kresol}. For $\varepsilon>2$, we can see that two soliton solutions exist for a fixed value of $\varepsilon$. We call the upper branch the large branch, and the lower branch the small branch. The large branch of solitons smoothly extends the growth of $K$, which increases without bound, indicating formation of a curvature singularity. Also we note that for the soliton, the metric component $g_{tt}$ tends to zero at the origin, possibly as $\varepsilon\rightarrow 2$ from above.

%%%%%%%%%%%%%%%%%%%%%%%%%
\subsection{\label{subsec:blres}Black Holes}
%%%%%%%%%%%%%%%%%%%%%%%%%
\subsubsection{\label{subsec:theh}The horizon geometry}
%%%%%%%%%%%%%%%%%%%%%%%%%
\begin{figure}[t]
\centering
  \begin{minipage}[t]{.9\textwidth}
    \includegraphics[width=\textwidth]{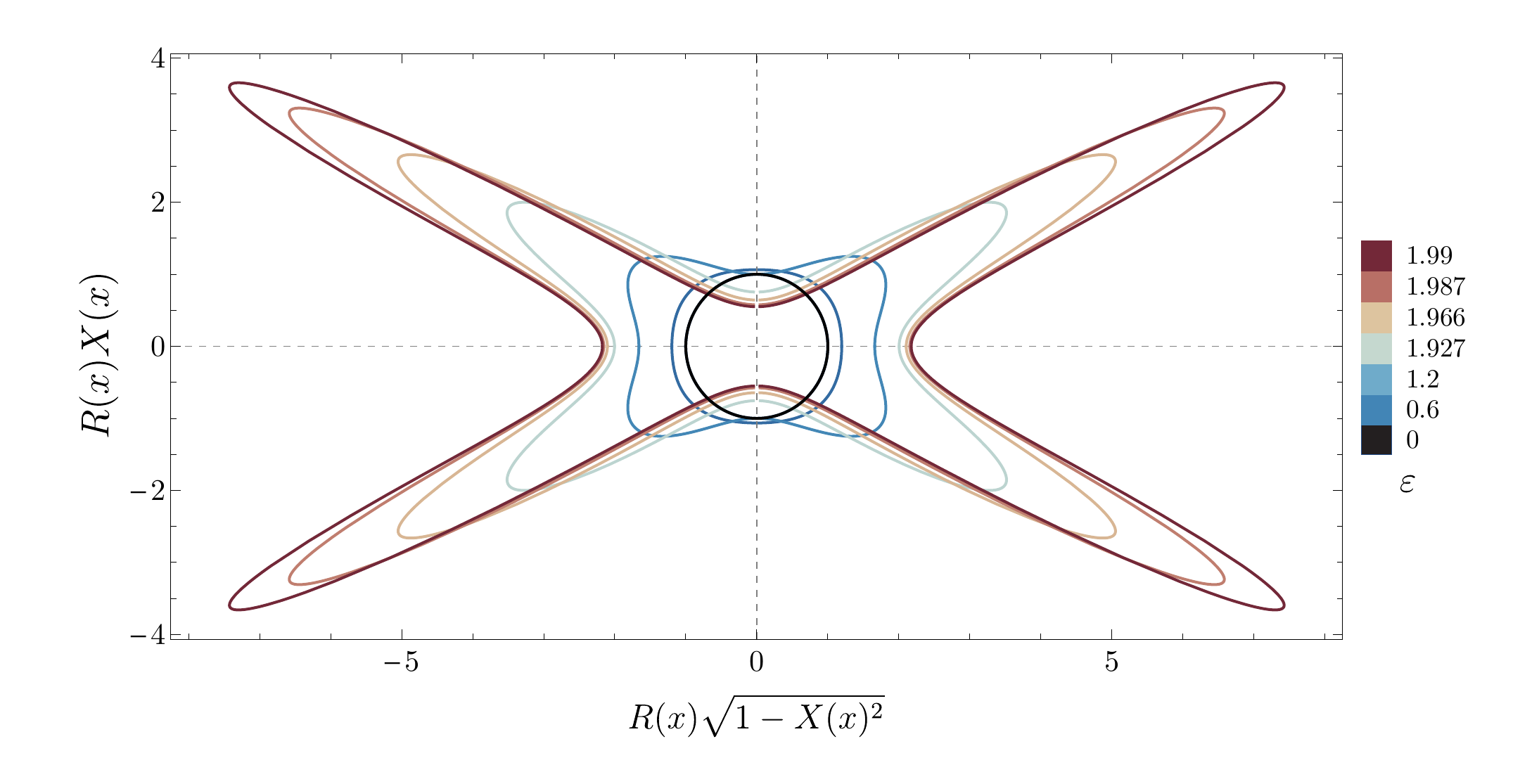}
  \end{minipage}
  \caption{Hyperbolic embedding of the cross section of the black hole horizons, for several values of the parameter $\varepsilon$ and a fixed temperature $T=1/\pi$. As we increase $\varepsilon$, the arms of the horizon cross section get stretched further apart. This picture is qualitatively the same for $T\geq 1/\pi$.}
  \label{fig:hype}
\end{figure}

 \begin{figure}[]
\centering
  \begin{minipage}[]{.55\textwidth}
    \includegraphics[width=\textwidth]{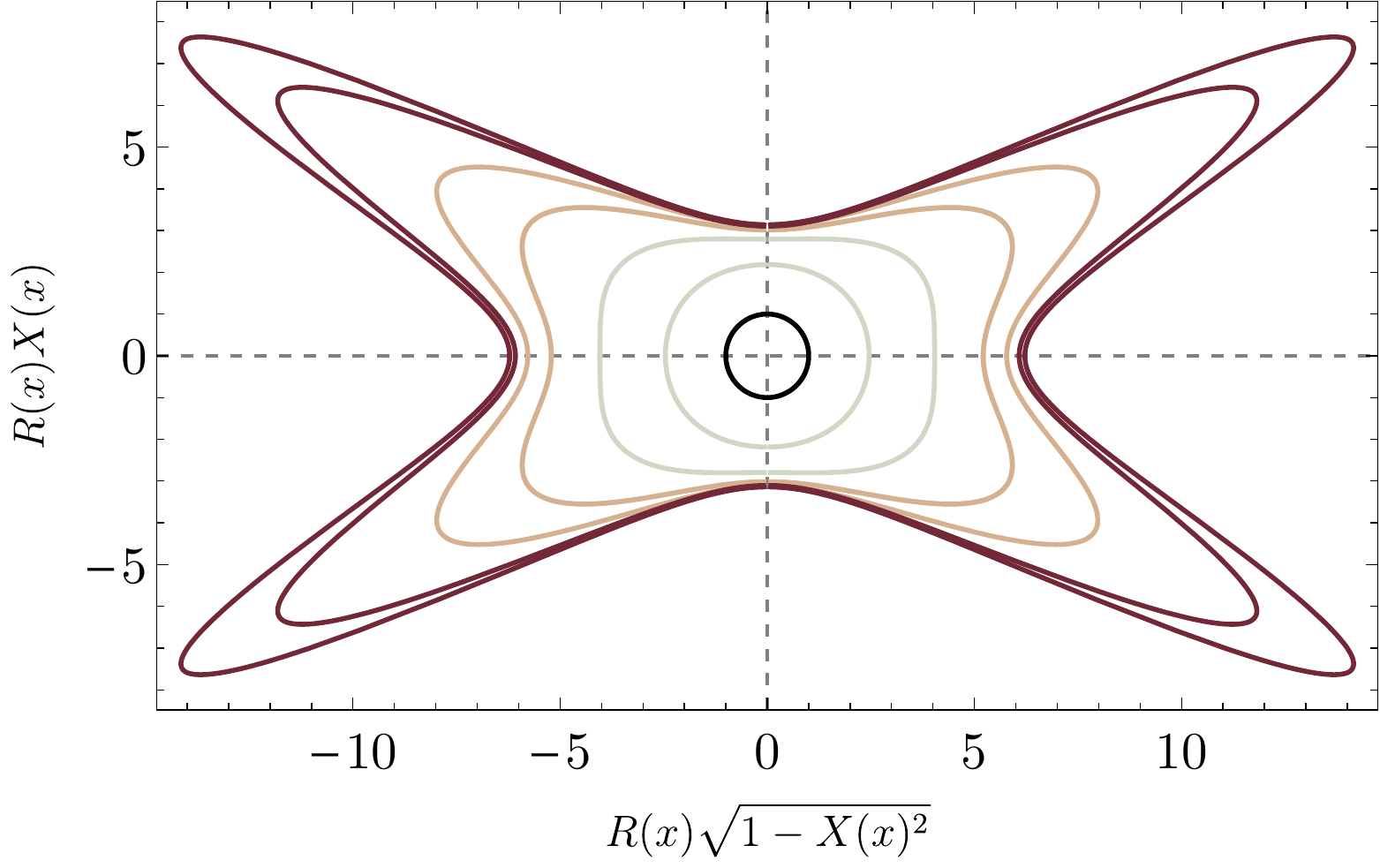}
  \end{minipage}
  \hfill
    \begin{minipage}[]{.35\textwidth}
    \includegraphics[width=\textwidth]{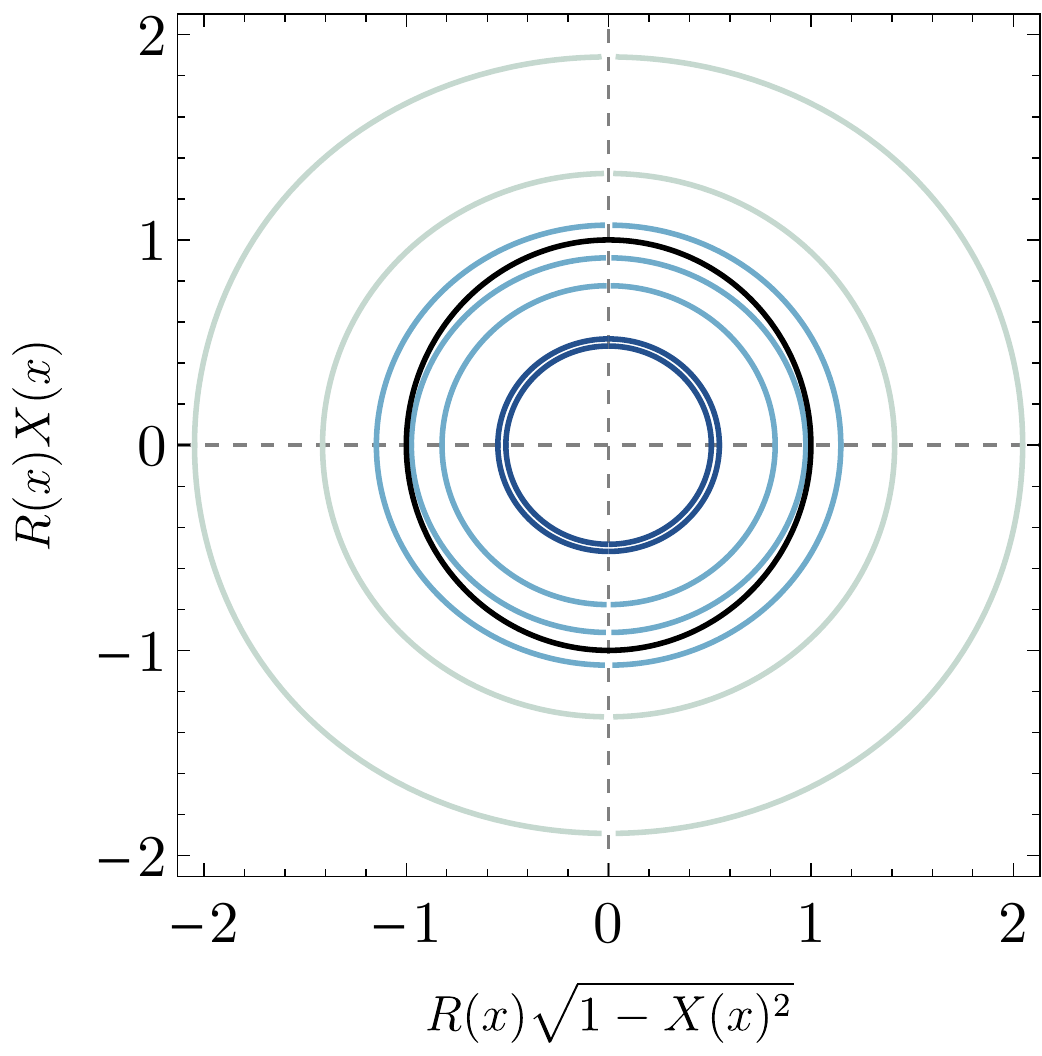}
  \end{minipage}
  \caption{Hyperbolic embedding of the cross section of the black hole solution horizons, for several values of the parameter $\varepsilon$ and a fixed low temperature $T=0.2506$. In this regime, there is a four-fold black hole non-uniqueness. \textit{Left}: Embeddings for the upper, thermodynamically dominant branch of solutions with $\varepsilon\in(1.520,2.059)$, and the largest entropy branch (two outer curves in red, $\varepsilon=2.055,2.05$). The black circle is the $T=1/\pi$, $\varepsilon=0$ black hole. \textit{Right}: Embeddings for the two small black hole branches. The higher entropy branch has $\varepsilon\in ( 1.520, 2.557)$, and the two inner circles in dark blue represent the lowest branch ($\varepsilon=2.55,2.53 $). The embeddings are qualitatively similar for the low $T$ regime.}
  \label{fig:hypeLOW}
\end{figure}
In order to obtain a better understanding of how the black hole horizon behaves under the increase of the boundary rotation amplitude we analyze horizon embedding diagrams. As with Schwarzschild-AdS$_4$ black holes, for a given temperature~(\ref{eq:tem}) there exist two solutions which we will call \textit{small} and \textit{large} black holes. In the case of the small black hole branch, we could successfully embed horizons into Eucliden space~$\mathbb{E}^3$ for all $\varepsilon$ - they are small and round spheres, slightly squashed through the poles. However, for the large branch, the scalar Gaussian curvature of the horizon spatial sections becomes too negative and the procedure is only successful for very small $\varepsilon$. Instead, we embed spatial cross sections of the horizon into hyperbolic~$\mathbb{H}^3$ space~\cite{Gibbons:2009qe} in global coordinates

\begin{equation}
\mathrm{d}s^2_{\mathbb{H}^3}=\frac{\mathrm{d}R^2}{1+R^2/\tilde{\ell}^2}+R^2\left[\frac{\mathrm{d}X^2}{1-X^2}+(1-X^2)\,\mathrm{d}\phi^2\right],
\label{eq:hyper3}
\end{equation}

\noindent where $\tilde{\ell}$ is the radius of the hyperbolic space. For sufficiently small values of $\tilde{\ell}$ the embedding always exists. The induced metric on the intersection of the horizon with a partial Cauchy surface of constant $t$ of the black hole line element (\ref{eq:ansatzbh}) is given by
\begin{equation}
\mathrm{d}s^2_H=L^2\left[\frac{4 y_+^2 Q_3(x,0)}{2-x^2}\,\mathrm{d}x^2+y_+^2(1-x^2)^2Q_5(x,0)\,\mathrm{d}\phi^2\right].
\label{eq:horizon}
\end{equation}

\noindent The embedding is given by a parametric curve $\{R(x),X(x)\}$. The pull back of (\ref{eq:hyper3}) to this curve induces a two-dimensional line element with the following form
\begin{equation}
\mathrm{d}s^2_{\mathrm{pb}}=\left[\frac{R'(x)^2}{\displaystyle 1+\frac{R(x)^2}{\tilde{\ell}^2}}+\frac{R(x)^2X'(x)^2}{1-X(x)^2}\right]\mathrm{d}x^2+R(x)^2(1-X(x)^2)\,\mathrm{d}\phi^2\,.
\end{equation}
Equating this line element with (\ref{eq:horizon}) gives the following first order differential equation for the polar coordinate
\begin{multline}
0=4 H(x) P(x) (X(x)^2-1)^2 \left[P(x)-\tilde{\ell}^2 (X(x)^2-1)\right]+4 \tilde{\ell}^2 P(x) X(x) (X(x)^2-1) P'(x)X'(x)\\-(X(x)^2-1)^2\tilde{\ell}^2P'(x)^2-4 P(x)^2 (\tilde{\ell}^2+P(x)) X'(x)^2
\end{multline} 
with $H(x)=(2-x^2)^{-1}(4 y_+^2 Q_3(x,0))$ and $P(x)=y_+^2(1-x^2)^2Q_5(x,0)$. The numerical results for several values of $\varepsilon$ are presented in Fig.~\ref{fig:hype} where we set $\tilde{\ell}=0.73$. Here we are fixing the black hole temperature to be $T=1/\pi$, and the results for other temperatures are qualitatively similar~(Fig.~\ref{fig:hypeLOW}).

For $\varepsilon<1$, the large black holes are round and the deformation is small. As we approach $\varepsilon\rightarrow 2^-$, the black hole horizon gets significantly distorted and develops remarkable extended features. These cross-section ``spikes" peak at $x=\sqrt{1-\sqrt{2}/2}$, \emph{i.e.} $\theta=\pi/4$, where the gravitational pull is the strongest\footnote{As measured by the conformal boundary metric.}. The horizons are squashed through the poles, and extended sideways, with the circumference slowly increasing. The black hole diameter through the poles tends to some non-zero value\footnote{This feature is not very apparent in the embedding diagrams.}, and for very large rotation parameter the thickness of the cross section arms is surprisingly uniform throughout its length, and as a function of $\varepsilon$. As we increase the temperature, the horizon deforms faster. We conjecture that for sufficiently high temperatures, these static, axially symmetric solutions exist for any $\varepsilon<2$, endlessly stretching the black hole horizon towards the AdS$_4$ boundary. This is difficult to check numerically, but our results indicate that there is no upper bound to how extended these black holes can become as we approach $\varepsilon\to2^-$.

Colder black holes turn out to be less elastic, and more extended through the equator. In Fig.~\ref{fig:hypeLOW} we plot embeddings for a low temperature $T=0.2506$, where for a fixed $\varepsilon$ we find two or four black holes. Even though this phase is inelastic, as $\varepsilon\rightarrow 2^+$, we expect the stretched solutions to enter a scaling regime as with the hot solutions.

%%%%%%%%%%%%%%%%%%%%%%%%%
\subsubsection{Curvature invariants}
%%%%%%%%%%%%%%%%%%%%%%%%%
As we tune the boundary rotation parameter, thus ``stirring" the bulk, the black hole becomes ever more stretched and we expect the curvature on the horizon to increase correspondingly.
\begin{figure}[t]
\centering
  \begin{minipage}[t]{0.45\textwidth}
    \includegraphics[width=\textwidth]{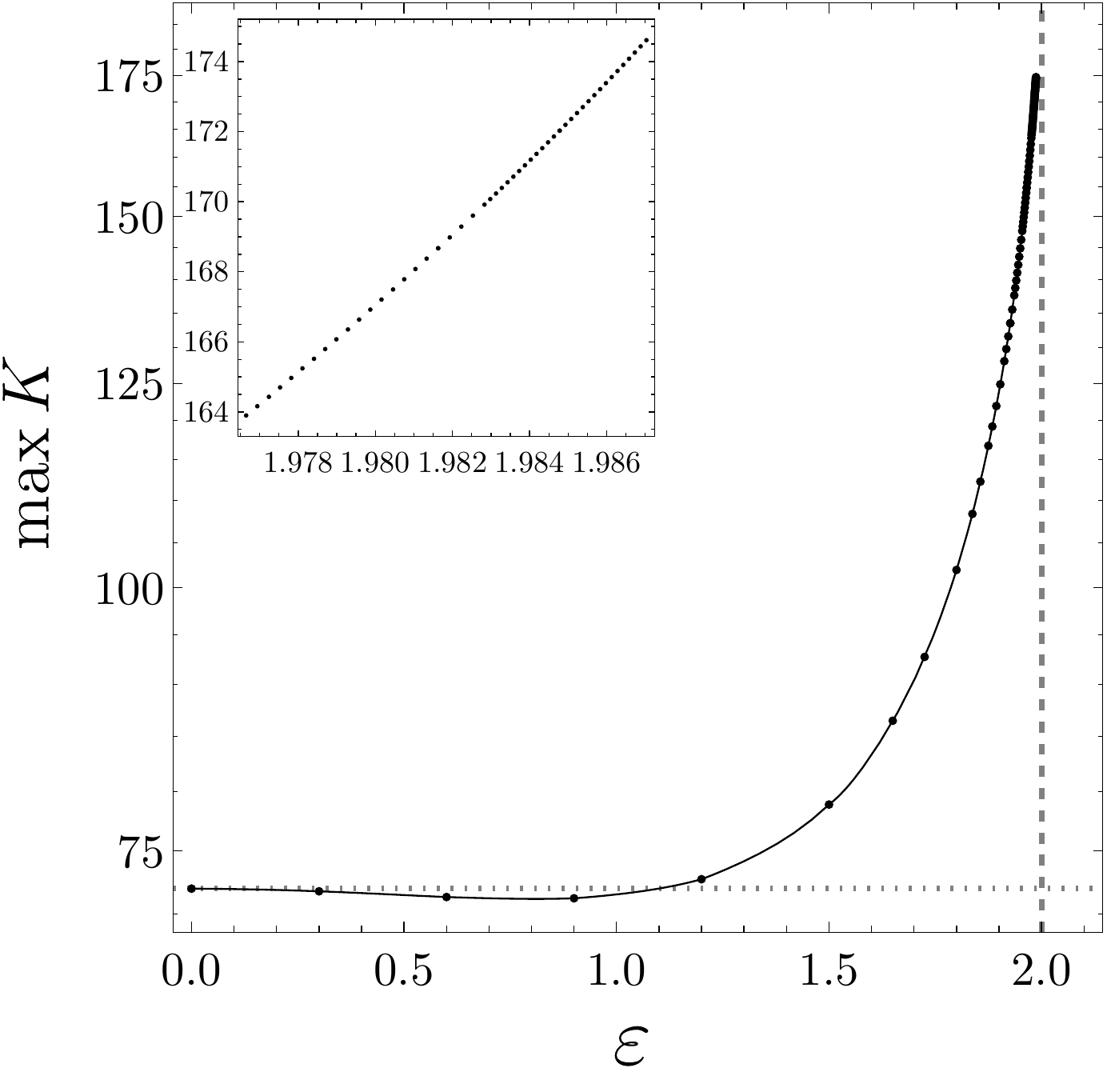}
  \end{minipage}
  \caption{Log-linear plot of the curvature scalar maximum against~$\varepsilon$, for a large black hole with $T=1/\pi$. The dashed gridline marks the critical value of $\varepsilon$, and the dotted gridline shows ${K_\mathrm{max}=72}$ for a Schwarzschild-AdS$_4$ black hole. The inset is a log-log plot for high values of the rotation amplitude.}
  \label{fig:kre}
\end{figure}

In Fig.~\ref{fig:kre}, we show the maximum of the Kretschmann scalar as a function of $\varepsilon$ for large black holes. For the black holes, the Kretschmann scalar is maximal on the horizon, located at $y=0$, and as $\varepsilon\rightarrow 2^-$, it peaks closer to $\theta=\pi/4$. For Schwarzschild-AdS$_4$, $L^4\,K|_\mathcal{H} =12/y_+^4+24/y_+^2+36$, and thus for $T=1/\pi$ the maximal value is $72/L^4$. As the boundary amplitude approaches $\varepsilon=2^-$, the Kretschmann scalar grows without bound (see Fig.~\ref{fig:kre}) and we expect the maximum of the Kretschmann scalar to enter a scaling regime. This quantity is difficult to extract accurately, however we find some numerical evidence that it behaves as a power law as $\varepsilon\rightarrow 2^-$.
 
For completeness, we present plots of the Kretschmann scalar of the thermodynamically dominant solutions in Appendix~\ref{sec:add}, Fig.~\ref{fig:kretfull}.

%%%%%%%%%%%%%%%%%%%%%%%%%%%%%%%%%%%%
\subsubsection{\label{subsec:ent}Entropy}
%%%%%%%%%%%%%%%%%%%%%%%%%%%%%%%%%%%%
 \begin{figure}[t]
\centering
  \begin{minipage}[t]{0.3\textwidth}
    \includegraphics[width=\textwidth]{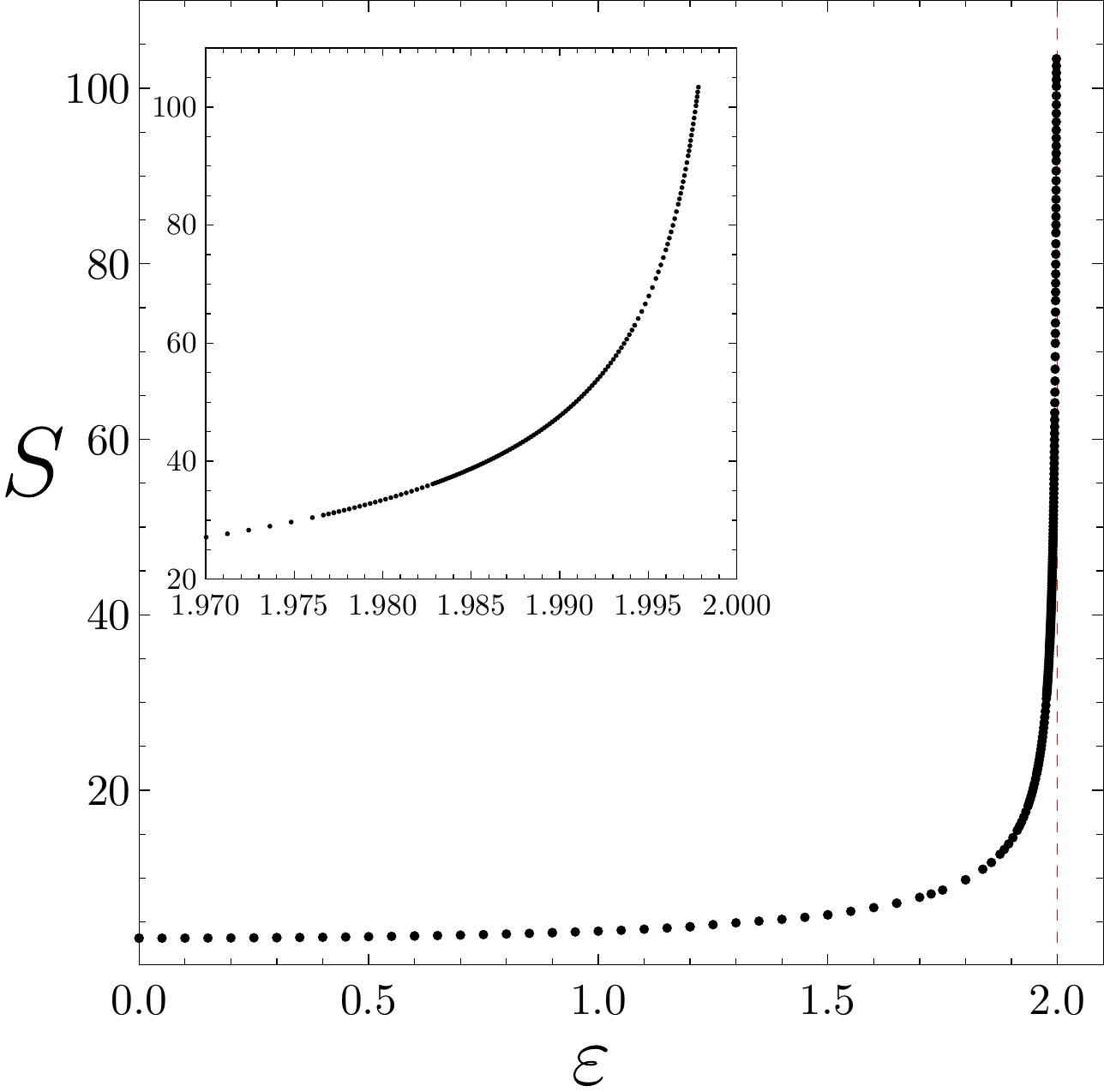}
  \end{minipage}
    \hfill
    \begin{minipage}[t]{0.32\textwidth}
    \includegraphics[width=\textwidth]{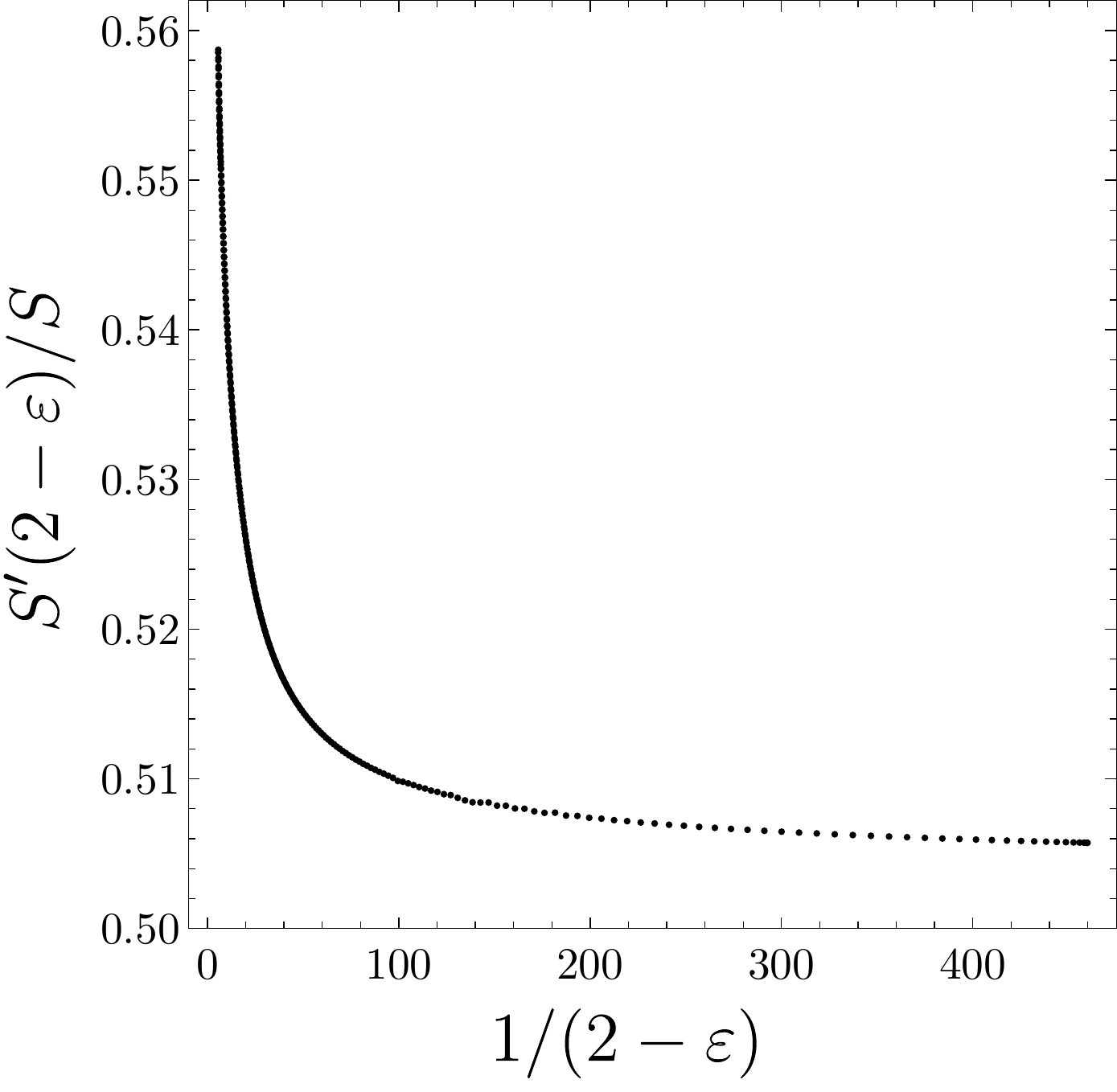}
  \end{minipage} 
    \hfill
    \begin{minipage}[t]{0.3\textwidth}
    \includegraphics[width=\textwidth]{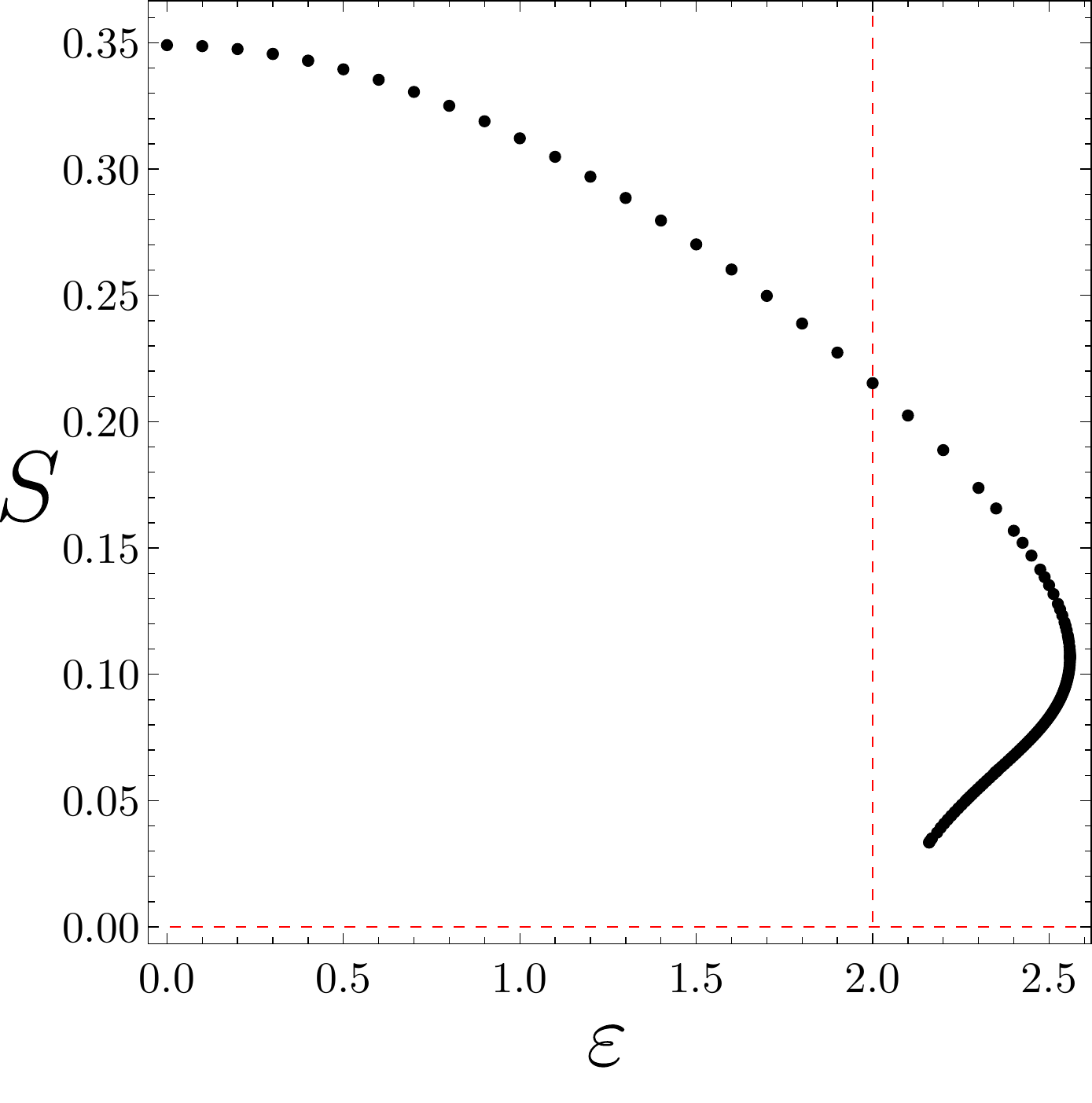}
  \end{minipage} 
  \caption{\textit{Left}: Entropy against the rotation parameter $\varepsilon$ for the large branch of black holes with $T=1/\pi$. The inset zooms around the limit, demonstrating the asymptotic behavior. Red dashed gridlines mark the $\varepsilon=2$ limit. \textit{Middle}: Logarithmic derivative of~$S$ for the large black hole with
   $T=1/\pi$. As $\varepsilon\rightarrow 2$, the entropy of the large black hole blows up as an inverse power law. \textit{Right}: Entropy of the small branch of black holes with the same temperature. Red dashed gridlines show where $\varepsilon=2$ and $S=0$.}
  \label{fig:ent}
\end{figure} 

The entropy associated with the black hole horizon is given by                                                                                                                                                                                                                                                                                                                                                                                                                                                                                                                                                                                                                                                                                                                                                                                                                                                                                                                                                                                                                                                                                                                                                                                                                                                                                                                                                                                                                                                                                                                                                                                                                                                                                                                                                                                                                                                                                                                                                                                                                                                                                                                                                                                                                                                                                                                                                                                                                                                                                                                                                            \begin{equation}
S=\frac{A}{4 G_N}=\frac{2\pi\,y_+^2\,L^2}{G_N}\int_0^1\mathrm{d}x\frac{1-x^2}{\sqrt{2-x^2}} \sqrt{Q_3(x,0)Q_5(x,0)}\,.
\end{equation}
In Fig.~\ref{fig:ent}, we present our results for the small and large branches of black hole solutions with a temperature $T=1/\pi$. For the large black holes, we find that the entropy is always an increasing function of $\varepsilon$, and as $\varepsilon\rightarrow 2^-$, it grows without bound. When $\varepsilon$ is sufficiently close to $2$, the entropy enters a scaling regime as we saw with the Kretschmann scalar, and $S\propto(2-\varepsilon)^{-\alpha}$ (see the middle panel of Fig.~\ref{fig:ent}). Numerically\footnote{Numerically, this is difficult to explore for temperatures much larger, and smaller than $T=1/\pi$. The hot, large black holes reach the limit $\varepsilon=2$ slowly, and the cold black holes are more difficult to resolve. However, we would expect that near the critical boundary value parameter $\varepsilon =2$, the scalar invariants, such as entropy and curvature blow up according to some scaling laws.} we find that $\alpha$ is consistent with being $1/2$ within a $1\%$ error.

For the small branch (see right panel of Fig.~\ref{fig:ent}), and fixed temperature, the entropy is a decreasing function of $\varepsilon$. In this case we can reach values $\varepsilon>2$, although we still find a maximum value of $\varepsilon$ beyond which we cannot find axially symmetric black hole solutions. We call this maximum value $\varepsilon_{\max}(T)$. As the temperature decreases, $\varepsilon_{\max}(T)$ approaches $\varepsilon_c$ from below, which is a turning point for the small branch of solutions, below which we obtain a second set of unstable solutions with lower entropy. In this region of the solution space numerical errors are increasingly large, and we were unable to extend the solutions by further decreasing $\varepsilon$. We conjecture that $S\rightarrow 0$ in the limit $\varepsilon\rightarrow 2^{+}$, and the singularity might be approached in a non-trivial way\footnote{In the literature, there have been examples where black hole solutions approaching a singular point in a phase space exhibit a non-trivial non-uniqueness, for instance in a spiralling manner (see \textit{e.g.} \cite{Dias:2011at,Markeviciute:2016ivy}).}.

 \begin{figure}[t]
\centering
  \begin{minipage}[t]{.425\textwidth}
    \includegraphics[width=\textwidth]{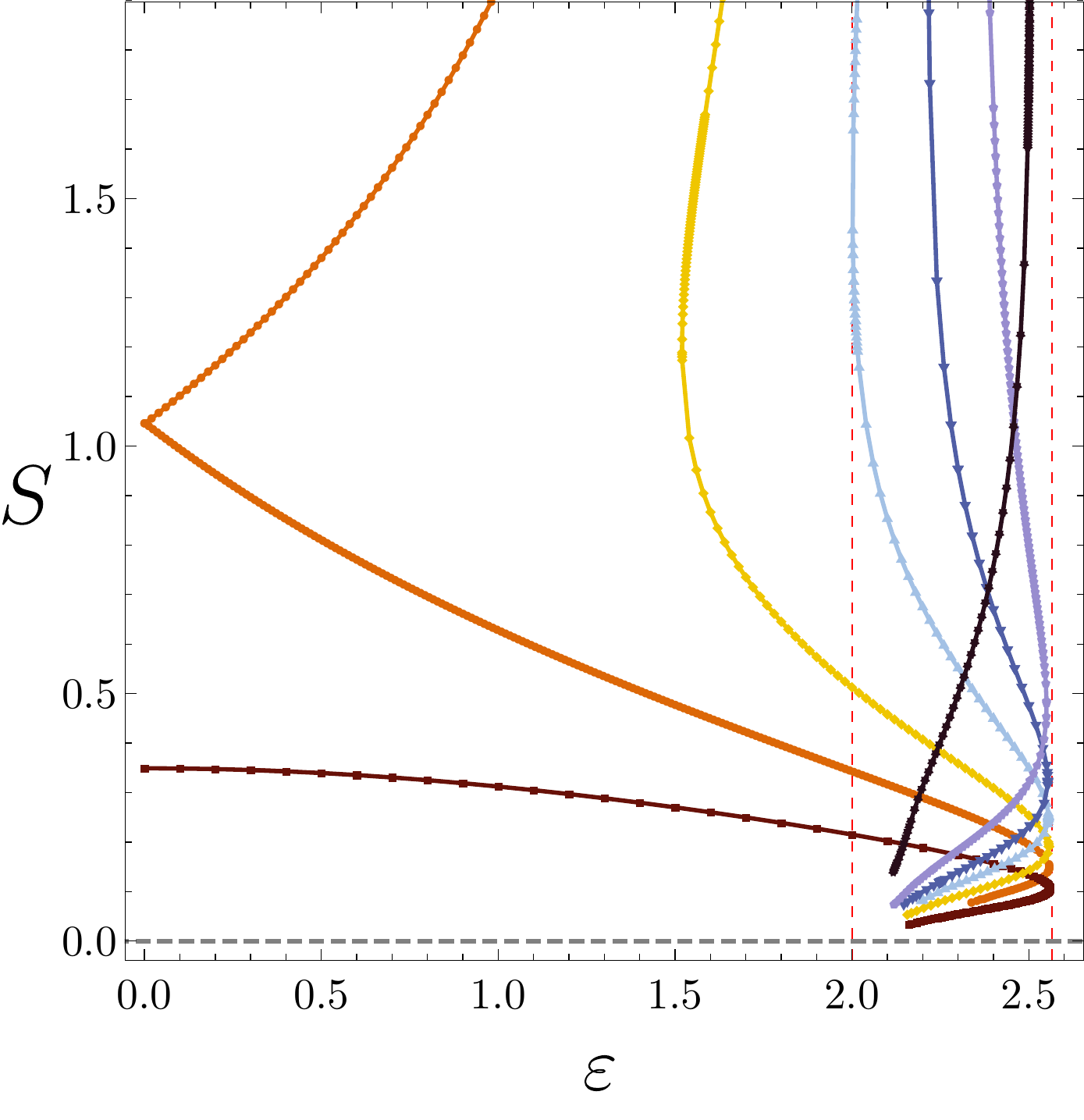}
  \end{minipage} 
  \hfill
    \begin{minipage}[t]{.52\textwidth}
    \includegraphics[width=\textwidth]{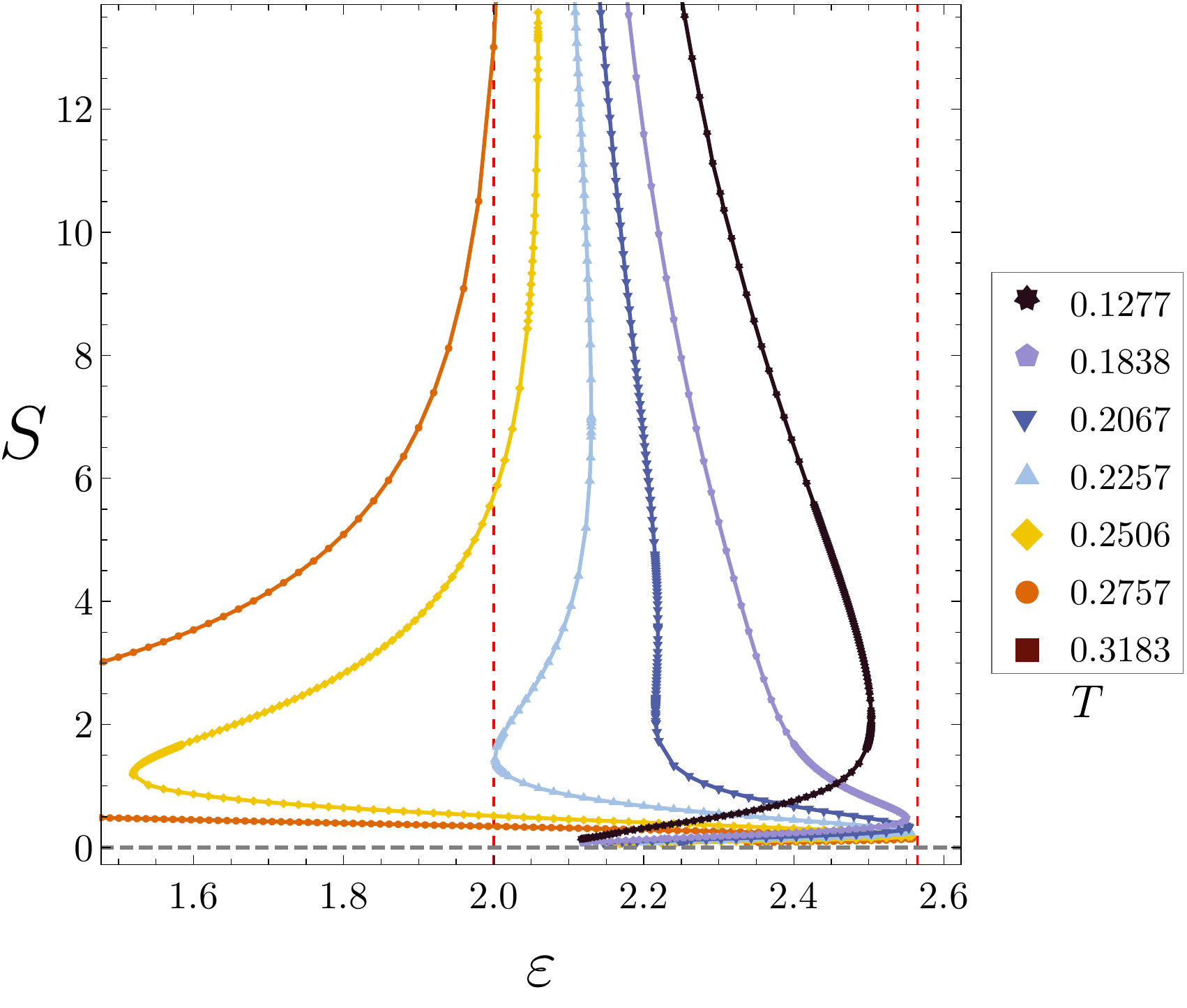}
  \end{minipage} 
  \caption{\textit{Left}: Entropy against the boundary parameter for low temperature black holes with $T<T^{\mathrm{Schw}}_{\min}\simeq 0.2757$. The orange disks show black holes with $T\simeq T^{\mathrm{Schw}}_{\min}$. The brown squares show the (hot) small black holes with $T=1/\pi$ for reference. As we lower the temperature, the large black holes begin to exist for $\varepsilon>2$. \textit{Right}: The asymptotic behavior of isothermal entropy curves. Both solution branches approach $\varepsilon=2$, with the large branch having a steeply increasing $S$, and the small branch having $S\rightarrow 0$. The red gridlines show $\varepsilon=2$ and $\varepsilon=\varepsilon_c$.}
  \label{fig:entropysmall}
\end{figure}

This picture looks qualitatively the same for black holes with $T>T^{\mathrm{Schw}}_{\min}$ (recall than $T^{\mathrm{Schw}}_{\min}\simeq 0.2757$), with large solutions expanding and small solutions shrinking down when we increase $T$. However, we can find black hole solutions with a minimal temperature $T_{\min}(\varepsilon)<T^{\mathrm{Schw}}_{\min}$. This implies that for fixed $T<T^{\mathrm{Schw}}_{\min}$, there are no black hole solutions for $\varepsilon<\varepsilon_{\min}$. As we cross the $T^{\mathrm{Schw}}_{\min}$, the two black hole branches at fixed temperature join up (see orange disks in Fig.~\ref{fig:entropysmall}), and as we lower the temperature further, the ``left'' turning point occurs for higher values of $\varepsilon$. The subsequent evolution of the phase space is somewhat intricate. It turns out, that for all fixed $T<T^{\mathrm{Schw}}_{\min}$, there exist large black holes with $\varepsilon>2$. There, we find yet another turning point, which bends back towards $\varepsilon=2$. For temperatures $T\lesssim 0.2257$, the ``left'' turning point crosses $\varepsilon=2$, and black holes exist only for $\varepsilon>2$. Soon after, once we reach $T\simeq 0.207$, the two upper turning points converge and disappear, leaving only the lower turning point, which appears to exist for all finite temperatures.

This analysis reveals interesting non-uniqueness of the black hole solutions. For temperatures ${T>T^{\mathrm{Schw}}_{\min}}$, we have two black hole solutions: for $\varepsilon<2$, there is a large and a small black hole, and for $\varepsilon>2$ there are two small black holes, if we assume that we are approaching $\varepsilon\rightarrow 2^+$ as $S\rightarrow 0$. For ${T<T^{\mathrm{Schw}}_{\min}}$, we have two solutions for $\varepsilon_{\mathrm{min}}(T)<\varepsilon<2$, four solutions for $\varepsilon>2$ until $T\simeq 0.2067$, when these extra phases merge, and then we again have two branches. All these solutions exist up to some value of $\varepsilon_{\mathrm{max}}(T)<\varepsilon_c$.

The low temperature regime is hard to access numerically, so we can only conjecture the very low temperature behaviour. The only remaining turning point is decreasing slowly with the temperature, separating the two branches of solutions the lower of which has vanishing horizon area as $\varepsilon\rightarrow 2^{+}$. These solutions look qualitatively very similar to the soliton solutions, and could be thought of as a small black hole placed in a soliton background. The upper branch has increasing entropy, which blows up as $\varepsilon\rightarrow 2^{+}$.

In the following subsections we present physical quantities and carry out an extensive thermodynamic analysis of both black holes and horizonless AdS$_4$ solutions, further supporting our conjectures.

%%%%%%%%%%%%%%%%%%%%%%%%%
\subsection{\label{subsec:str}Stress-Energy tensor}
%%%%%%%%%%%%%%%%%%%%%%%%%
The bare on-shell gravity action in AdS$_4$ is divergent and in order to obtain the boundary stress tensor we need to regularise the action using holographic renormalization techniques~\cite{Brown:1992br,Henningson:1998gx,Balasubramanian:1999re}. The counterterms that we must add to the bulk action~\eqref{eq:action} are 
 \begin{equation}
 S_{ct}=-\frac{1}{4\pi G_N L}\int_{\partial \mathcal{M}}\mathrm{d}^3x\sqrt{-\gamma}\left(1-\frac{L^2}{4}R[\gamma]\right),
 \end{equation}
and the expectation value of the stress-energy tensor of the dual field theory can be found to be
\begin{equation}
\langle\tilde{T}\rangle_{\mu\nu}=\frac{1}{8\pi G}\left(\Theta_{\mu\nu}-\Theta\gamma_{\mu\nu}+L G_{\mu\nu}[\gamma]-\frac{2}{L}\gamma_{\mu\nu}\right),
\end{equation} 
from the on-shell action.

As an exercise, following~\cite{deHaro:2000vlm}, we also compute the holographic stress tensor by recasting the metric~(\ref{eq:ansatzbh}) into Fefferman-Graham coordinates. Firstly, we solve the equations of motion about the conformal boundary $y=1$, order by order, by expanding the metric functions as shown below
\begin{equation} 
Q_i(x,y)=\sum_{n=0}^{3}(1-y)^nq_{i,n}(x)+\log{(1-y)}(1-y)^4\tau_{i,4}(x)+\smallO\left((1-y^2)^3\right).
\end{equation}

 \begin{figure}[t]
\flushleft
  \begin{minipage}[t]{1\textwidth}
    \includegraphics[width=1\textwidth]{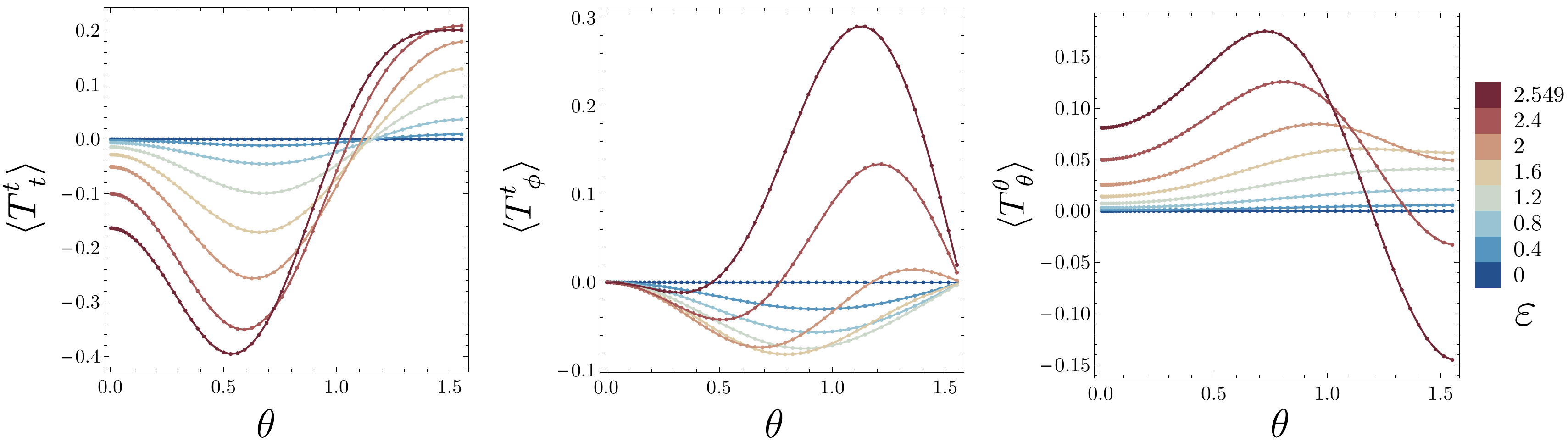}
  \end{minipage}
  \caption{The stress-energy tensor component $\langle T^t_{\,\,t} \rangle$ (\textit{left}), component $\langle T^t_{\,\,\phi} \rangle$ (\textit{middle}), and component $\langle T^\theta_{\,\,\,\theta} \rangle$ (\textit{right}) for solitonic solutions as a function of~$\theta$, the angle on the $\mathbb{S}^2$. Colours represent different values of the parameter $\varepsilon$, with larger values corresponding to larger amplitudes.}
  \label{fig:sols} 
\end{figure}

Next we perform an asymptotic coordinate change near the AdS$_4$ boundary to the Fefferman-Graham form\footnote{Here and further in this subsection we set $G_N L^2=1$.}
\begin{equation}
\mathrm{d}s^2=\frac{1}{z^2}\left[\mathrm{d}z^2+\mathrm{d}s^2_\partial+z^2\mathrm{d}s^2_1+z^3\mathrm{d}s^2_2+\mathcal{O}(z^4)\right]\,.
\label{eq:boundary0}
\end{equation}
To achieve this, we take
\begin{align}
y=1+\sum_{n=1}^{4}z^n\,Y_n(\theta)+\smallO(z^4)\nonumber,
\\
x=\sqrt{1-\sin\theta}+\sum_{n=1}^{4}z^n\,X_n(\theta)+\smallO(z^4),
\end{align}
and determine $\{Y_n(\theta),X_n(\theta)\}$ order by order in $z$ so that our line elements can be cast in the form (\ref{eq:boundary1}) with conformal boundary metric
\begin{equation}
\mathrm{d}s^2_\partial=-\mathrm{d}t^2+\mathrm{d}\theta^2+\sin^2\theta\left[\mathrm{d}\phi+\mathrm{d}t\,\cos\theta\,\varepsilon\right]^2\,.
\label{eq:boundary1}
\end{equation}

The holographic stress-energy tensor is then given by
\begin{equation}
\langle T_{ij}\rangle=\frac{3}{16\pi}g_{(2)ij},
\end{equation}
where $g_{(2)ij}$ is the metric associated with $\mathrm{d}s_2^2$. As expected, this stress-energy tensor is traceless and transverse with respect to~(\ref{eq:boundary1}). The energy density is simply given by $\rho(\theta)=-\langle T^t_{\phantom{t}t}\rangle$, whilst the angular momentum density is simply $j(\theta)=\langle T^t_{\phantom{t}\phi}\rangle$.
%%%%%%%%%%%%%%%
\subsubsection{Soliton}
%%%%%%%%%%%%%%%
The independent boundary stress-energy tensor components for the soliton solution are given by
\begin{align}
\langle T^t_{\phantom{t}t}\rangle&= \frac{1}{768 \pi }\left[18 q_1(\theta)+\varepsilon  \sin ^2\theta \left\{-65 \varepsilon +\sin ^2\theta \left(3 \varepsilon ^3 \sin ^22\theta+46 \varepsilon +18 q_6(\theta)\right)-18 q_6(\theta)\right\}\right],\\
\langle T^t_{\phantom{t}\phi}\rangle&=\frac{1}{128 \pi}  \sin ^2\theta \cos\theta \left[2 \varepsilon ^3 \sin ^4\theta+3 \varepsilon -3 q_6(\theta)\right] ,\\
\langle T^\theta_{\phantom{\theta}\theta}\rangle&=\frac{1}{768 \pi }\left[\varepsilon ^2 \sin ^2\theta(16 \cos{2\theta}-27)+18 q_3(\theta)\right].
\end{align}
Numerical results are presented in Fig.~\ref{fig:sols} where we plot these components against $\theta$ for several values of the boundary rotation amplitude $\varepsilon$.  The stress-energy tensor component $\langle T^t_{\phantom{t}t}\rangle$ is positive at the equator where it is also maximal; it becomes negative at a value which depends on $\varepsilon$, and is negative at the poles. As the rotation amplitude increases towards $\varepsilon_c$, $\langle T^t_{\phantom{t}t}\rangle$ develops a large negative minima in the upper quarter of the sphere. $\langle T^t_{\phantom{t}\phi}\rangle$, on the other hand, is negative for small values of $\varepsilon$, and is vanishing on both the poles and the equator. As $\varepsilon\rightarrow\varepsilon_c$, $\langle T^t_{\phantom{t}\phi}\rangle$ develops a broad, large maxima close to the equator, and for the second soliton branch, it becomes entirely positive. The component $\langle T^\theta_{\phantom{\theta}\theta}\rangle$, measuring the pressure along $\theta$, is positive and peaked at the equator for small values of $\varepsilon$. As the amplitude is increased, the pressure becomes very negative at the equator, increases towards the poles and develops a maxima close to $\theta=\pi/4$.
 \begin{figure}[t]
\flushleft
    \begin{minipage}[t]{1\textwidth}
    \includegraphics[width=1\textwidth]{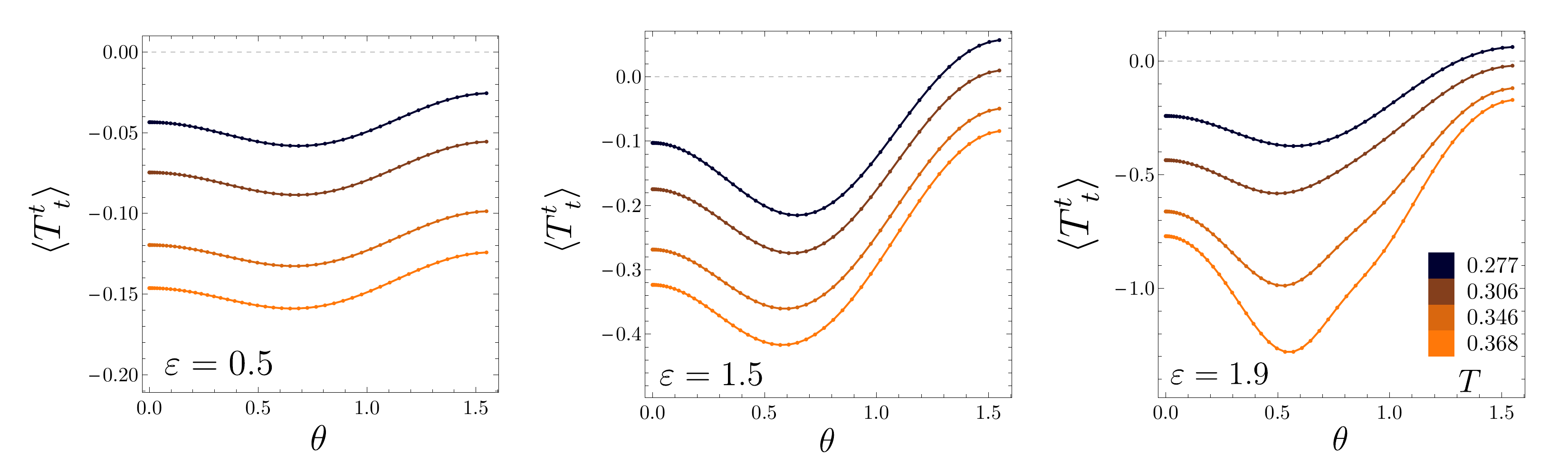}
  \end{minipage}
  \begin{minipage}[t]{1\textwidth}
    \includegraphics[width=1\textwidth]{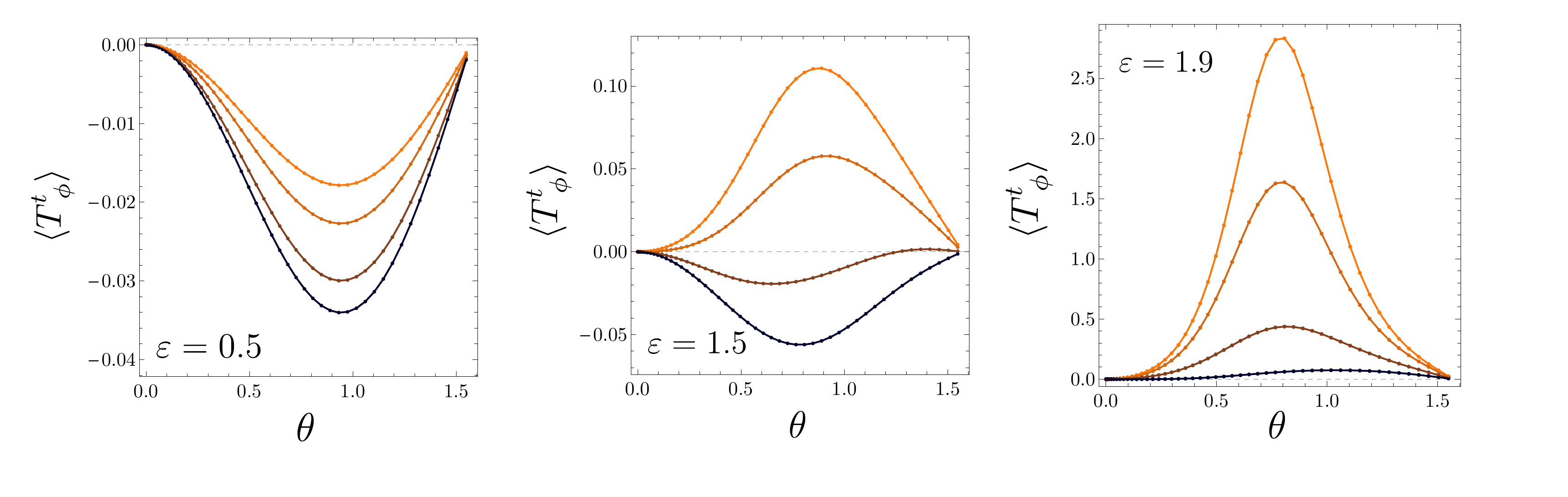}
  \end{minipage}
    \begin{minipage}[t]{1\textwidth}
    \includegraphics[width=1\textwidth]{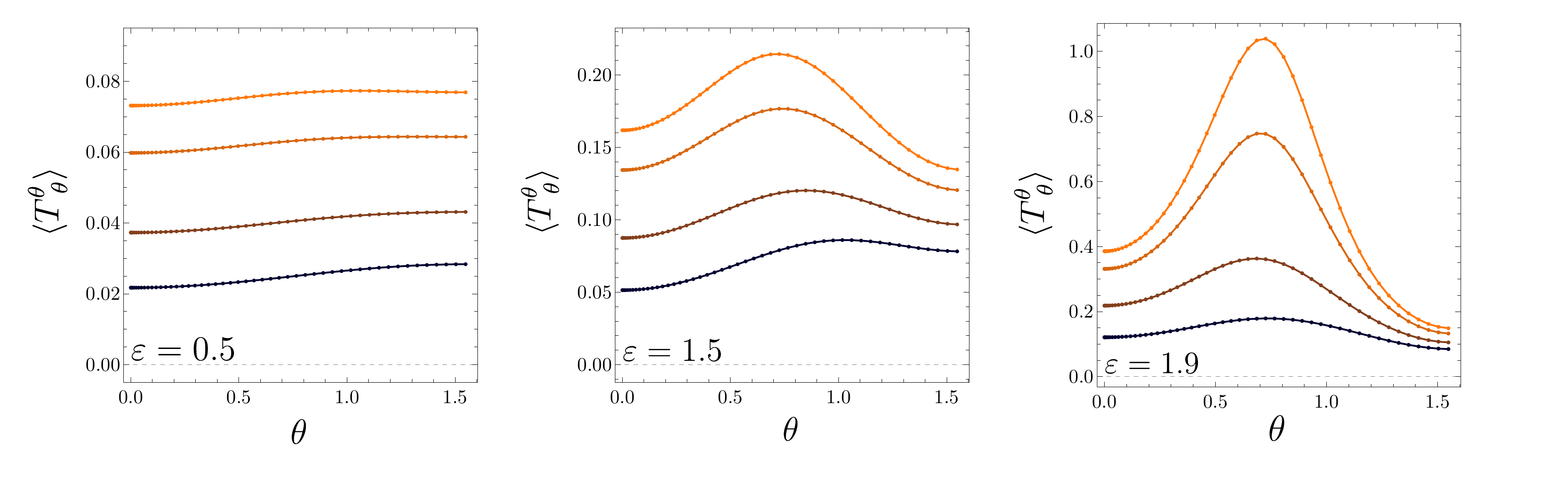}
  \end{minipage}
  \caption{Independent stress-energy tensor components of the dual to the large black holes against~$\theta$, for several values of the parameter $\varepsilon$. Colouring codes black hole temperature with darker colours representing temperatures closer to $T_\mathrm{min}(\varepsilon)$.}
  \label{fig:bhtensor}
\end{figure}
%%%%%%%%%%%%%%%
\subsubsection{Black Holes}
%%%%%%%%%%%%%%%
The independent holographic stress-energy tensor components for the black hole metric~(\ref{eq:ansatzbh}) are given by
\begin{align}
\begin{split}
\langle T^t_{\phantom{t}t}\rangle=\frac{y_+}{49152 \pi} \bigg[ 3y_+^2 \left(635 \delta ^3-1533 \delta ^2-887 \delta +384 q_1(\theta)-263\right)-192 \left(29 \delta ^2-14 \delta
   +17\right) \\
   +16 \varepsilon  \sin ^2{\theta} \left\{\sin ^2{\theta} \left(-12 \varepsilon ^3 \cos {4 \theta}+72q_6(\theta)y_+^2+\varepsilon  \left(-1113 \delta +12 \varepsilon ^2-288 y_+^2+2285\right)\right) \right.\\ \left.
   -72 q_6(\theta) y_+^2+32 \varepsilon  \left(27 \delta +9 y_+^2-65\right)\right\} \bigg],\\
\end{split}  
\end{align}
\vspace{-2em} 
\begin{flalign}
\langle T^t_{\phantom{t}\phi}\rangle&=\frac{y_+}{128 \pi} \sin ^2{\theta} \cos{\theta} \left[4 \varepsilon ^3 \sin ^4{\theta}-3 q_6(\theta) y_+^2+12 \varepsilon \left(3 \delta +y_+^2-5\right)\right],&
\end{flalign}
\vspace{-2em} 
\begin{align}
\begin{split}
\langle T^\theta_{\phantom{\theta}\theta}\rangle=\frac{y_+}{49152\pi }\left[  192 \left(7
   \delta ^2+26 \delta -17\right)+3y_+^2 \left(-109 \delta ^3+315 \delta ^2+721 \delta +384q_3(\theta)+97\right)-\right. \\
   \left.16 (33 \delta +139) \varepsilon ^2 \sin ^4{\theta}-1024 \varepsilon ^2 \sin ^2{\theta}\right].
\end{split}
\end{align} 
Numerical results for several values of the black hole temperature and boundary rotation amplitude are presented in Fig.~\ref{fig:bhtensor} (and for low temperatures see Appendix~\ref{sec:add}, Fig.~\ref{fig:lowstress}). For small $\varepsilon$ and large $T$, $\langle T^t_{\phantom{t}t}\rangle$ is entirely negative. It is maximal at the equator and minimal at a value which depends on both $\varepsilon$ and $T$, where it also attains its largest absolute value. If we fix $T$ and increase $\varepsilon$, the energy density at the equator decreases, and becomes slightly negative. As expected, for a fixed boundary amplitude, the energy density is an increasing function of temperature.

The component $\langle T^t_{\phantom{t}\phi}\rangle$ is negative for small $\varepsilon$ and temperatures close to $T_\mathrm{min}(\varepsilon)$, and vanishes at the poles and at the equator, similarly to the soliton case. Generally,~$\langle T^t_{\phantom{t}\phi}\rangle$ is a complicated function of $T$ and $\varepsilon$. For a fixed value of $\varepsilon$, it is an increasing function of temperature, and for large enough temperatures it becomes large and entirely positive. In particular, when $\varepsilon\rightarrow 2$, the density develops a large positive maxima peaked at $\theta=\pi/4$.

The anisotropic pressures capture some qualitative behaviour of the system. The normal stress component along $\theta$ is positive for all $\varepsilon$ and $T$, and for a fixed rotation amplitude is an increasing function of temperature. Furthermore, for a fixed temperature, it is an increasing function of $\varepsilon$. In the large temperature, large $\varepsilon$ regime, $\langle T^{\theta}_{\phantom{\theta}\theta}\rangle$ develops a broad positive maxima, peaked in the vicinity of $\theta=\pi/4$. The normal stress component in the direction of $\phi$ (not pictured) can be negative around the equator, and in the large $\varepsilon$, $T$ regime, it is positive with a sharp dip to negative values at $\theta=\pi/4$. 
%%%%%%%%%%%%%%%%%%%%%%%%%
%%%%%%%%%%%%%%%%%%%%%%%%%
\subsection{\label{subsec:Ther}Thermodynamics}
%%%%%%%%%%%%%%%%%%%%%%%%%
%%%%%%%%%%%%%%%%%%%%%%%%%
\subsubsection{Energy}
%%%%%%%%%%%%%%%%%%%%%%%%%
\begin{figure}[t]
\centering
  \begin{minipage}[t]{0.45\textwidth}
    \includegraphics[width=\textwidth]{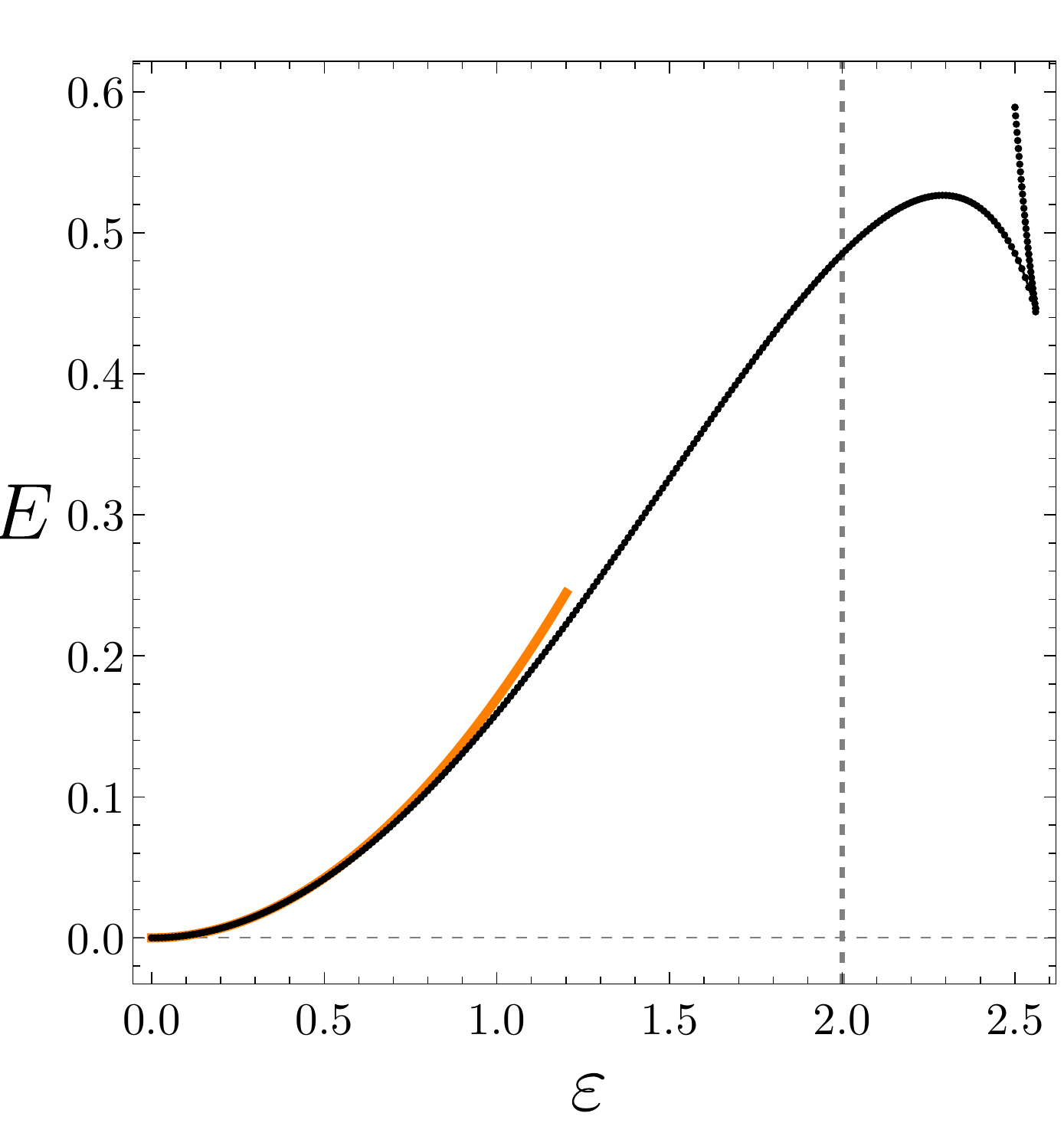}
  \end{minipage}
  \hfill
    \begin{minipage}[t]{.45\textwidth}
    \includegraphics[width=\textwidth]{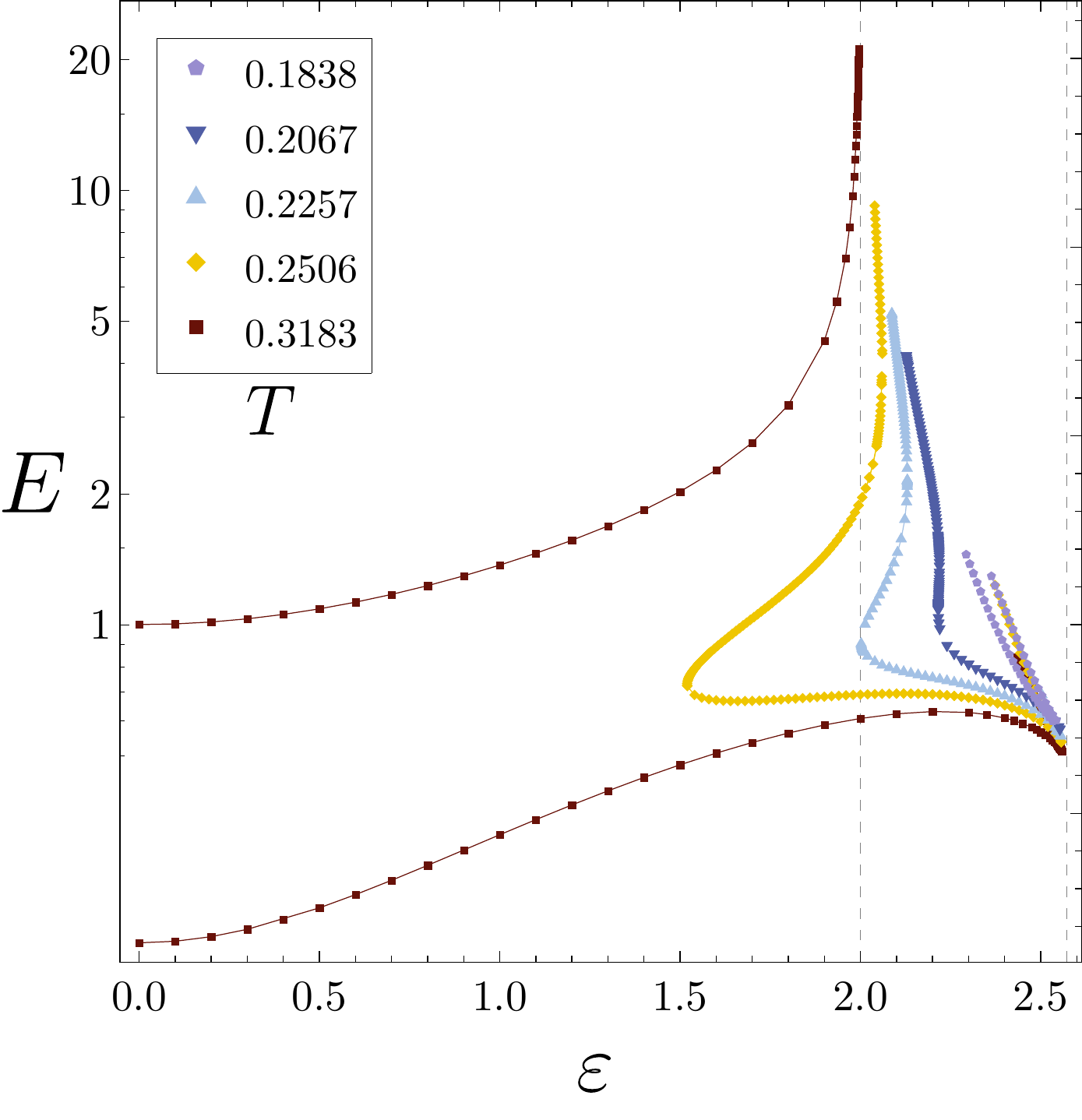}
  \end{minipage} 
  \caption{\textit{Left}: Soliton solution boundary energy against rotation parameter. The dashed vertical gridline marks the critical amplitude $\varepsilon=2$, and the bold orange line is the perturbative result. \textit{Right}: Log-linear plot of the energy for the black hole solutions against $\varepsilon$, for several values of $T$.}
  \label{fig:sole}
\end{figure}

 \begin{figure}[t] 
\centering
  \begin{minipage}[t]{1\textwidth}
    \includegraphics[width=\textwidth]{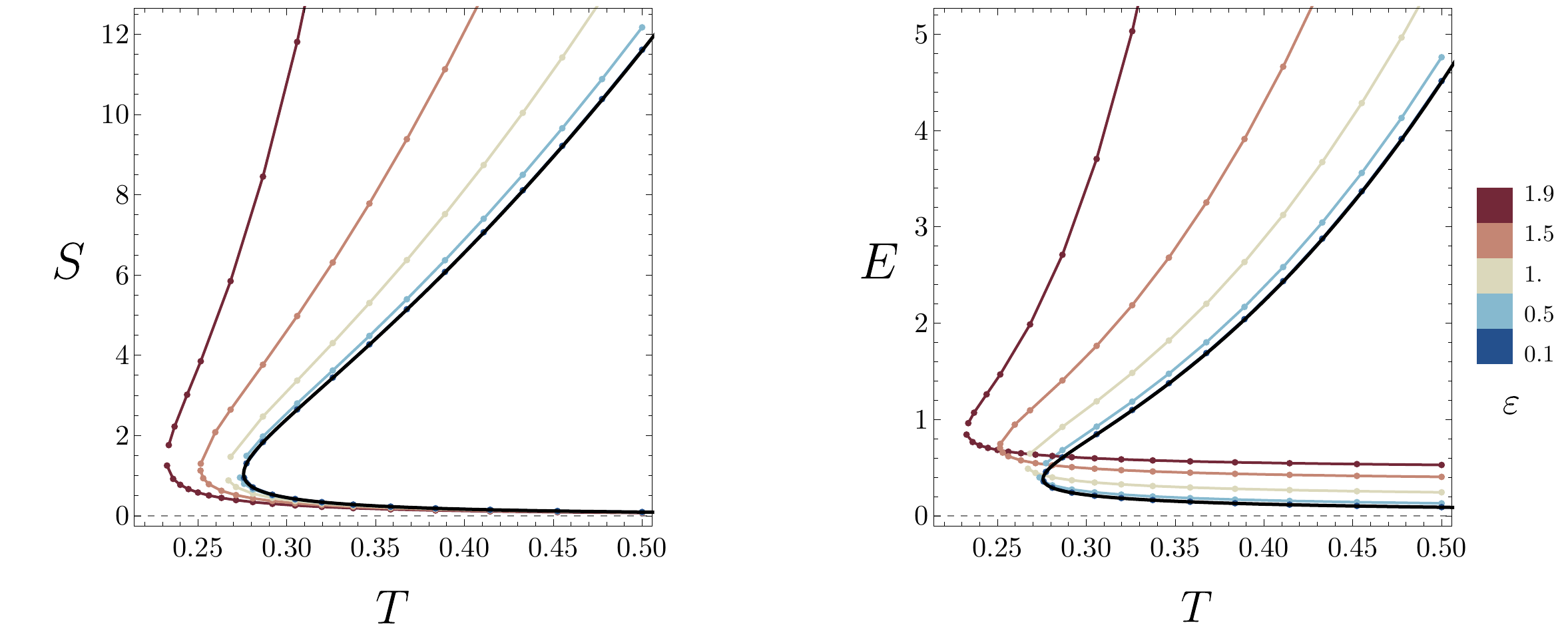}
  \end{minipage} 
    \caption{\textit{Left}: Black hole entropy against temperature for several values of the boundary rotation parameter $\varepsilon$, for $\varepsilon<2$. The upper branches correspond to the large solutions and the lower branches to the small solutions. \textit{Right}: Black hole energy as a function of temperature. Colouring codes the value of the boundary amplitude. The bold black curve is the Schwarzschild-AdS$_4$ solution with $\varepsilon=0$.}
  \label{fig:bhentmass}
\end{figure}

We obtain conserved charges in the usual way by integrating the angular momentum and energy densities~\cite{Brown:1992br}. We note that by construction, $\varepsilon=0$ solutions take the Schwarzschild-AdS$_4$ values, $T=(1+3y_+^2)/(4\pi y_+)$, $S=\pi y_+^2$, $M=y_+(1+y_+^2)/2$ and $G=y_+(1-y_+^2)/4$, discussed below.

As expected, the total angular momentum is identically zero. We present the energy\footnote{We note that strictly speaking the quantity that we are computing is not an energy, since $\partial_t$ is changing as we change $\epsilon$. A better defined quantity would be $\Delta E \equiv E_{\mathrm{black\;hole}}-E_{\mathrm{soliton}}$, which could have been regarded as the energy above the vacuum for each value of $\varepsilon$.} for black holes as a function of $\varepsilon$ and $T$ on the right panels of Fig.~\ref{fig:sole} and Fig.~\ref{fig:bhentmass}, respectively. For a fixed $\varepsilon$ and varying $T$, the large black holes increase their energy, and the small black holes decrease it, asymptotically approaching $E\rightarrow 0$. Further, for a fixed $T$ and increasing $\varepsilon$, the large black hole energy increases monotonically, as does the small black hole energy for $\varepsilon<2$, which is the opposite behaviour to that of the entropy (see left panel of Fig.~\ref{fig:bhentmass}). 

For values $2<\varepsilon<\varepsilon_\mathrm{max}(T)$, where $\mathrm{max}(T)$ was identified in the discussion of Fig.~\ref{fig:ent} and Fig.~\ref{fig:entropysmall}, the solution space has extra phases associated with cold temperatures. As we lower the temperature, the black hole energy approaches $\varepsilon=2$ from above, but unlike entropy, it appears to blow up for both small and large solution branches. As we lower $T$ further, black holes eventually exist only for $\varepsilon>2$, and the two branches appear to have similar values of energy.

The holographic energy for the soliton is presented on the left panel of Fig.~\ref{fig:sole}. With increasing boundary parameter $\varepsilon$, the energy is increasing until it peaks at $\varepsilon\simeq 2.29$ with maximum energy $E_\mathrm{max}\simeq 0.527$ for the small branch. The large branch has larger energy, which increases steeply when $\varepsilon\rightarrow 2^+$. We constructed these solutions numerically up to $\varepsilon=2.3$, however numerical errors in this region of the parameter space become increasingly large.
%%%%%%%%%%%%%%%%%%%%%%%%%
\subsubsection{Grand-canonical ensemble}
%%%%%%%%%%%%%%%%%%%%%%%%%
 \begin{figure}[t]
\centering
  \begin{minipage}[t]{1\textwidth}
    \includegraphics[width=\textwidth]{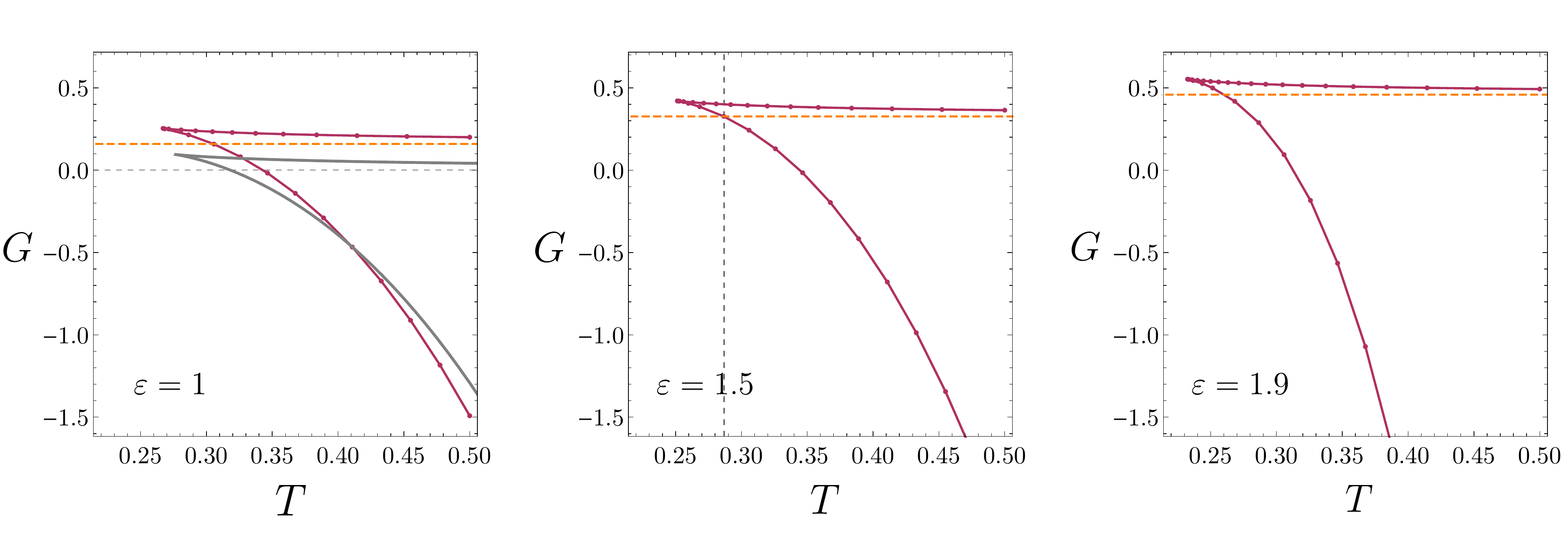}
  \end{minipage} 
  \caption{\textit{Left}: Gibbs free energy for the black holes against temperature for a fixed boundary rotation parameter $\varepsilon=1$ (red data points). The orange gridline shows the value for the corresponding AdS$_4$ soliton with the same $\varepsilon$. The grey solid curve traces the values of Schwarzschild-AdS$_4$, and the grey-dashed gridline shows $G_\mathrm{AdS}=0$. \textit{Middle}: Similar plot for $\varepsilon=1.5$, with the horizontal gridline marking the intersection temperature $T_\mathrm{HP}(\varepsilon)$. To the right of this line the large black holes will be the dominant solutions, whilst solitons dominate to the left. \textit{Right}: As we approach the critical value of $\varepsilon$, the phase transition temperature decreases.}
  \label{fig:bhgibbs}
\end{figure}
 
 \begin{figure}[t]
\centering
    \begin{minipage}[t]{1\textwidth}
    \includegraphics[width=\textwidth]{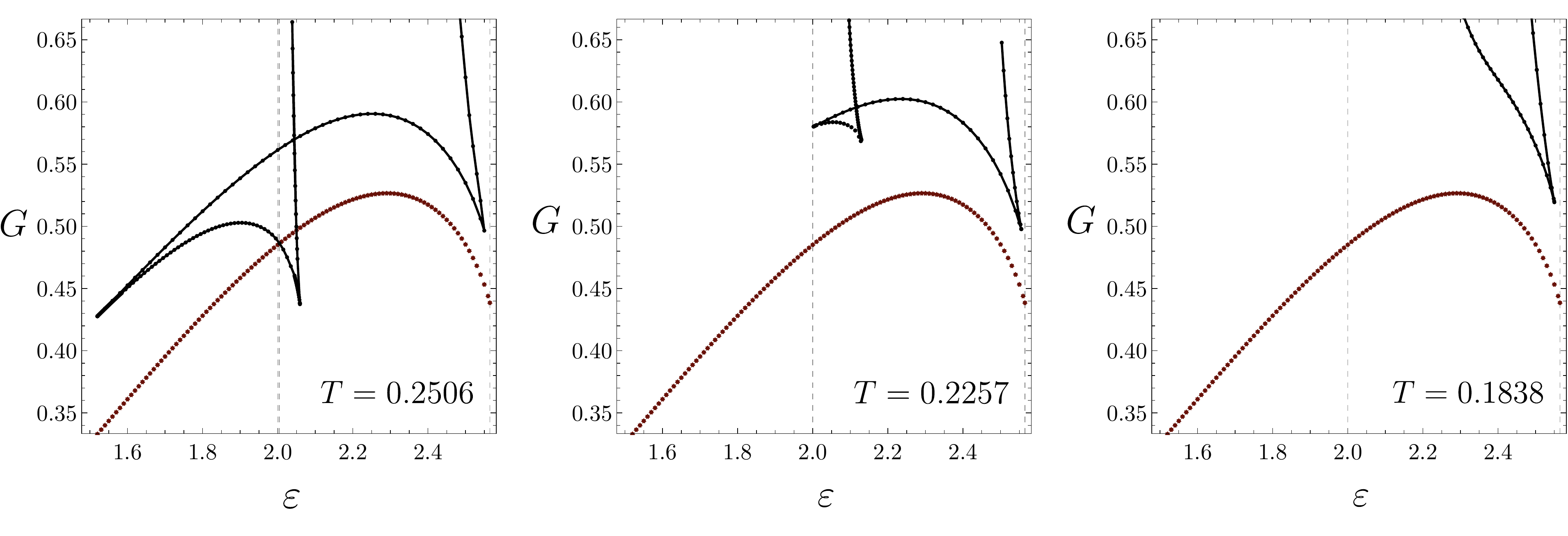}
  \end{minipage}
  \caption{Gibbs free energy against the boundary parameter for low temepreature black holes. It exhibits a swallowtail phase transition between the inelastic black hole phase and the soliton (brown stars).}
  \label{fig:Gsmall}
\end{figure}

Because we can fix the boundary chemical potential and temperature $T$, the appropriate thermodynamic ensemble for analysing our solutions is the grand-canonical ensemble. In this ensemble the preferred phase minimises the Gibbs free energy
\begin{equation}
G=E-TS\,.
\end{equation}
Note that in this system the energy, $E$, absorbs the chemical potential term, and in particular for solitons we have $G=E$. This is further supported by the first law of thermodynamics  for these solutions, derived following~\cite{Iyer:1994ys,Papadimitriou:2005ii,Compere:2007vx}, and given by
\begin{equation} 
\mathrm{d}E=T\,\mathrm{d}S+4\pi\int_0^{\pi/2} \mathrm{d}\Omega(\theta) j(\theta)\,\sin\theta\,\mathrm{d}\theta,
\label{eq:law}
\end{equation}
where $\Omega(\theta)=\varepsilon \cos\theta$ is the boundary angular velocity density, and $j(\theta)$ is the angular momentum density. In order to check the first law we constructed soliton solutions perturbatively to second order, the procedure for which is detailed in Appendix~\ref{sec:per}. We use the second order expansion to calculate the boundary free energy to be $E=8 \varepsilon ^2/(15 \pi)$, which matches our numerical solutions to small $\varepsilon$ amplitude~(see left panel of Fig.~\ref{fig:sole}) and satisfies the first law~(\ref{eq:law}) to first order in $\varepsilon$. Numerically, the first law is satisfied to at least to $0.1\%$ for the small soliton branch, to $2\%$ for the large branch, and to $0.1\%$ for the black holes (with $2\%$ for the second set of the small black holes), while varying~$\varepsilon$. For black holes\footnote{Large black holes, and small black holes up to the turning point in the parameter space.} it is also satisfied to $0.1\%$, while varying temperature and keeping the boundary profile fixed.

We next discuss our results in terms of free energy. In Fig.~\ref{fig:bhgibbs}, we present $G$ for the black hole solutions as a function of the temperature for some values of the rotation amplitude. For a fixed $\varepsilon$, the corresponding soliton energy (see Fig.~\ref{fig:sole}) is constant, which is pictured as a line. The intersection of the two marks the generalised Hawking-Page~transition~\cite{Hawking:1982dh}\footnote{The Hawking-Page transition occurs for $\varepsilon=0$ and $T=1/\pi$ (see the left panel of  Fig.~\ref{fig:bhgibbs}).},~and the transition temperature $T_\mathrm{HP}(\varepsilon)$ is a decreasing function of $\varepsilon$. For temperatures higher than the transition value, the large black hole phase is the dominant one, whilst for ${T<T_\mathrm{HP}(\varepsilon)}$ the soliton is the preferred phase. 

For $T<T^\mathrm{Schw}_\mathrm{HP}=1/\pi$, we observe curious phase transitions in $G$, characterised by a ``swallowtail'' behaviour in the Gibbs free energy~\cite{Chamblin:1999tk,Chamblin:1999hg} (Fig.~\ref{fig:Gsmall}). This occurs because large black holes start to exist past $\varepsilon=2$, where there is a turning point in the parameter space resulting in a four-fold non-uniqueness, and thus there exists a moduli space for which we have four possible black hole phases (and two soliton phases). We find that for $0.24<T<1/\pi$, there is a temperature dependent  interval $\varepsilon_\mathrm{HP}(T)<\varepsilon<\varepsilon_\mathrm{\star}(T)$ with $\varepsilon_\mathrm{\star}(T)>2$, where the cold, inelastic black holes dominate the ensemble. Finally, we note that the small black hole branch is never favoured thermodynamically, as it always has higher free energy than both the large black hole and the soliton. The corresponding phase diagram is illustrated in Fig.~\ref{fig:phase}. We also present $G$ as a function of $\varepsilon$ for a fixed temperature in Appendix~\ref{sec:add} (see the left panel of Fig.~\ref{fig:massS}).
 
 \begin{figure}[t]
\centering
    \begin{minipage}[t]{0.45\textwidth}
    \includegraphics[width=\textwidth]{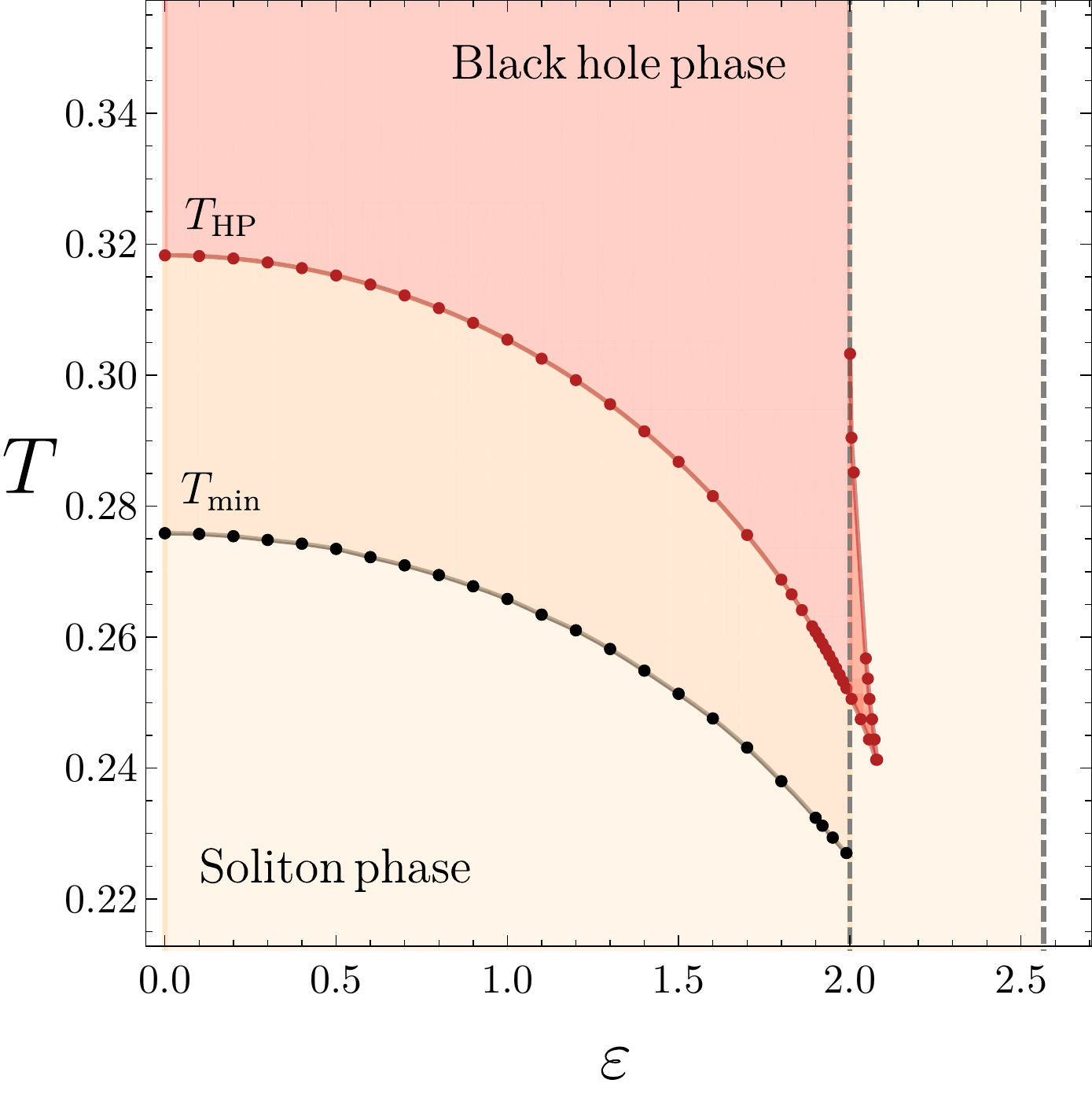}
  \end{minipage}
  \caption{ Phase diagram for the rotationally polarised solutions. For $\varepsilon<2$, black holes exist above $T_\mathrm{min}(\varepsilon)$ (black data points) and are the dominant phase above the $T_\mathrm{HP}(\varepsilon)$ curve (red area) which marks the corresponding Hawking-Page transition. Below the transition line (red data points), the dominant phase is the soliton. For $\varepsilon>2$, there is a cold inelastic black hole phase (red wedge). The elsewhere subdominant black holes exist up to some maximum $\varepsilon_\mathrm{max}(T)$, but are hidden by the soliton phase.}
  \label{fig:phase}
\end{figure} 
%%%%%%%%%%%%%%%%%%%%%%%%
%%%%%%%%%%%%%%%%%%%%%%%%
\section{\label{sec:sta}Stability}
%%%%%%%%%%%%%%%%%%%%%%%%
%%%%%%%%%%%%%%%%%%%%%%%%
\subsection{\label{subsec:quasol}Quasinormal modes}
%%%%%%%%%%%%%%%%%%%%%%%%
To investigate the stability, we consider a wave equation for a neutral, minimally coupled scalar field
\begin{equation}
\Box\Phi=0\,.
\label{eq:box} 
\end{equation}
Since both solitons and black holes have two commuting Killing fields, $\partial_t$ and $\partial_\phi$, we can decompose the perturbations as follows
$$
\Phi=\widehat{\Phi}_{\omega,m}(x,y)e^{-i\,\omega\,t+i\,m\,\phi}\,.
$$
The resulting equation for $\widehat{\Phi}_{\omega,m}$ is linear and can be regarded as a Sturm-Liouville problem for the eigenpair $\{\widehat{\Phi}_{\omega,m},\omega\}$, once suitable boundary conditions are imposed. The set of all $\omega$ are called the quasinormal mode (QNM) frequencies. For each value of $m$, there is an infinite number of QNM frequencies labelled by two integers. These are in one to one correspondence with the number of nodes along the $x$ and $y$ directions. If $\varepsilon=0$, we can use the background spherical symmetry to decompose the angular part in terms of spherical harmonics. These are labeled by $\ell$ and $m$, with $\ell-|m|$ counting the number of nodes along the polar angular direction. We will be interested in following the modes with $\ell=|m|$ as a function of $\varepsilon$, that is to say, modes with no nodes along the polar direction. Furthermore, we will also be focusing on the modes which have no nodes along the radial direction, \emph{i.e.} with zero overtone. We focus on these modes since they have the largest $|\mathrm{Im}(\omega)|$ for fixed~$m$.

We follow the method of~\cite{Cardoso:2013pza,Dias:2013sdc,2016CQGra..33m3001D} and solve the resulting eigenvalue equation system using a Newton-Raphson method, on a fixed numerical solution background with a corresponding numerical grid.

The QNM spectrum with $\varepsilon=0$ is well--known: the Schwarzschild AdS$_4$ black hole QNM frequencies are given in~\cite{Horowitz:1999jd}, and the scalar normal mode frequencies of AdS$_4$ are given by integers $L\,\omega=3+\ell$. As we vary the boundary rotation parameter, the imaginary part of the black hole frequency $\omega$ becomes positive, indicating the presence of an instability.

The boundary conditions imposed at $y=0$ will depend strongly on the background solution we wish to perturb.
%%%%%%%%%%%%%%%%%%%%%%%%
\subsubsection{\label{subsec:qua}Soliton}
%%%%%%%%%%%%%%%%%%%%%%%%
\begin{figure}[t]
\centering
    \begin{minipage}[t]{0.48\textwidth}
    \includegraphics[width=\textwidth]{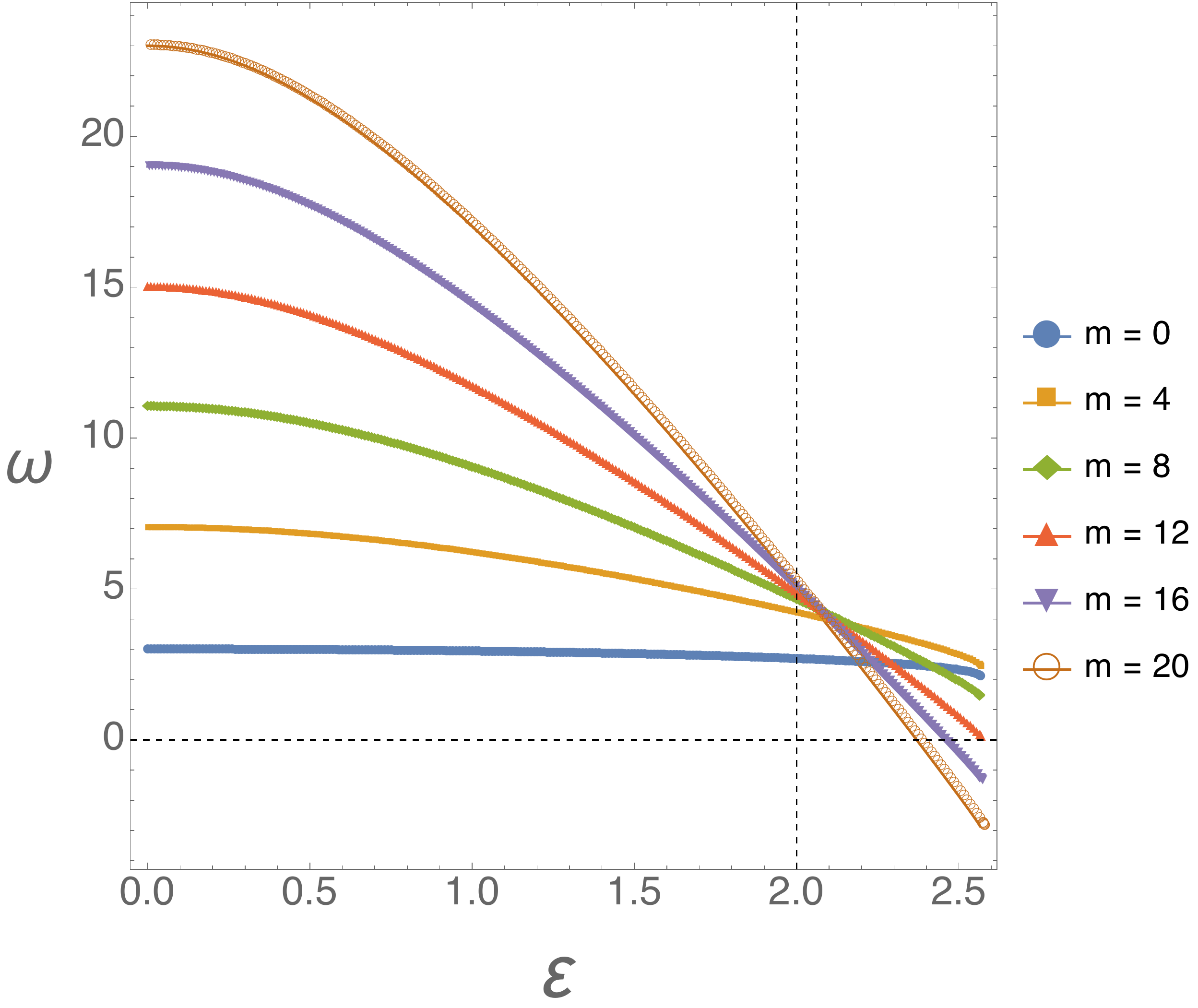}
  \end{minipage}
  \hfill
      \begin{minipage}[t]{0.41\textwidth}
    \includegraphics[width=\textwidth]{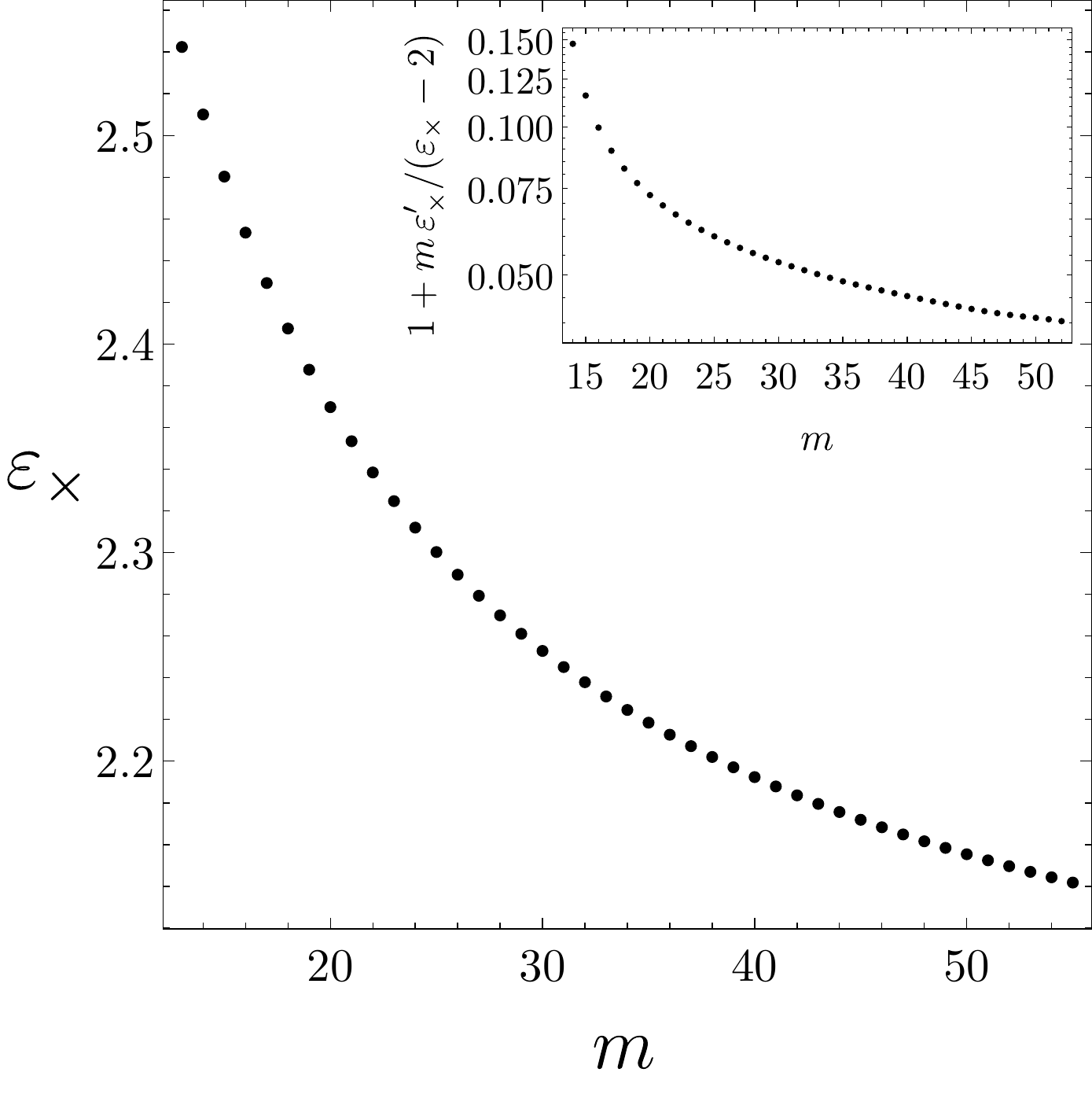}
  \end{minipage}
  \caption{\textit{Left:} Normal mode frequencies with $\ell=m$ vs the boundary rotation parameter~$\varepsilon$ for the soliton metric. At $\varepsilon=0$ these reduce to the usual AdS$_4$ frequencies $L\,\omega=3+\ell$. Dashed gridlines show where $\varepsilon=2$ and $\mathrm{Re}\,\omega=0$. \textit{Right:} Rotation amplitude $\varepsilon_\times$, where $\omega^2$ becomes negative, against~$m$. The inset shows the logarithmic derivative of $\varepsilon_\times$. We expect that in the limit $m\rightarrow\infty$, $\varepsilon_\times\rightarrow 2$.}
  \label{fig:solqnm}
\end{figure}

We first investigate scalar perturbations~(\ref{eq:box}) on soliton backgrounds, and consider the following decomposition
\begin{equation}
\Phi (t,x,y,\phi)=e^{-i \omega t}e^{i m\phi}y^{|m|} (1-y^2)^3 (1-x^2)^{|m|}\psi(x,y),
\end{equation}
where the factorisation ensures regularity at the origin for a smooth function $\psi(x,y)$. The boundary conditions at $y=1$ and $x=1$ are obtained by solving the equation~(\ref{eq:box}) at the corresponding boundaries and are given by
\begin{subequations}
\begin{equation}
|m|\, \psi(x,y)+\partial_y\psi(x,y)=0 \quad\text{at}\quad y=1\,,
\end{equation}
and
\begin{equation}
\partial_x\psi(x,y)=0\quad\text{at}\quad x=1\,.
\end{equation}
\end{subequations}
\noindent We also require $\partial_x\psi(x,y)=0$ at $x=-1$ and $\partial_y\psi(x,y)=0$ at $y=0$.

On the left panel of Fig.~\ref{fig:solqnm}, we plot the normal mode frequencies $\omega$ with $\ell=m$, against~$\varepsilon$. The $\omega$ first becomes negative when $m=13$, and with each subsequent mode, $\omega$ becomes negative at a lower value of $\varepsilon$. We denote the value at which $\omega$ changes sign by~$\varepsilon_\times$, and we find that $\varepsilon_{\times}-2\sim1/m$ at large $m$\footnote{Higher modes are more difficult to obtain numerically, and thus are less reliable, however, we find that the scalar eigenfunctions for these solutions have good convergence properties. For instance, we observe the variation with the grid size in a mode with $m=40$, $\varepsilon=2.5$ to be less than $10^{-7}$.} (see right panel of Fig.~\ref{fig:solqnm}).

In order to better resolve where $\omega$ becomes negative for a given $m$, we can also set $\omega=0$ and look for zero--modes directly, and the results for this approach are presented in Fig.~\ref{fig:solqnm}, right. We find that this method gives us the same values of $\varepsilon_\times$ as the ones inferred from computing $\omega$ directly. For each pair of values of $\{m,\varepsilon_\times\}$, we expect a given hairy family with nontrivial $\Phi$ to condensate. Whether this new family of hairy solutions will extend to large or small values of $\varepsilon$, is likely to depend on the model. We have constructed soliton solutions with complex scalar hair within cohomogeneity two, and will present these solutions elsewhere. These were built by minimally coupling a massless complex scalar field to gravity, and are similar in spirit to the black holes with a single Killing field of \cite{Dias:2011at} and to the holographic Q-lattices of \cite{Donos:2013eha}.

%%%%%%%%%%%%%%%%%%%%%%%%
\subsubsection{\label{subsec:quabh}Black hole}
%%%%%%%%%%%%%%%%%%%%%%%%

\begin{figure}[t]
\centering
    \begin{minipage}[t]{.45\textwidth}
    \includegraphics[width=\textwidth]{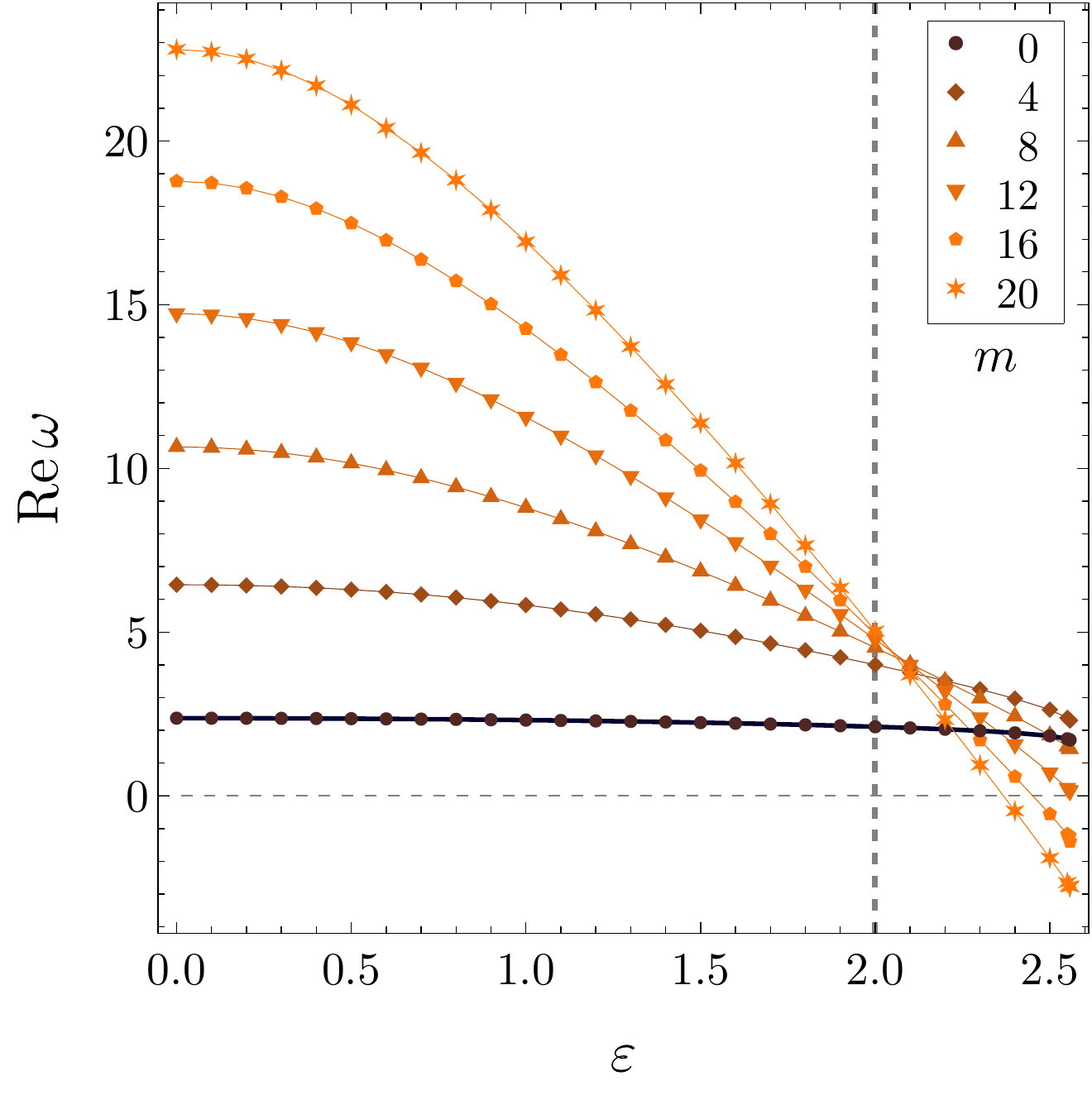}
  \end{minipage}
    \hfill
      \begin{minipage}[t]{0.45\textwidth}
    \includegraphics[width=\textwidth]{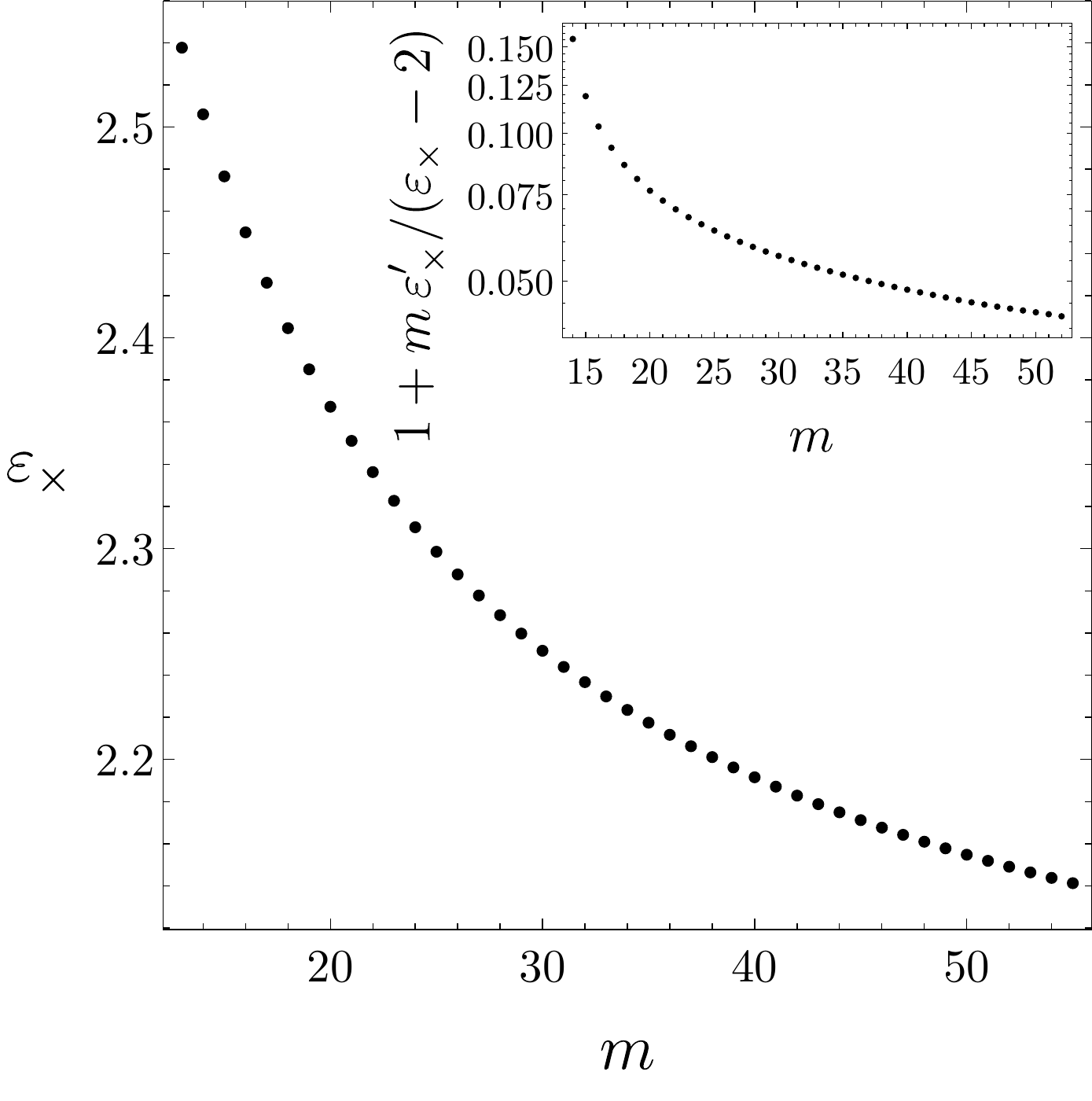}
  \end{minipage}
  \caption{\textit{Left:} Real part of quasinormal mode frequency with $\ell=m$ against the boundary rotation parameter~$\varepsilon$, for small black holes with $T=1/\pi$. At $\varepsilon=0$, these reduce to the Schwarzschild-AdS$_4$ values. The bold dashed gridline shows the critical $\varepsilon=2$ value. \textit{Right:} Rotation amplitude where the zero--mode is for a given $m$, against $m$. The inset shows the logarithmic derivative of $\varepsilon_\times$. We expect that in the limit $m\rightarrow\infty$, $\varepsilon_\times\rightarrow 2$.}
  \label{fig:bhqnm}
\end{figure} 

Next, we consider the black hole solutions. A Frobenius analysis at the integration boundaries motivates the following separation ansatz
\begin{equation}
\Phi (t,x,y,\phi)=e^{-i \omega t}e^{i m\phi}y^{-i\frac{2 \omega y_+}{1 + 3 y_+^2}} (1-y^2)^3 (1-x^2)^{|m|}\psi(x,y)\,,
\end{equation}
where we have implicitly imposed regularity in ingoing Eddington-Finkelstein coordinates~\cite{Horowitz:1999jd}.

The boundary conditions for $\psi(x,y)$ at $y=1$ and $x=1$ are then found by solving (\ref{eq:box}) and demanding that $\psi(x,y)$ has a regular Taylor expansion at $y=1$ and $x=1$. This in turn gives:
\begin{subequations}
\begin{equation}
-2 i\,y_+ \omega\psi(x,y)+(1+3 y_+^2)\partial_y \psi(x,y)=0\quad \text{at}\quad y =1\,,
\end{equation}
and
\begin{equation}
\partial_x\psi(x,y)=0\quad\text{at}\quad x=1\,.
\end{equation}
\end{subequations}
\noindent As with the soliton, we require $\partial_x\psi(x,y)=0$ at $x=-1$ and $\partial_y\psi(x,y)=0$ at $y=0$.

Numerical results are presented in Fig.~\ref{fig:bhqnm}. On the left panel, we plot the real part of the QNM frequencies with $\ell=m$ against $\varepsilon$, for a fixed $T=1/\pi$ small black hole branch. We find that the behaviour of $\mathrm{Re}\,\omega$ is very similar to that of the soliton, and the first unstable mode appears at $m=13$. The imaginary part, $\mathrm{Im}\,\omega$, becomes positive exactly where $\mathrm{Re}\,\omega$ crosses the zero axis, indicating a zero--mode. If we set $\omega=0$ and look for these directly, we again find a $1/m$ fall--off (see right panel of Fig.~\ref{fig:bhqnm}) at large $m$. The growth rate of the instability is less than $\Gamma\sim 10^{-20}$, so it sets in incredibly slowly\footnote{We believe that the rate will increase for low temperatures, but these backgrounds require very high resolutions.}. We expect that these solutions will become unstable in modes with high $m$ as we increase the boundary rotation past $\varepsilon=2$ and, if we assume the $1/m$ decay, the instability should manifest itself slowly with increasing $\varepsilon$. Finally, for hot, large black holes we do not see any sign of instability, as these solutions never cross~$\varepsilon=2$.

This is a surprising result - the ergoregion appears to induce hairy solutions in both
phases. This opens up new possibilities for the phase diagram, some of which we will
discuss in the following sections. This section was devoted entirely to the stability analysis
of scalar perturbations, and the issue of the stability of our backgrounds with respect to
gravitational perturbations is an open problem. However, if we assume that the physics involved in this instability is any similar to the usual superradiance instability, we expect the gravitational perturbations to be unstable and to have larger instability growth rates.

%%%%%%%%%%%%%%%%%%%%%%%%
%%%%%%%%%%%%%%%%%%%%%%%%
\section{\label{sec:fie}Field Theory}

Our calculation in this section follows \emph{mutatis mutandis} that of Section 4.1 of \cite{Costa:2015gol}. We will first consider a conformally coupled scalar field satisfying the Klein-Gordon equation on $\mathbb{R}_t\times S^2$. The idea is to determine the eigenfunctions on this background, and use those to determine the spectrum of a conformally coupled scalar on the rotating background (\ref{eq:boundary}).

If $\varepsilon=0$, the Klein-Gordon equation for a conformally coupled scalar in $1+2$ reads (where we set the radius of the $S^2$ to unity):
$$
-\frac{\partial^2}{\partial t^2}\varphi+\nabla^2_{S^2}\varphi=\frac{1}{4}\varphi\,.
$$

The Hamiltonian eigenstates are given by spherical harmonics $|\ell,m\rangle$, with wave-functions\footnote{Note that Spherical harmonics are normalised such that $\int_0^{2\pi}\mathrm{d}\phi\int_{0}^{\pi}\mathrm{d}\theta\,\sin\theta\,|Y_{\ell m}(\theta,\phi)|^2=1$, for all $\ell$ and $|m|\leq1$.}
$$
\langle \theta,\phi|\ell,m\rangle =Y_{\ell m}(\theta,\phi)\,.
$$
In this basis, the Hamiltonian $\hat{H}=i\partial/\partial t$ has the following spectrum
$$
\hat{H}|\ell,m\rangle = \left(\ell+\frac{1}{2}\right)|\ell,m\rangle\,.
$$
After turning on the electric field, the Hamiltonian is deformed into
$$
H = \sqrt{\frac{1}{4}-\nabla_{S^2}+\frac{\varepsilon^2}{16}\sin^4\theta}+i\,\varepsilon\,\cos\theta\,\frac{\partial}{\partial \phi}\,.
$$
Since $H$ commutes with $\partial/\partial \phi$, it will remain diagonal in the azimuthal quantum number $m$. However, it is no longer diagonal in $\ell$, and its elements have to be computed numerically:
\begin{multline}
\langle \ell^\prime m^\prime|H|\ell m \rangle = \delta_{m,m^\prime}\Bigg\{\int_0^{2\pi}\mathrm{d}\,\varphi\int_0^{\pi}\mathrm{d}\theta \,\sin\theta\,\bar{Y}_{\ell^\prime m}\,\sqrt{\ell\left(\ell+\frac{1}{2}\right)+\frac{\varepsilon^2}{16}\sin^4\theta}\,Y_{\ell m}\\ -m\,\varepsilon\,\left[\delta_{\ell,\ell^\prime-1}\sqrt{\frac{(\ell+1+m)(\ell+1-m)}{(2\ell+1)(2\ell+3)}}+(\ell\leftrightarrow \ell^\prime)\right]\Bigg\}.
\end{multline}

\begin{figure}[t]
\centering
    \begin{minipage}[t]{0.45\textwidth}
    \includegraphics[height=0.3\textheight]{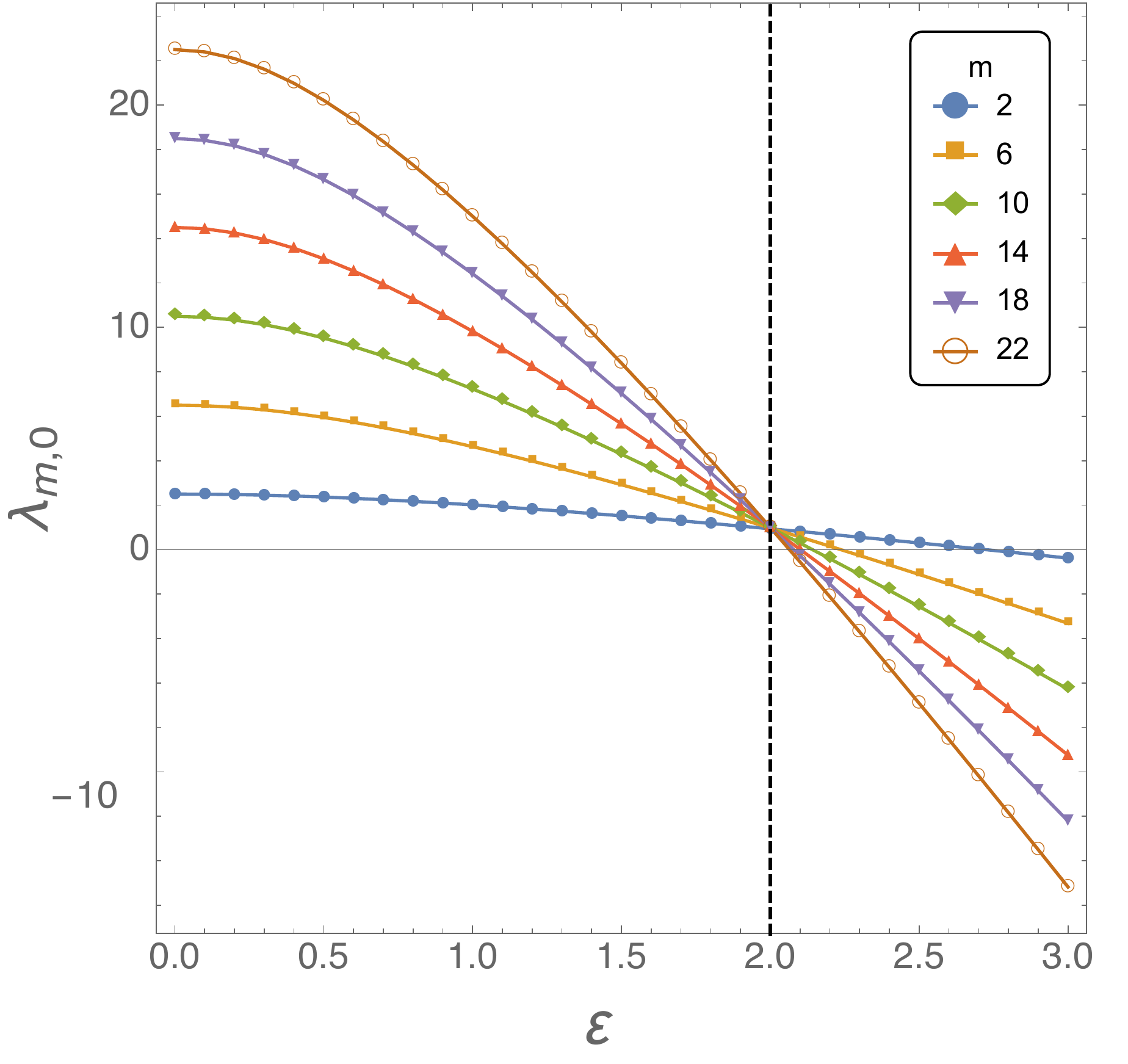}
  \end{minipage}
  \hfill
      \begin{minipage}[t]{0.45\textwidth}
    \includegraphics[height=0.3\textheight]{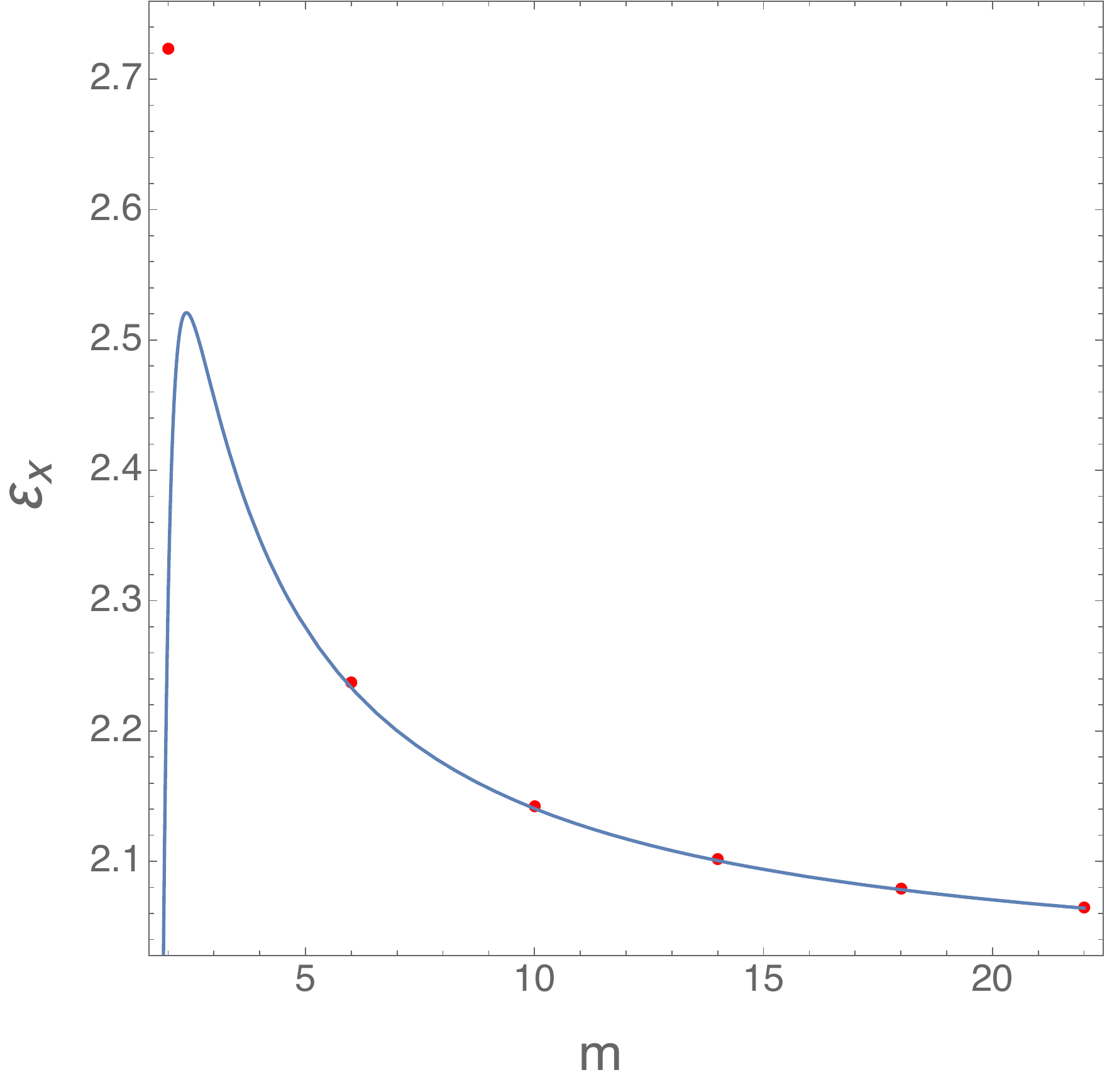}
  \end{minipage}
  \caption{\textit{Left:} Lowest lying mode eigenvalue of the boundary hamiltonian as a function of the boundary rotation parameter~$\varepsilon$. At $\varepsilon=0$ these reduce to $\ell+1/2$. \textit{Right:} $\varepsilon$ above which $\lambda_{m,0}$ becomes negative, plotted as a function of $m$. The solid line represents our analytic approximation (\ref{eq:critepsilon}) valid at large $m$.}
  \label{fig:cft}
\end{figure}

For each value of $m$, we can truncate the values of $\ell$ and $\ell^\prime$ up to some maximum value $\ell_{\max}$ and determine the eigenvalues numerically. One can show that if we are only interested in a few low lying modes, the convergence is exponential in $\ell_{\max}$. The results are illustrated in Fig.~\ref{fig:cft} where we plot the lowest lying eigenvalue of $\langle \ell^\prime m |H|\ell m \rangle$, which we denote by $\lambda_{m,0}$, for several values of $m=2,6,10,14$. The similarity between the left panel of Fig.~\ref{fig:solqnm} and the left panel of Fig.~\ref{fig:cft} is striking.

One can go further and compute the critical value of $\varepsilon$ above which $\lambda_{m,0}$ becomes negative, and compare it with the right panel of Fig.~\ref{fig:solqnm}. We call this special value $\varepsilon$ $\varepsilon_{\times}$, in accordance to section \ref{sec:sta}. In the large $m$ limit, the Hamiltonian can be diagonalised using a WKB approximation, and can be used to determine $\varepsilon_\times$. We have performed this calculation to order $m^{-12}$, and it turns out that
\begin{multline}
\varepsilon_{\times}=2+\frac{\sqrt{2}}{m}-\frac{3}{32 m^2}+\frac{1}{32 \sqrt{2}m^3}-\frac{85}{2048 m^4}-\frac{5427}{32768 \sqrt{2} m^5}-\frac{132879}{524288 m^6}-\frac{8720967}{8388608 \sqrt{2}m^7}\\-\frac{322048277}{134217728 m^8}-\frac{109314890677}{8589934592 \sqrt{2}m^9}-\frac{5202962510427}{137438953472 m^{10}}-\frac{550282627773161}{2199023255552 \sqrt{2}m^{11}}\\-\frac{32005862571264331}{35184372088832 m^{12}}+\mathcal{O}(m^{-13})\,.
\label{eq:critepsilon}
\end{multline}
The validity of this approximation is tested on the right panel of Fig.~\ref{fig:cft}, where we can see that for $m\geq10$, the error is below $0.1\%$ and for $m\geq22$ it is below $0.02\%$.
%%%%%%%%%%%%%%%%%%%%%%%%
%%%%%%%%%%%%%%%%%%%%%%%%
\section{\label{sec:eqv}Equilibrium of spinning test particles}
%%%%%%%%%%%%%%%%%%%%%%%%
%%%%%%%%%%%%%%%%%%%%%%%%
The spacetimes we consider are stationary, however, the black hole solution has zero angular velocity on the horizon, and vanishing total angular momentum. If we place a small Kerr hole on the axis of symmetry, there will be a general relativistic ``spin--spin" force acting on the spinning black hole, which for stationary fields can be thought of as analogous to the electromagnetic ``dipole--dipole" force with opposite sign \cite{PhysRevD.6.406,Natario:2007pu,PhysRevD.93.104006}. In this section, we will investigate whether it is possible to have a spinning test particle hovering above the central black hole, balanced by the pull towards the black hole and the ``spin--spin" interaction with the boundary.

Equilibria of spinning test particles can be studied using the Mathisson-Papapetrou (MP)  equations~\cite{Mathisson2010,Papapetrou248,Pirani:1956tn,592311013,doi10.1063/1.1704055,Dixon1964,Dixon499} (for a review see \cite{1999MNRAS.308..863S}), which are derived by taking a multipole expansion about a suitably defined reference worldline. Truncated at dipole order, they are given~by
\begin{align}
\begin{split}
\frac{D p^\mu}{Ds}&=-\frac{1}{2}R^\mu_{\nu\rho\sigma}v^\nu S^{\rho\sigma}, \\
\frac{D S^{\mu\nu}}{Ds}&=p^\mu v^\nu-p^\nu v^\nu,
\end{split}
\label{eq:mp}
\end{align}
where $p^\mu$ is the four-momentum of the particle, which in general is not co-linear with the kinematic four-velocity  $v^\mu$. They are related by  $p^\mu=m u^\mu - u_\nu \dot{S}^{\mu\nu}$, where $S^{\mu\nu}$ is the anti-symmetric spin matrix and $m$ is a mass parameter\footnote{Here the dot is differentiation with respect to the proper time.}. We also require a supplementary spin condition to fix the center of mass, which we choose to be the Tulczyjew-Dixon condition\footnote{This condition states that  momentum is perpendicular to the covariant four-spin vector $s=\star(S\wedge p)$.} $p_\nu S^{\mu\nu}=0$. On the rotation axis, it is equivalent to the Mathisson-Pirani condition.

We will be concerned with the static case, thus setting $\dot{S}^{\mu\nu}=0$ and $\dot{p}^\mu=0$. This implies $m=M$ and $p^\mu=M u^\mu$, where $M$ is the rest mass of the particle~$-M^2=p_\mu p^\mu$. Together with the spin magnitude, defined as $S^2=S_{\mu\nu}S^{\mu\nu}/2$, these are constant throughout the motion. These conditions, in conjunction with the MP equations~(\ref{eq:mp}), are sufficient to determine the static ratio $S/M$ as a function of the radial coordinate. Equivalently, we can instead consider a conserved quantity which includes the spin potential energy
\begin{equation}
E=-p^\rho \xi_\rho+\frac{1}{2}\xi_{\rho;\sigma}S^{\rho\sigma},
\end{equation}
where $\xi=\partial_t$ is the time Killing vector field, and $E>0$ is the energy at infinity of a spinning particle. The first term is conserved for a non-spinning particle, and is positive when the Killing vector is future directed and timelike. Using the normalization $-M^2=p_\mu p^\mu$, we can find an effective potential for radial on-axis motion of a spinning particle. In order to find equilibrium points, we look for local minima of the potential $\dot{r}^2+V(r)=0$, which constrains both the energy $E$, and the ratio $S/M$.

 \begin{figure}[t]
\centering
  \begin{minipage}[t]{.45\textwidth}
    \includegraphics[width=\textwidth]{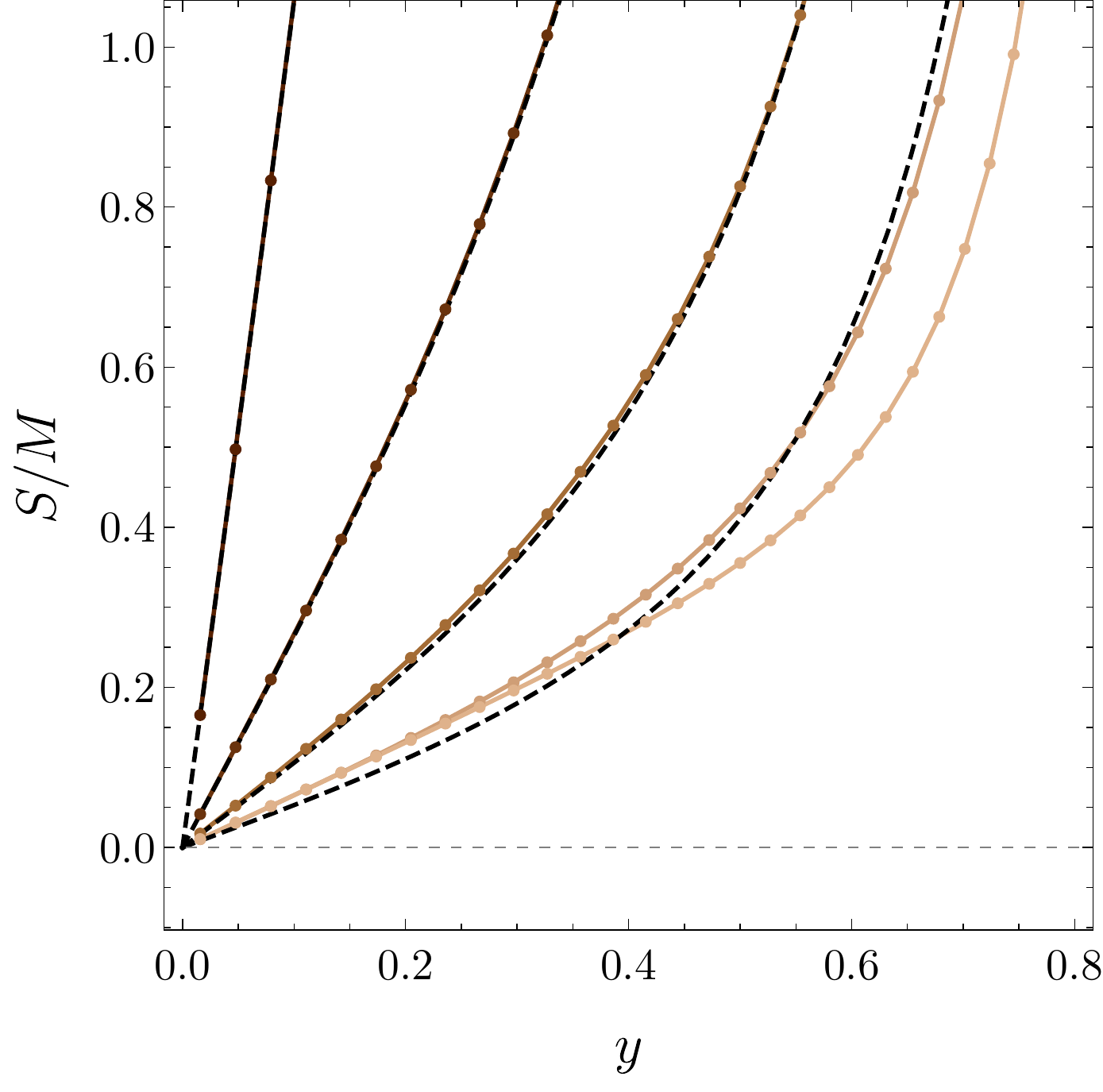}
  \end{minipage} 
  \hfill
    \begin{minipage}[t]{.43\textwidth}
    \includegraphics[width=\textwidth]{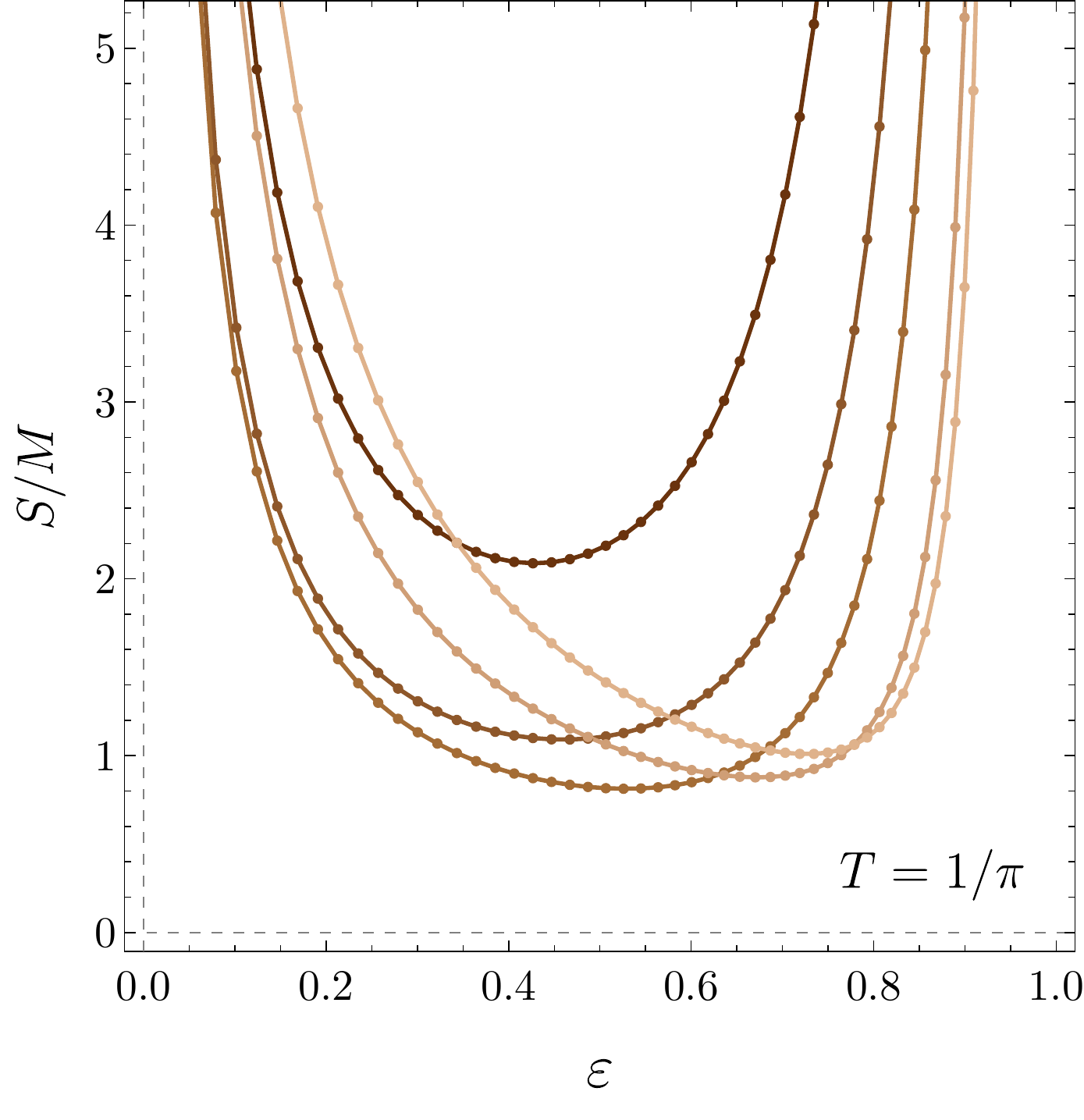}
  \end{minipage} 
  \caption{\textit{Left}: Spin to mass ratio for a spinning test particle in the soliton background with boundary parameter $\varepsilon=0.1,0.4,1,2,2.5$ (corresponding colours vary from dark to light). Dashed black lines show analytic perturbative approximations~(\ref{eq:smpert}) (pictured for $\varepsilon \leq 2$) which, for small $\varepsilon$, are in a very good agreement with numerical solutions. \textit{Right}:~Spin to mass ratio for large black holes with $T=1/\pi$, for $\varepsilon=0.1,0.5,1,1.5,1.93,1.99$ (corresponding colours vary from dark to light).}
  \label{fig:smratio}
\end{figure}

We calculate these quantities for the black hole metric~(\ref{eq:ansatzbh}) and the soliton metric~(\ref{eq:ansatzsol}). Because our coordinates are not well-defined on the symmetry axis $\theta=0$, we first perform a local coordinate change
\begin{equation}
\begin{split}
1-x^2=x_1^2+x_2^2, \qquad\qquad \tan{\phi}=x_2/x_1,
\end{split}
\end{equation}
and the symmetry axis is now given by $x_1=0$ and $x_2=0$. The supplementary condition implies that the only non-vanishing component of the spin matrix is $S^{x_1 x_2}$, making the on-axis calculation considerably easier. The effective potential is then found to be
\begin{equation}
\begin{split}
&V(r)=(g_{rr}g_{tt})^{-1}\left[g_{tt}+\left(\frac{E}{M}+\frac{S}{M}y^{\alpha}Q_6(1,y)\right)^2 \right],
\end{split}
\label{eq:eff}
\end{equation}
where $\alpha=1$ for the soliton, and $\alpha=2$ for the black hole. The spin potential energy term depends on the metric ansatz, and in the expression~(\ref{eq:eff}) it is evaluated explicitly. The total energy of a spinning particle is
\begin{equation}
\frac{E}{M}=\sqrt{-g_{tt}}-\frac{S }{M}y^{\alpha} Q_6(1,y),
\label{eq:emsol}
\end{equation}
\noindent and the spin to mass ratio of a static spinning test particle on the symmetry axis is
\begin{equation}
\frac{S}{M}=\frac{-\partial_y g_{tt}(1,y)}{2y^{\alpha-1}\sqrt{-g_{tt}(1,y)}(2^{\alpha-1} Q_6(1,y) + y \partial_y Q_6(1,y))}.
\label{eq:spin}
\end{equation}
\\
Numerical results for both solutions are presented in Fig.~\ref{fig:smratio}. For the test body approximation, the M\o ller radius describing the minimum size of a rotating body is $R\geq S/M$~\cite{587743163}, and it should be possible to make it arbitrarily small~\cite{PhysRevD.6.406}. For the black holes, we do not find such a limiting process. For the  soliton, this quantity vanishes as $y\rightarrow 0$, thus it is possible to place a spinning particle at this point, whose backreaction would lead to the formation of a small Kerr-AdS$_4$ black hole.
%%%%%%%%%%%%%%%%%%%%%%%%
\subsection{\label{subsec:eqs}Equilibrium for non-spinning test particles}
%%%%%%%%%%%%%%%%%%%%%%%%
For static spacetimes the equilibrium positions are independent of spin, and can be  achieved by balancing electromagnetic and gravitational forces, or coincide with the static radius due to a repulsive cosmological constant $\Lambda>0$. For Kerr black holes~\cite{carter1968}, it is well known that no equilibrium positions exist~\cite{PhysRevD.6.406}. Generally, conditions for equilibrium on the rotation axis are independent of the spin of the test particle. From~(\ref{eq:spin}), we can see that the requirement $S/M\rightarrow 0$ implies that the equilibrium conditions coincide with the minima of the static potential $-g_{tt}$.
 
 \begin{figure}[t]
\centering
  \begin{minipage}[t]{.45\textwidth}
    \includegraphics[width=\textwidth]{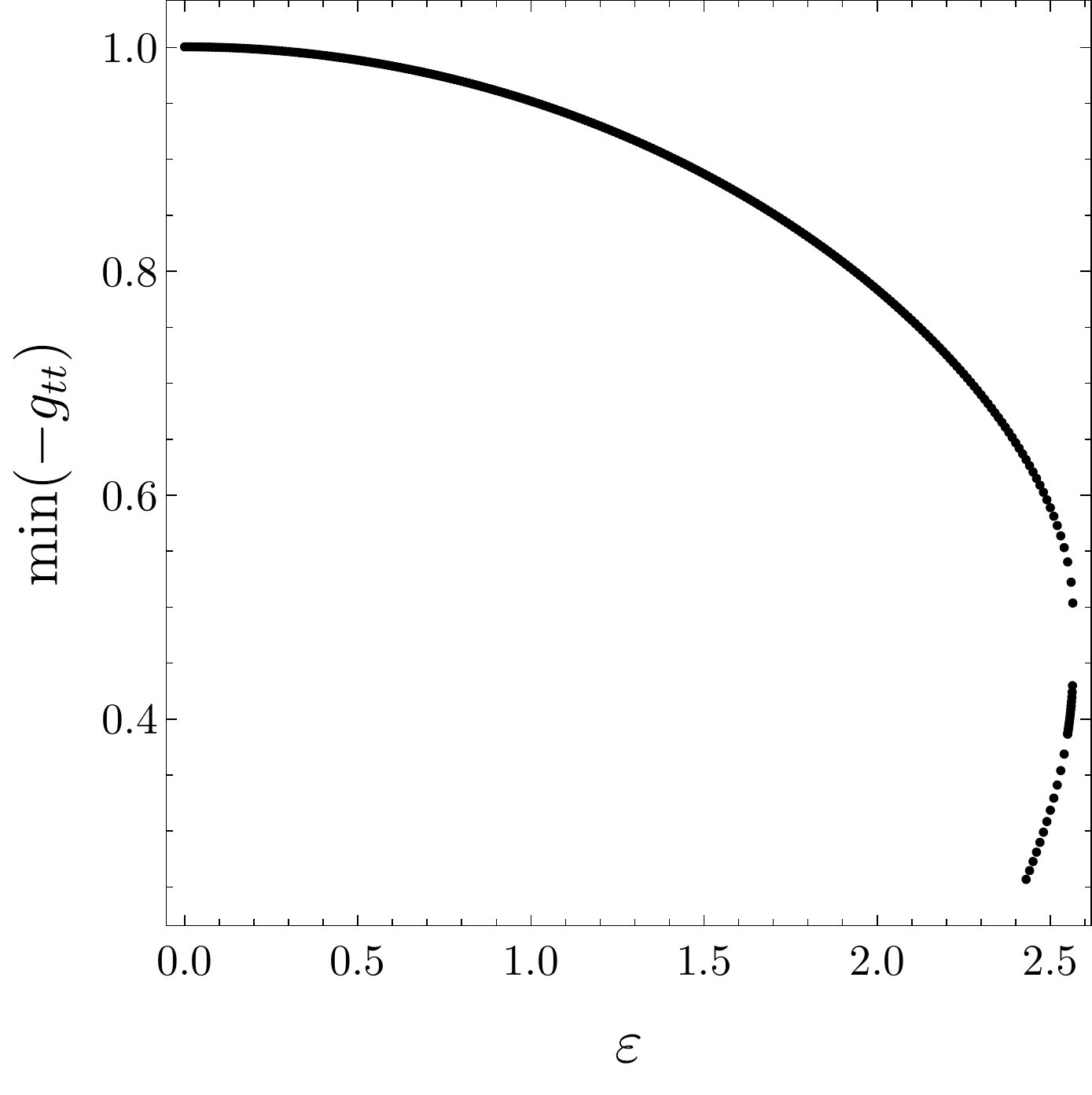}
  \end{minipage} 
  \hfill
      \begin{minipage}[t]{.45\textwidth}
    \includegraphics[width=\textwidth]{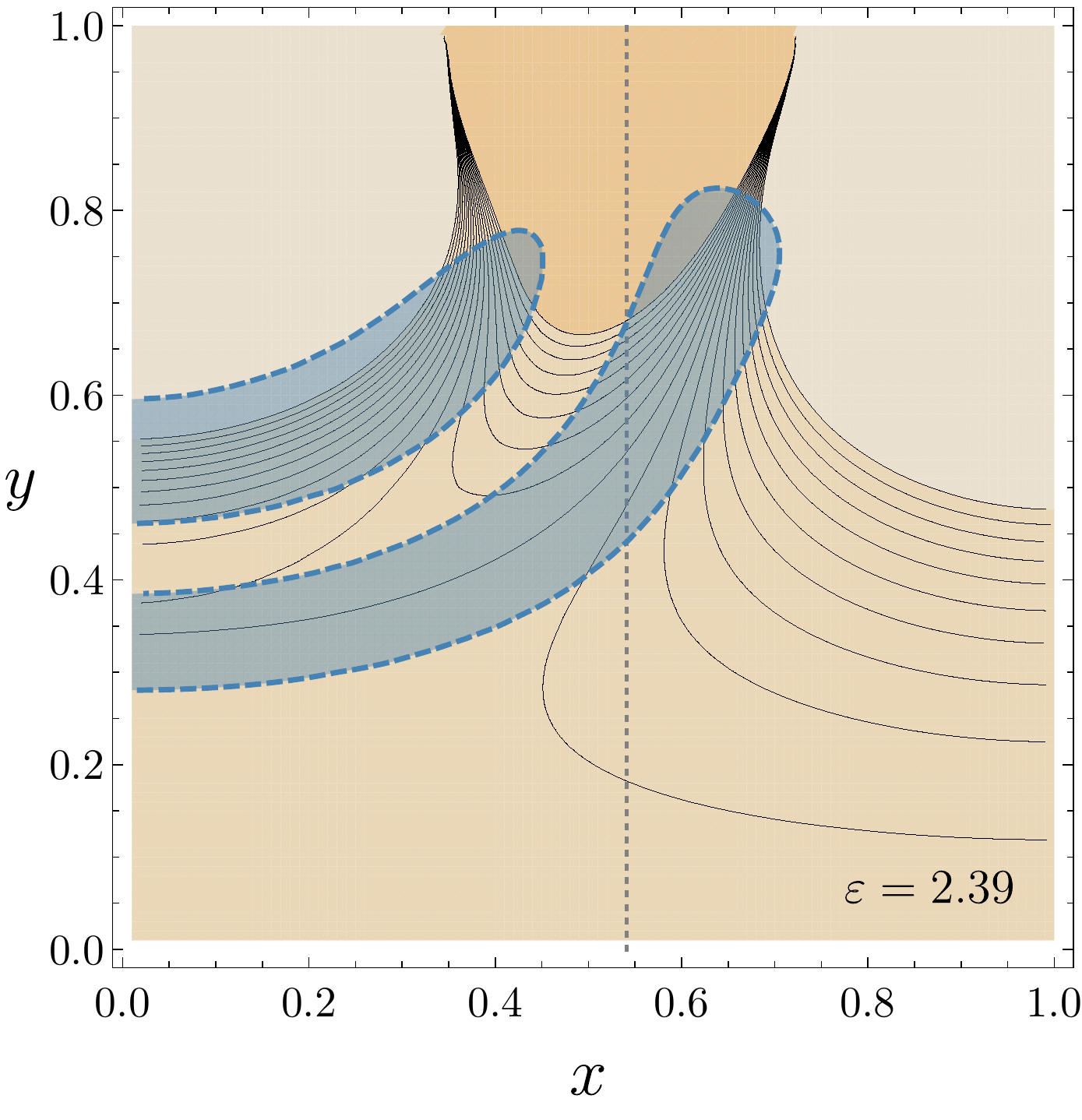}
  \end{minipage} 
  \caption{\textit{Left}: Minima of static potential, $-g_{tt}$, for the soliton against the boundary parameter. For $\varepsilon<2$, there is a global minima at $y=0$. For~${\varepsilon>2}$, the minima is only local, owing to the fact that there is an ergoregion attached to the boundary. For the second branch, the local minima is on the $x=0$ axis, and for $\varepsilon\gtrsim 2.43$, the minima is no longer local. \textit{Right}: Contour plot for $-g_{tt}$ for the second branch soliton, at $\varepsilon=2.39$. Black lines mark the contours, with the orange area around $\theta=\pi/4$ (black dashed gridline) showing the ergoregion, and the fainter orange being the positive areas, where the function increases rapidly. Overplotted in blue are the zero contours of the Kretchmann scalar, $K$, whilst the shaded areas are the two minima with a steep maxima between them.}
  \label{fig:mingtt}
\end{figure}

The evolution of $-g_{tt}$ with the boundary parameter is non-trivial, and is related to the changes in the Kretchmann scalar $K$. For the soliton with~$\varepsilon<2$, there is a global minima at $y=0$ (see the left panel of Fig.~\ref{fig:mingtt}, and~\textit{c.f.}~\cite{2017JHEP...06..024C},~Fig.~6). When $\varepsilon\geq 2$, $-g_{tt}$ becomes very negative near the boundary due to the ergoregion, and the central minima is then local. For the second branch, the minima moves up the equator towards the boundary, and as $\varepsilon\rightarrow 2$, the minima is found closer to where $K$ has the largest gradient. The ergosurface becomes visibly deformed towards the $x=0$ boundary, and the local minima eventually connects to the ergoregion at $\varepsilon\gtrsim 2.43$ (see the right panel of Fig.~\ref{fig:mingtt}). 
 
For the large hot black hole solutions, the minima of $-g_{tt}$ is on the horizon, tending to $\theta=\pi/4$ as $\varepsilon\rightarrow 2$. The corresponding small branch has a turning point at $\varepsilon>2$, where the entropy $S\rightarrow 0$ (see the right panel of Fig.~\ref{fig:ent}). The small black holes, for all temperatures, are similar to soliton solutions. When the black hole entropy is small, $-g_{tt}$ looks very similar to the soliton (see the right panel of Fig.~\ref{fig:mingtt}), but with the local minimum along the horizon.

Finally, we comment on the stability of the test particles. For the stable soliton branch ($\varepsilon>2$), the (spinning) particles can be put at the global minimum of the static potential, however, such solutions are not thermodynamically favoured. Let us place a small (rotating) black hole at the origin, approximated by a small Kerr black hole. If we parametrise the thermodynamic quantities by the outer horizon radius $r_+$, and temperature~$T$, the entropy is given by
\begin{equation}
S=\frac{2 \pi r_+^2}{4 \pi r_+ T+1}.
\end{equation}
In the grand-canonical ensemble, at finite temperature $T$ and fixed chemical potential $\mu$, we find that  the change in the free energy is $\delta G=M E+\mathcal{O}(r_+^2)$, and thus adding (spinning) particles is not advantageous\footnote{In the probe limit $r_+\rightarrow 0$.}. In order to achieve a preferred phase, we would like to have $V_{\mathrm{min}}<0$, however, we find that the energy $E/M$ is positive\footnote{The ratio is also independent of the spin sign, which agrees with the sign of $S$.}. This is in agreement with the fact that the small black hole phase is always hidden by the soliton phase. 

Once we cross $\varepsilon=2$, any particle will eventually fall into the ergoregion near the boundary. Similarly, this should happen for the small black holes which exist past the critical value $\varepsilon=2$.

It is interesting to contrast these results with a polarised black hole in global AdS$_4$ \cite{Costa:2015gol,2016CQGra..33k5011C}. In these papers, solutions with a dipolar chemical potential on the boundary are considered. For the black holes, they find static charged particle orbits, which could allow for multi-black hole solutions. In our scenario, we find that the purely gravitational on--axis ``spin--spin'' interaction is not sufficient to support stationary solutions with more than one black hole.
%%%%%%%%%%%%%%%%%%%%%%%%%
%%%%%%%%%%%%%%%%%%%%%%%%%
\section{\label{sec:noergo}No Ergoregions}
%%%%%%%%%%%%%%%%%%%%%%%%%
%%%%%%%%%%%%%%%%%%%%%%%%%
\begin{figure}[t]
\centering
    \includegraphics[width=0.45\textwidth]{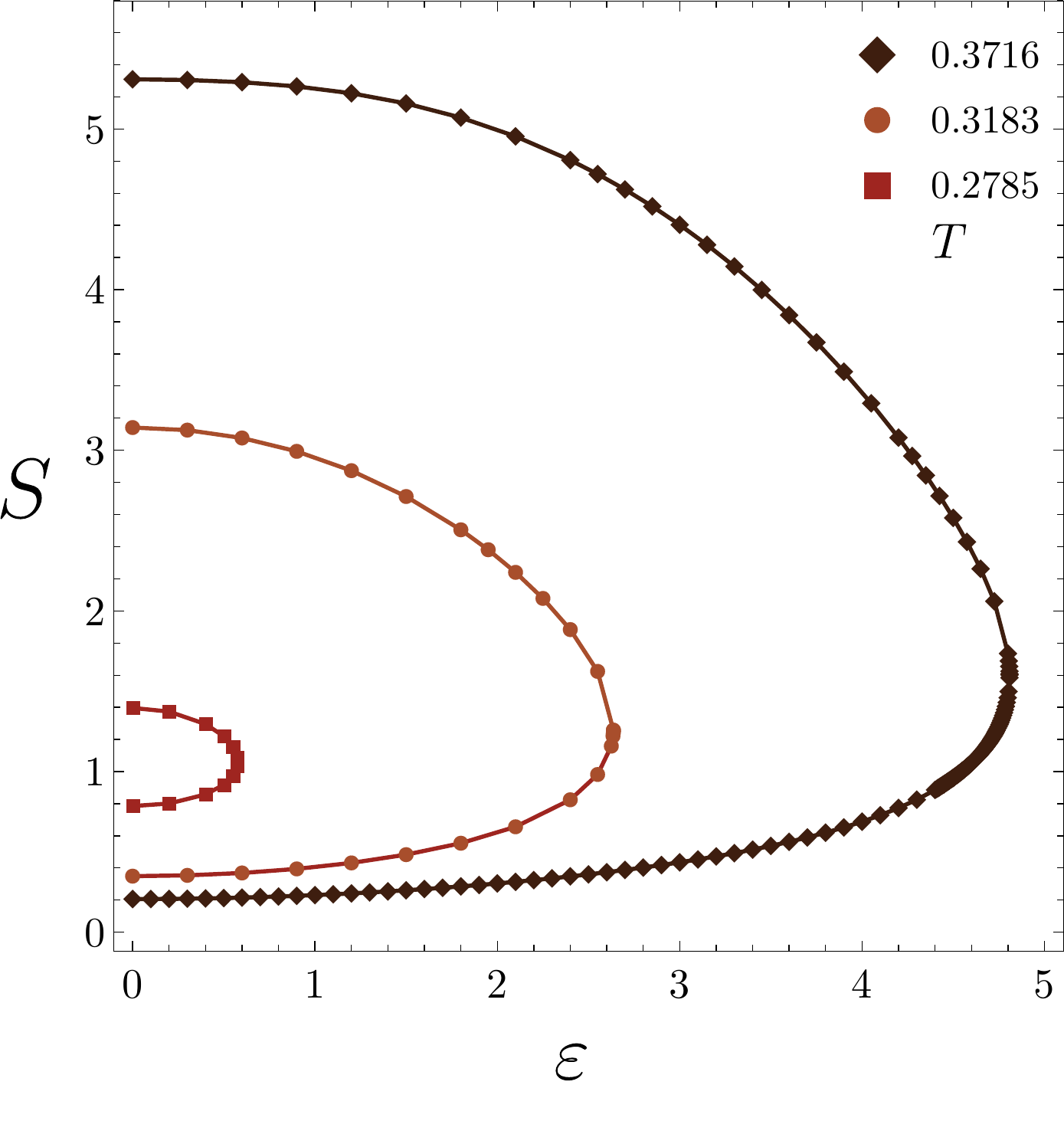}
  \caption{Entropy vs the boundary rotation parameter, when the metric is adjusted so that there is no ergoregion on the boundary. We observe two branches of black hole solutions which join up at some maximum value of $\varepsilon$.}
  \label{fig:phasediag}
\end{figure}

The instabilities that we observe in the rotationally polarised AdS$_4$ system are primarily due to the induced ergoregion. It is interesting to ask, whether we still see instabilities if we isolate the boundary interaction term in the metric, in such a way that it does not cause the ergoregion to form. We consider the adjustment to the line elements (\ref{eq:ansatzsol}) and (\ref{eq:ansatzbh}), such that~$g_{tt}$ is always negative near the boundary, by choosing the reference metric in (\ref{eq:deturck}) to have $Q_1(x,y)= 1 + \varepsilon^2 x^2 (1 - x^2)^2  (2 - x^2) y^2$. In this case, at any given fixed temperature $T>T^{\mathrm{Schw}}_{\min}=\sqrt{3}/(2\pi)$, we also find a maximum rotation amplitude past which we do not find  stationary black hole solutions~(see right panel of Fig.~\ref{fig:phasediag}). As the system heats up, this value of $\varepsilon$ appears to increase without a bound. We also find a soliton solution, which appears to exist for any value of the rotation amplitude\footnote{The Kretschmann scalar exhibits a complicated structure, and for large $\varepsilon$ has a slowly increasing central maximum, and two opposite sign extrema on the $x=0$, approaching the boundary.}. We have been able to reach $\varepsilon=32$, and believe that the solution exists for any value of $\varepsilon$.

%%%%%%%%%%%%%%%%%%%%%%%%%
%%%%%%%%%%%%%%%%%%%%%%%%%
\section{\label{sec:con}Conclusions and Outlook}

We have used the gauge/gravity duality to study the behaviour of a strongly coupled field theory on a fixed rotating spacetime. Since our bulk field theory depends only on the metric, it lies in the universal sector of AdS/CFT. In particular, it can be embedded into a full top-down model for AdS/CFT, such as ABJM \cite{Aharony:2008ug}. We restricted attention to boundary spacetimes taking the following form
$$
\mathrm{d}s^2_\partial=-\mathrm{d}t^2+\mathrm{d}\theta^2+\sin^2\theta(\mathrm{d}\phi+\varepsilon\,\cos \theta\,\mathrm{d}t)^2\,,
$$
and studied the behaviour of several one point functions, such as the expectation value of the stress energy tensor, for several values of $\varepsilon$. We have considered both the vacuum state and the thermal state for each value of $\varepsilon$. The holographic dual of the latter contains a black hole solution in the interior of the bulk spacetime, setting the temperature of the field theory \cite{Witten:1998qj}.

We have seen that both the thermal and the vacuum states are non-unique; for a fixed angular boundary profile and fixed energy, more than one solution exists in the bulk. For boundary profiles containing ergoregions, we have observed a rather interesting behaviour. Specifically, singular black hole solutions exist, extending from the bulk all the way to the boundary, whenever the boundary profile has an evanescent ergosurface, \emph{i.e.} when $\partial/\partial t$ becomes null along the equator at the boundary, but is everywhere else timelike. Furthermore, we have given ample numerical evidence that above a certain value of ${\varepsilon=\varepsilon_{c}\simeq 2.565}$, no axially symmetric stationary bulk solution exists. This value happens to always be larger than the critical value needed to generate hair. Finally, we provided a first principle derivation of the superradiant bound, \emph{i.e.} we determined the minimum value of $\varepsilon$ that makes our system develop hair, purely from CFT data.

The most interesting question raised by our work does not yet have a definite answer. Namely, what happens if we start in the vacuum of the theory, \emph{i.e.} pure global AdS$_4$, and promote $\varepsilon$ to be a function of time. In the most interesting scenario, we would increase $\varepsilon$ to a value above $\varepsilon_c$ and monitor the bulk dynamical evolution. We have provided numerical evidence that there are no axially symmetric and stationary solutions above this critical value. We have further seen that both the solitonic and the black hole phases can develop hair for $\varepsilon>2$. There are a number of possibilities for the nonlinear evolution of our system:
\begin{enumerate}
\item The superradiant instability quickly controls the dynamics, and the subsequent evolution proceeds just as described in \cite{Dias:2011at,Dias:2015rxy,Niehoff:2015oga}.
\item A bulk horizon forms, becomes hot and expands all the way to the boundary, reaching the boundary in a finite amount of time. This is reminiscent of what was observed in \cite{Bosch:2017ccw,Buchel:2017map}.
\item A bulk horizon forms, the curvature grows large in the bulk as a power law in time, until the superradiant instability kicks in and controls the dynamics. This scenario is very similar to the one presented in \cite{Crisford:2017zpi,2017arXiv170907880C}.
\item No horizon forms, and the curvature grows large in the bulk as a power law in time, until the superradiant instability kicks in and controls the dynamics. This scenario is again very similar to the one presented in \cite{Crisford:2017zpi,2017arXiv170907880C}.
\end{enumerate}
We do not know which one of these possibilities is most likely, but were possibility 2 to be the correct one, we would suspect that for our class of boundary metrics no positivity of energy theorem could be proven.

There are several extensions of our work that deserve attention. First, it would be interesting to consider a generalisation of this work to the Poincar\'e patch of AdS$_4$. Work in this direction will be presented elsewhere. Second, it would be interesting to understand whether the behaviour we observe depends on the rotation profile we choose. We think that most of what we reported should not be dependent on the choice of the boundary profile. For instance, we certainly believe that no solution should exist for arbitrarily large amplitudes of the boundary profile, so long as $\partial_t$ becomes spacelike.

%%%%%%%%%%%%%%%%%%%%%%%%%
%%%%%%%%%%%%%%%%%%%%%%%%%
\acknowledgments

JM is supported by an STFC studentship. JES was supported in part by STFC grants PHY-1504541 and ST/P000681/1. It is a pleasure to thank J.~Penedones for collaboration at an early stage of this work. The authors would like to thank C. R. V. Board and \'O. J. C. Dias for reading an earlier version of this manuscript. JES would like to thank M.~S.~Costa, L.~Greenspan and G.~T.~Horowitz for helpful discussions. The authors thankfully acknowledge the computer resources, technical expertise and assistance provided by CENTRA/IST. Part of the computations were performed at the cluster €œBaltasar-Sete-S\'ois€ and supported by the H2020 ERC Consolidator Grant ``Matter and strong field gravity: New frontiers in Einstein's theory" grant agreement no. MaGRaTh-646597. Part of this work was undertaken on the COSMOS Shared Memory system at DAMTP, University of Cambridge operated on behalf of the STFC DiRAC HPC Facility. This equipment is funded by BIS National E-infrastructure capital grant ST/J005673/1 and STFC grants ST/H008586/1, ST/K00333X/1.
 
\appendix
\section{\label{sec:numval}Numerical validity}

We verified that our solutions satisfy $\xi^\mu=0$ to sufficient precision. 
In order to compute the stress energy tensor, adequate precision is required as we need to obtain third derivatives about the conformal boundary. We found that double precision was adequate thus significantly speeding up our computation. Nevertheless, we checked that the results converge as expected when the precision is increased.

\begin{figure}[t]
\centering
  \begin{minipage}[t]{0.45\textwidth}
    \includegraphics[width=\textwidth]{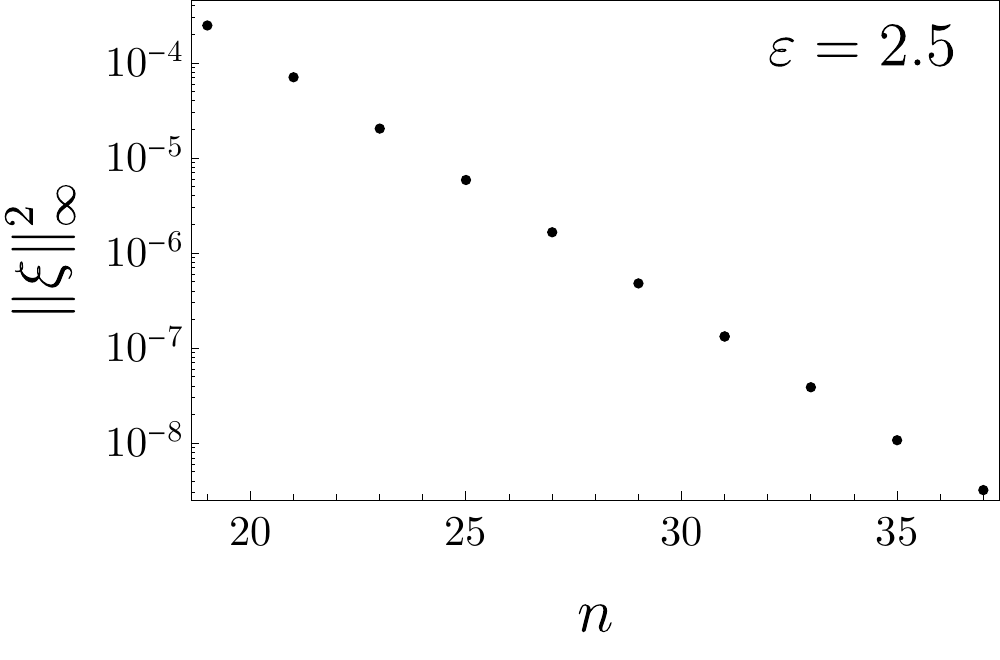}
  \end{minipage}
  \hfill
    \begin{minipage}[t]{0.45\textwidth}
    \includegraphics[width=\textwidth]{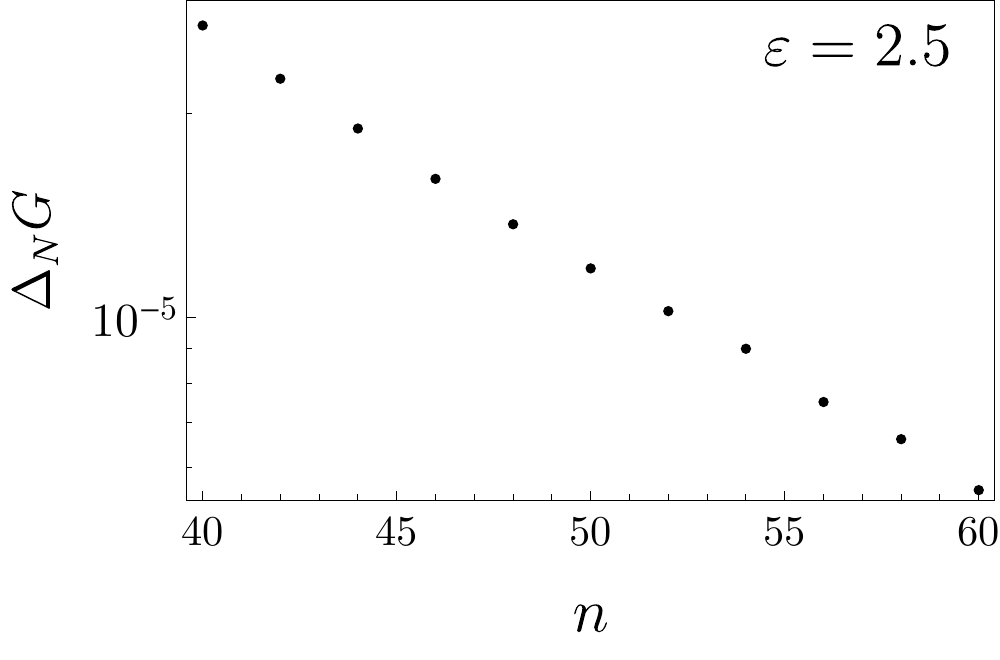}
  \end{minipage}
    \begin{minipage}[t]{0.45\textwidth}
    \includegraphics[width=\textwidth]{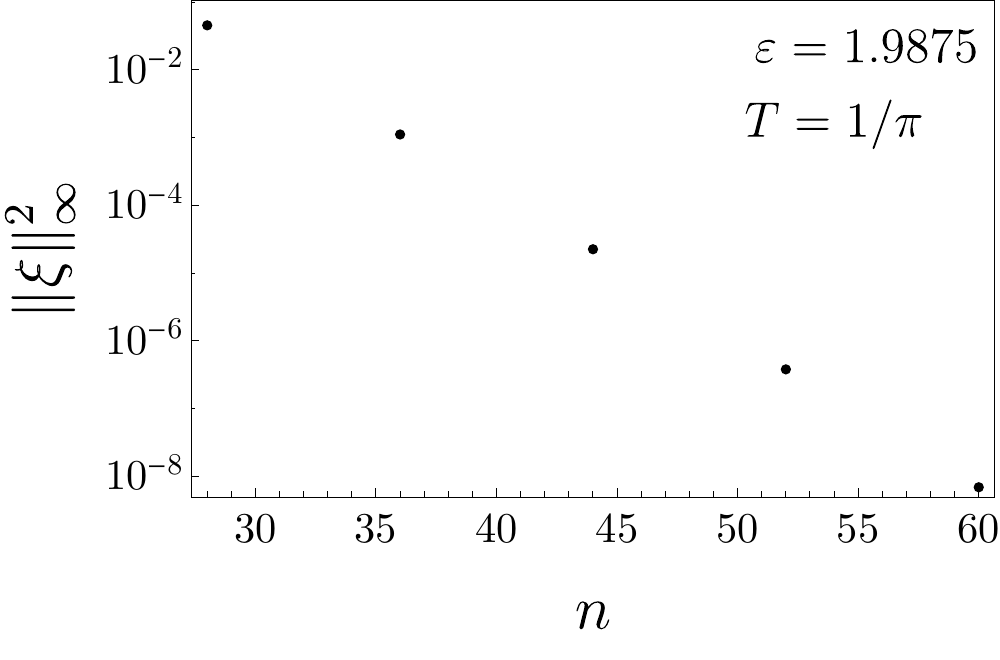}
  \end{minipage}
  \hfill
    \begin{minipage}[t]{0.45\textwidth}
    \includegraphics[width=\textwidth]{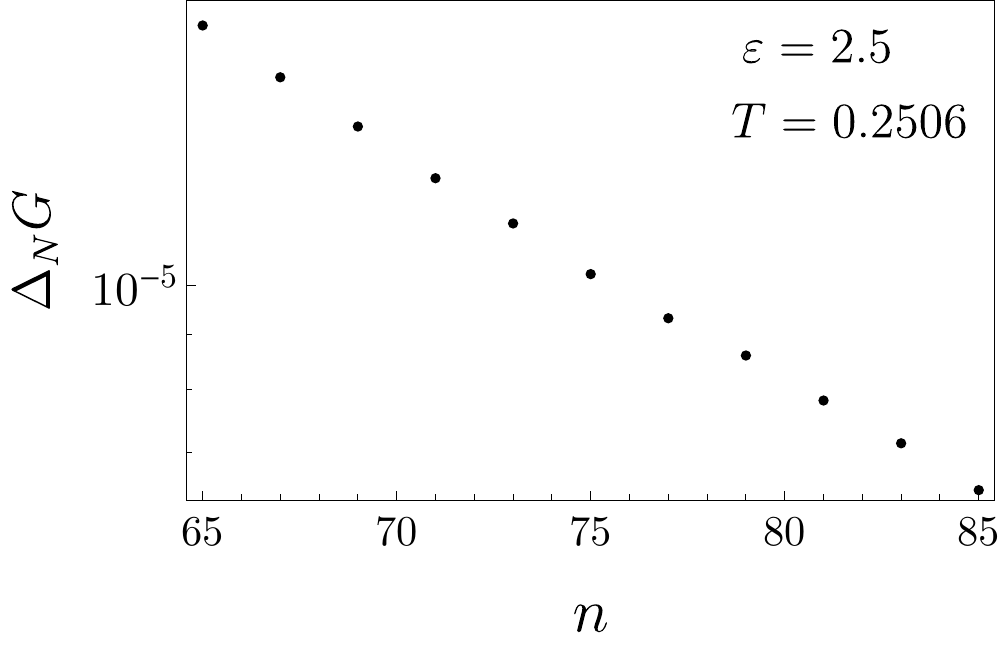}
  \end{minipage}
  \caption{\textit{Top left}: Log--linear plot of the DeTurck norm squared for the dominating soliton branch, at $\varepsilon=2.5$. Straight line indicates exponential convergence. \textit{Top right}: Log--linear plot of the relative error in the free energy $G$ vs the grid size $n$,  for the same data set as top left. \textit{Bottom left:} Log--linear plot of the DeTurck norm squared vs grid size for a large black hole with $T=1/\pi$ and $\varepsilon=1.9875$. \textit{Bottom right:} Log--linear plot of $\Delta_N G$ against $n$, for a cold ($T=0.2506$), small black hole, upper small branch ($\varepsilon=2.5$).}
  \label{fig:solc}
\end{figure}

We monitor convergence of our numerical method by computing the infinity norm of the DeTurck vector $\norm{\xi}_\infty$, as well as relative errors of physical quantities of interest defined by $\Delta_N G=\left|1-G_{N+1}/G_N\right|$, where $G_N$ denotes the quantity computed with $N$ grid points on each integration domain. This provides a good check of the numerics, as well as a measure of the error. We present convergence results for free energy in Fig.~\ref{fig:solc} which, in this work, is the physical quantity which exhibits the largest error. As~$\varepsilon$ increases, the grid size has to be increased accordingly to maintain satisfactory error and typically a $60\times 60$ to $100\times 100$ grid was used. For the \textit{dominant} soliton branch, $\norm{\xi}^2_\infty$ is never above $10^{-10}$ and errors in quantities never above $0.01\%$. For the \textit{large} branch, $\norm{\xi}^2_\infty < 10^{-7}$, and the errors are below $1\%$. For the \textit{large} black holes, we keep $\norm{\xi}^2_\infty < 10^{-9}$ and $\Delta_N G<0.01\%$, and for the \textit{small} branch, $10^{-7}$ and $1\%$ respectively. Generally, we found that errors increase with both increasing and decreasing the temperature. 

For analytic functions, we expect the pseudospectral methods to offer exponential convergence with an increasing grid size. However, we cannot attain these rates of convergence if the error drops below the machine precision. The error settles to an exponential decay until it starts to pivot around the machine error, and this interval is shown in Fig.~\ref{fig:solc}. In order to improve convergence it is then necessary to increase numerical precision.

\section{\label{sec:add}Additional figures} 

 \begin{figure}[t]
\centering
  \begin{minipage}[t]{0.45\textwidth}
    \includegraphics[width=\textwidth]{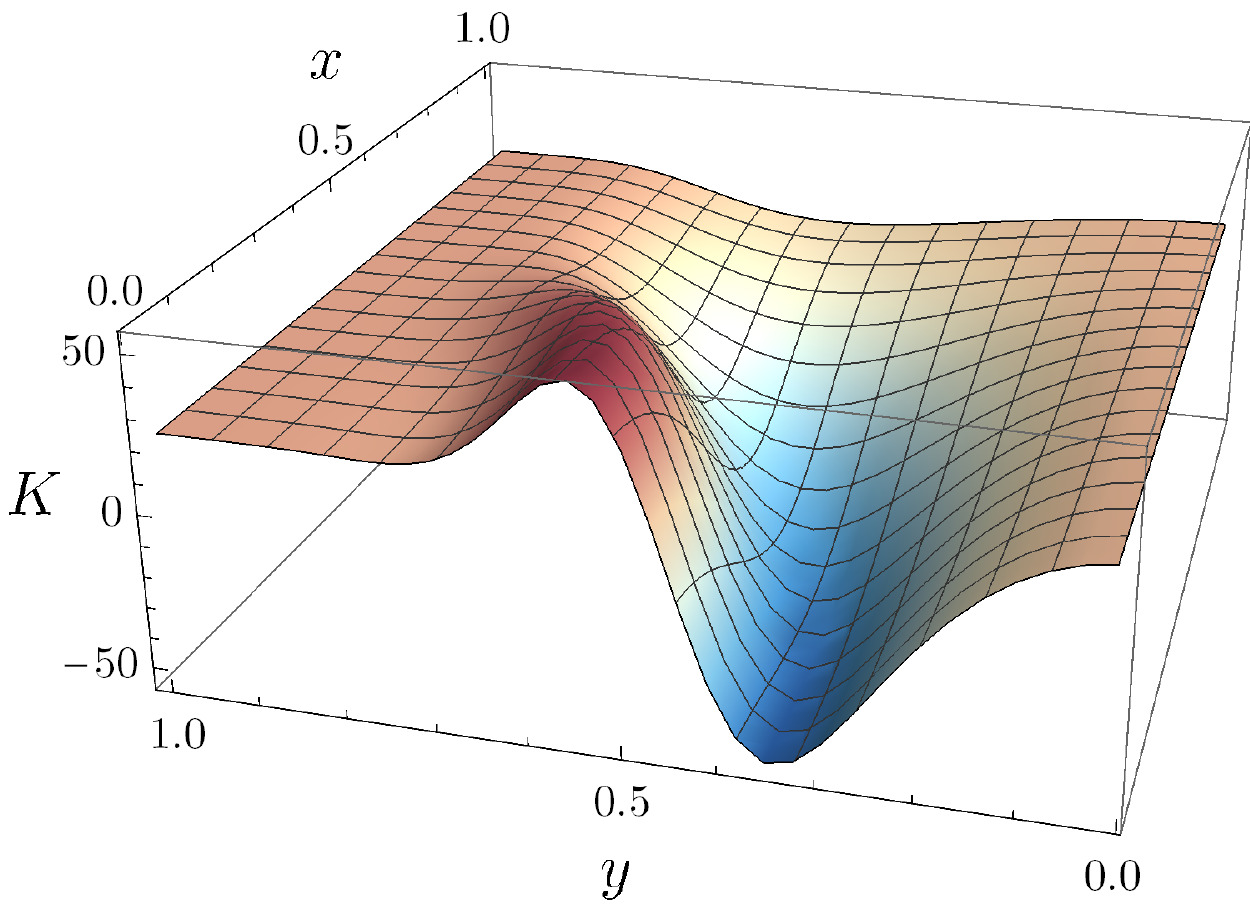}
  \end{minipage}
  \hfill
    \begin{minipage}[t]{0.45\textwidth}
    \includegraphics[width=\textwidth]{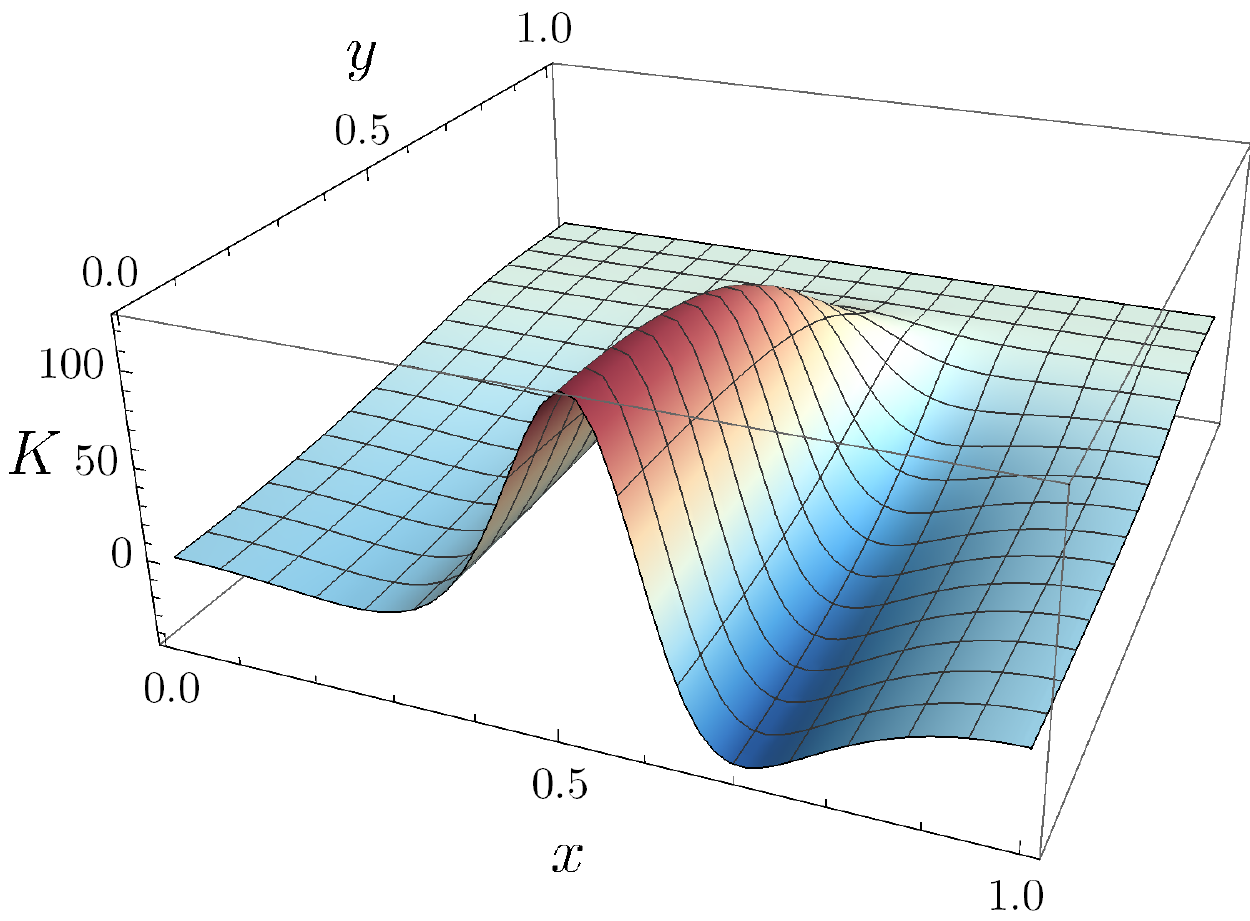}
  \end{minipage}
  
  \caption{\textit{Left}: Kretschmann scalar $K$ for the dominant soliton branch, $\varepsilon=2.564$. \textit{Right}: $K$ for a large black hole with $\varepsilon=1.9$.}
  \label{fig:kretfull}
\end{figure}

In Fig.~\ref{fig:kretfull}, we present the Kretschmann scalar $K$ for a fixed value of the boundary parameter $\varepsilon\simeq\varepsilon_c$, where $\varepsilon_c$ is the maximal value up to which stationary solutions exist. The evolution of $K$ with the boundary deformation is non-trivial. In the left panel, we plot $K$ for the small soliton branch. For $\varepsilon < 1.9$, the scalar has a positive minima at $y=0$, and for $\varepsilon>1.9$, it starts to display interesting features: the scalar is maximal at the equator $x=0$, displaying two large extrema close to each other, the magnitude of which are growing rapidly with the increase of the boundary rotation amplitude. At the boundary ${y=1}$, it reduces to the value of $K_{AdS}=24$, as expected for the asymptotically AdS$_4$ spacetimes. We also observe a local slowly increasing maxima at $y=0$, when $\varepsilon$ is large. The large, unstable, soliton branch shows a similar behavior, with magnitudes of the both extrema growing further and the distance between them decreasing, and the extrema are slowly moving towards the origin, as $\varepsilon\rightarrow 2$. There are also now two minima on each side of the maximum.

 \begin{figure}[t]
\centering
    \begin{minipage}[t]{0.45\textwidth}
    \includegraphics[width=\textwidth]{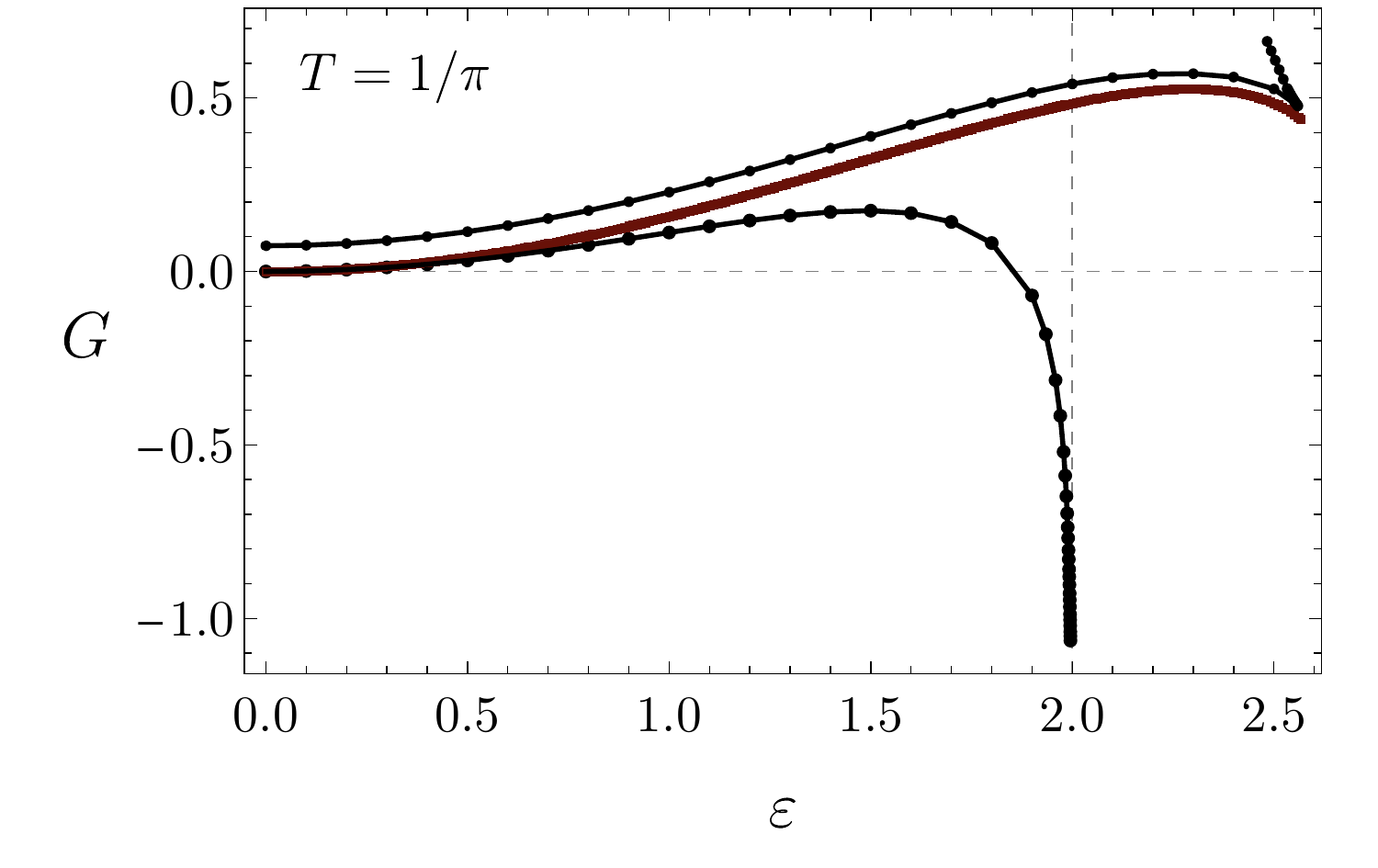}
  \end{minipage}
  \hfill
    \begin{minipage}[t]{0.45\textwidth}
    \includegraphics[width=\textwidth]{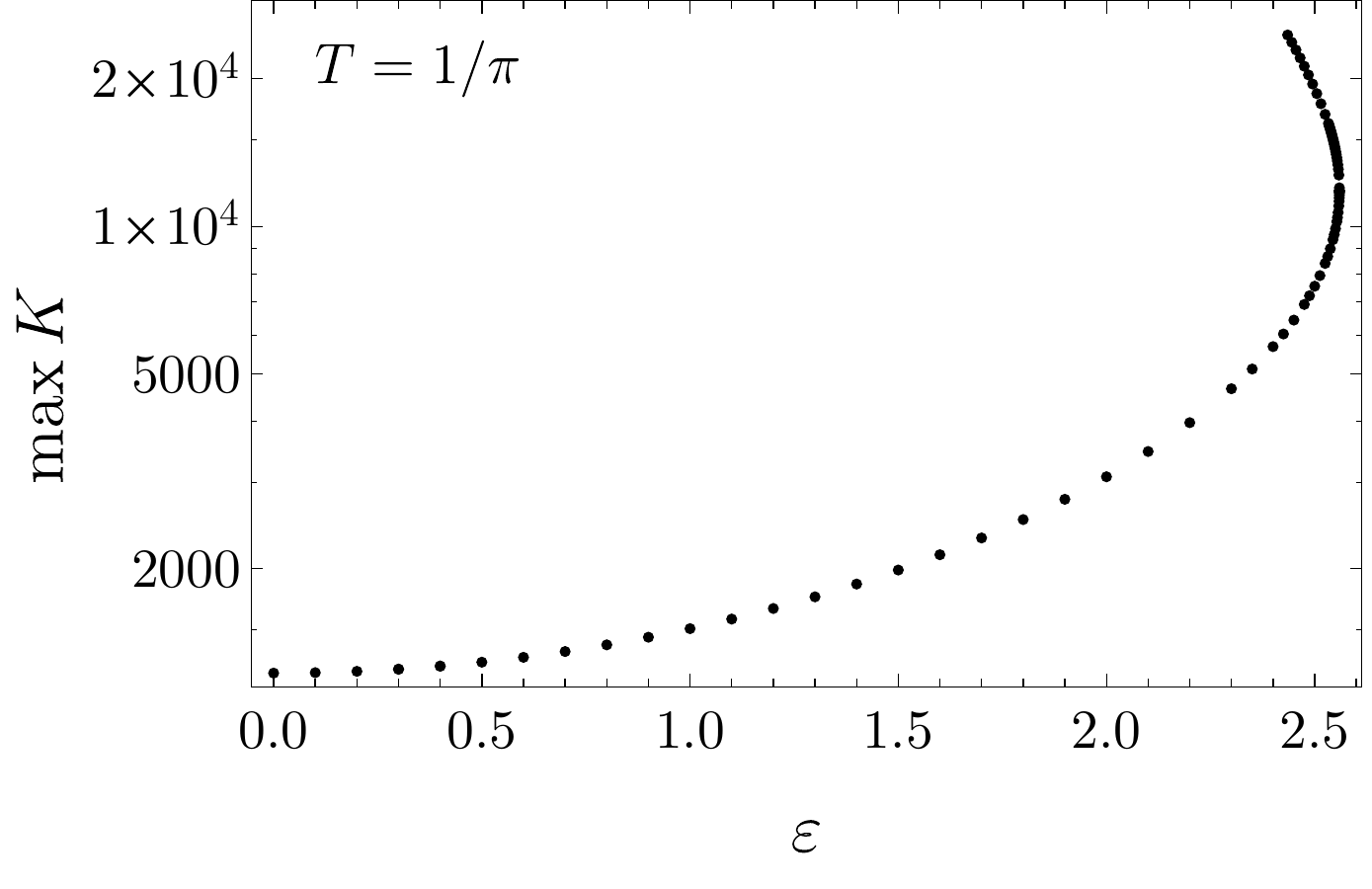}
  \end{minipage}
  \caption{\textit{Left}: Gibbs free energy against the boundary profile for black holes with a fixed temperature $T=1/\pi$ (black disks), and soliton (brown squares). The dashed gridline marks the $\varepsilon\rightarrow 2$ limit. \textit{Right}: Maximum of the Kretschmann invariant against $\varepsilon$ for the small black hole with a fixed temperature. As $\mathrm{max}\,K$ increases, the $K$ looks very similar to that of the soliton.}
  \label{fig:massS}
\end{figure}

 \begin{figure}[t]
\centering
  \begin{minipage}[t]{1\textwidth}
    \includegraphics[width=\textwidth]{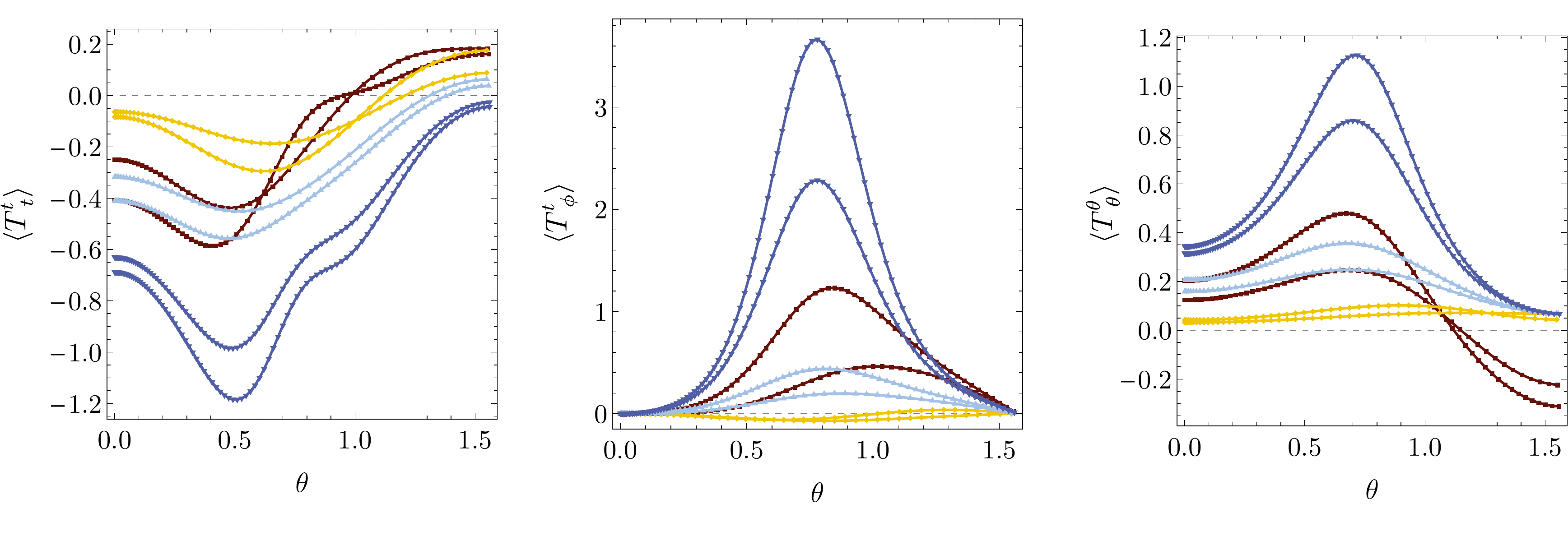}
  \end{minipage}
  \caption{Boundary energy-stress tensor components for a fixed low temperature (${T=0.2506}$), where there are four possible black hole phases: lowest entropy black hole branch (brown squares), small black hole branch (yellow diamonds), thermodynamically dominant middle branch (faint blue triangles) and largest entropy large black holes (blue upside-down triangles).}
  \label{fig:lowstress}
\end{figure}

In the right panel we present the curvature scalar for the large black hole with $\varepsilon=1.9$ and $T=1/\pi$. It is maximal on the black hole horizon $y=1$, and is peaking at $\theta\rightarrow\pi/4$ as $\varepsilon\rightarrow 2$. On either side of the maximum there are two minima one of which is global. As we approach the critical amplitude, the peak, and the minima, become narrower. For~$\varepsilon$ sufficiently large, the $K$ is small and positive on the poles and the equator.

\section{\label{sec:per}Perturbative expansion}

In this section we present the perturbative expansion of the solitonic solution to order $\varepsilon^2$, where~$\varepsilon$ is the boundary rotation parameter defined in section~\ref{sec:setup}. The following procedure can be easily generalized to yield higher orders of the expansion, however the solutions become increasingly complicated.
We will work in null quasi-spherical gauge~\cite{Bartnik:1996js} in which the solution reads
\begin{align}
\begin{split}
\mathrm{d}s^{2} = -\left(1+\frac{r^2}{L^2}\right)Q_1(r,\theta)\,\mathrm{d}t^2 & +\left(1+\frac{r^2}{L^2}\right)^{-1}Q_2(r,\theta)\,\mathrm{d}r^2 \\
& +r^2Q_3(r,\theta)\left[\mathrm{d}\theta^2+\sin{\theta}^2\left(\mathrm{d}\phi+\Omega(r,\theta)\,\mathrm{d}t\right)^2\right]
\end{split}
\label{eq:ansatzpert}
\end{align}
where $\{\theta,\phi\}$ are the standard polar coordinates of unit two-sphere, and $L$ is the AdS$_4$ length. The gauge requires that the functions $Q_i$ ($i=1,2,3$) and $\Omega$ depend only on $r$ and~$\theta$, and we will also use the fact that if $Q_i=Q_i(r)$ (and $\Omega=0$), we can additionally fix $Q_3=1$. As the induced stress-energy tensor is of order $\varepsilon^2$ in the angular velocity, consider an expansion in $\varepsilon$ given by
\begin{align}
Q_i(r,\theta)=\sum_{i=0}^{+\infty}\varepsilon^{2i}q_i^{(2i)}(r,\theta),\qquad\qquad \Omega(r,\theta)=\sum_{i=0}^{+\infty}\varepsilon^{2i+1}\Omega^{(2i+1)}(r,\theta)
\end{align}
where we are expanding around a pure AdS$_4$ background given by $q_i^{(0)}=1$. At linear order we obtain a second order partial differential equation for $\Omega^{(1)}$ which can be solved using separation of variables subject to regularity at the origin. The solution is given by
\begin{align}
\Omega^{(1)}(r,\theta)=\frac{2}{3\sqrt{\pi }}\sum_{l=0}^{+\infty} c^{(1)}_l\frac{\Gamma \left(\frac{l+3}{2}\right) \Gamma \left(\frac{l+5}{2}			\right)}{\Gamma\left(l+\frac{5}{2}\right)}\,
P_{l+1}'(\cos{\theta})\left(\frac{r}{L}\right)^{l}\!_2F_1\left(\frac{l}{2},\frac{l+3}{2};l+\frac{5}{2};-\frac{r^2}{L^2}\right),
\end{align}
where $l$ is the harmonic number, $c^{(1)}_l$ are real numbers depending on the boundary profile, $P_l$ is Legendre polynomial of degree $l$ and $\!\,_2F_1$ is ordinary hypergeometric function. At the boundary the perturbation reduces to
\begin{align}
\lim_{r\rightarrow +\infty}\Omega^{(1)}(r,\theta) = \sum_{l=0}^{+\infty} c^{(1)}_l P_{l+1}'(\cos{\theta}),
\end{align}
and at first order the first harmonic $l=1$ gives us the dipolar boundary rotation profile ${\Omega(r,\theta)=\varepsilon \cos{\theta}}$. The final expression is
\begin{align}
\Omega^{(1)}(r,\theta)=\frac{2 L^4}{\pi  r^4}\cos{\theta} \left[\frac{r}{L}\left(\frac{r^2}{L^2}+3\right)+\left(\frac{r^2}{L^2}-3\right) \left(\frac{r^2}{L^2}+1\right) \arctan\left(\frac{r}{L}\right)\right].
\end{align}
The first order angular velocity perturbation effectively sources the stress energy tensor at~$\varepsilon^2$. We decompose the second order perturbations as~\cite{Kodama:2003jz}
\begin{align}
q_i^{(2)}(r,\theta)=\gamma_i(r)+\alpha_i(r)P_2(\cos{\theta})+\beta_i(r)P_4(\cos{\theta}),
\end{align}
where the remaining gauge freedom is used to set $\gamma_3(r)=0$. We solve the equations requiring that the functions are regular at the origin $r=0$ and, that up to a diffeomorphism, they asymptotically satisfy AdS$_4$ boundary conditions.
Below we provide the full perturbative functions:
 
\begin{align*}
\begin{split}
&\gamma_1(r)=\frac{2 L^2}{15 \pi ^2 r^6 (L^2+r^2)}\left.\bigg( 24 L^6r^2+20 L^4 r^4-9(\pi ^2-4) L^2 r^6-8 \left(6 L^7 r+7 L^5 r^3-6 			L^3 r^5\right.\right.\\
&\quad\bigg.\left.-9 L r^7\right)\arctan\left(\frac{r}{L}\right)+4 \left(6 L^8+9 			L^6 r^2+L^4 r^4+7 L^2 r^6+9r^8\right) \arctan\left(\frac{r}{L}\right)^2-9\pi ^2 r^8\bigg),\\
&\gamma_2(r)=-\frac{8 L^4}{15 \pi ^2 r^6(L^2+r^2)}\left((3 L^2-r^2)(L^2+r^2) \arctan\left(\frac{r}{L}\right)-L r(3L^2+r^2)\right)
\end{split}
\end{align*} 
\begin{align}
\begin{split}
&\hspace{+23em}\times\bigg((3 L^2+r^2) \arctan\left(\frac{r}{L}\right)-3 L r\bigg),\\
&\alpha_1(r)=\frac{L^2}{84 \pi ^2 r^6 (L^2+r^2)}\left.\bigg(96 L^6r^2+L^4 (-752+63 \pi ^2) r^4+7 L^2(-16+15 \pi ^2)r^6\right.\\
&\qquad+\arctan\left(\frac{r}{L}\right) \left(-192 L^7 r+7 L^5 (176-9 \pi^2) r^3+2 L^3(368-63 \pi ^2) r^5+L(80-63 \pi ^2) r^7 \right.\\
&\hspace{+12em}\Bigg. \bigg.+32(L^2+r^2) \left(3L^6-18 L^4 			r^2+L^2 r^4+6r^6\right) \arctan\left(\frac{r}{L}\right)\bigg)\Bigg),\\
&\alpha_2(r)=\frac{L^3}{84 \pi ^2 r^6 (L^2+r^2)}\left(3936 L^5 r^2+(4208-63 \pi ^2) L^3 r^4+\arctan\left(\frac{r}{L}\right) \left(-7872 L^6r+\right.\right.\\
&\qquad\left.(63 \pi ^2-10832) L^4 r^3+2(63 \pi^2-944)L^2 r^5+32 L(L^2+r^2) \left(123 L^4+84 L^2 r^2-23r^4\right)\right.\\
&\hspace{+14em}\Bigg.\bigg.\times\arctan\left(\frac{r}{L}\right)+(304+63 	\pi ^2)r^7\bigg)+(304-105 \pi ^2) Lr^6\Bigg),\\
&\alpha_3(r)=\frac{L^2}{84 \pi ^2 r^6}\left(-384 L^4 r^2-7  (64+9 \pi  ^2 ) L^2 r^4+\arctan\left(\frac{r}{L}\right) \left(768 L^5r+ (1360+63 \pi ^2 ) \right.\right.\\
&\times L^3 r^3-48 \left(8 L^6+19 L^4 r^2-10 L^2 r^4+3 r^6\right) \arctan\left(\frac{r}{L}\right)- (592+63 \pi ^2 ) L r^5\bigg)+84 \pi^2r^6\Bigg),\\
&\beta_1(r)=\frac{L^2}{1680 \pi ^2 r^6 (L^2+r^2 )}\left.\Bigg(3  (5408+735 \pi^2 ) L^6 r^2+38  (608+105 \pi ^2 ) L^4r^4\right.\\
&\qquad\quad+21 (224+81\pi^2 ) L^2 r^6+\arctan\left(\frac{r}{L}\right) \left(3 (1504-735 \pi ^2 ) L^7
    r+63 (352-75 \pi ^2 ) L^5 r^3\right.\\
    &\qquad    \left. +3 (7712-945 \pi ^2 ) L^3r^5-256(L^2+r^2 ) \left(81 L^6+144 L^4 r^2+48 L^2 					r^4+r^6\right) \arctan\left(\frac{r}{L}\right) \right.\\
&\hspace{+28.5em}\Bigg. + (4448-315 \pi ^2 ) 		L r^7\Bigg),\\    
& \beta_2(r)=\frac{L^3}{1680 \pi ^2 r^6 (L^2+r^2 )} \Bigg(-3  (32+735 	\pi ^2 ) L^5 r^2-2  (3104+1995 \pi ^2 ) L^3 r^4 \Bigg.\\
&\hspace{+11.5em}	+\arctan\left(\frac{r}{L}\right) \bigg(3  (735 \pi ^2-12256 ) L^6 r+15  (315 \pi^2-4448 ) 			L^4 r^3   \\
&\hspace{+9em}\left.+ (2835 \pi ^2-35936 ) L^2 r^5+256 L  (L^2+r^2 )\left(144 L^4+189 L^2 r^2+61 r^4\right) 	\right.\\
&\hspace{+12.7em}\Bigg.\bigg.\times \arctan					\left(\frac{r}{L}\right)+5 (63 \pi^2-992 ) r^7\bigg)- (4960+1701 \pi ^2 ) L r^6\Bigg),\\
&\beta_3(r)=\frac{L^2}{1680 \pi ^2 r^6}\left(-3 L^4  (7072+735 \pi ^2 ) 		r^2+7 L^2  (544-45 \pi ^2 ) r^4+224 \pi ^2 r^6\right.\\
&\qquad\quad+\arctan\left(\frac{r}{L}\right)\left(9 L^5  (608+245 \pi ^2 ) r+2 L^3 					 (-5536+525 \pi^2 ) r^3+L  (2656-315 \pi ^2 ) r^5\right.\\
&\hspace{+13em}\Bigg.\bigg.+128 \left(123 L^6+153 L^4 r^2-19 L^2 r^4-9r^6\right) \arctan			\left(\frac{r}{L}\right)\bigg)\Bigg).
\end{split}
\end{align} 
 Using this second order expansion we compute thermodynamic quantities, and test the first law~(see subsec.~\ref{subsec:Ther}). We also find equilibrium conditions for spinning test particles on the rotation axis $\theta=0$. For the metric~(\ref{eq:ansatzpert}), the spin to mass ratio~(see sec.~\ref{sec:eqv}) is given~by
%\begin{equation}
%\frac{S}{M}=\frac{\left(r^2+L^2\right)\partial_r Q_1(r,0)+2 r Q_1(r,0)}{2L%\sqrt{\left(r^2+L^2\right) Q_1(r,0)}\,\partial_r Q_4(r,0)}.
%\end{equation} 

\begin{equation}
\frac{S}{M}=\frac{\pi  r^6}{16 \varepsilon  L^2 \sqrt{r^2+1} \left[\left(3 L^2+r^2\right) \tan^{-1}\left(r/L\right)-3 L r\right]}+\mathcal{O}(\varepsilon),
\label{eq:smpert}
\end{equation}
and vanishes when $r\rightarrow 0$. The comparison with the full non-linear numerical results are presented in Fig.~\ref{fig:smratio}.

%%%%%%%%%%%%%

%%%%%%%%%%%%%

%\nocite{*}
\bibliography{stirring}{}
\bibliographystyle{JHEP}
\end{document}